\pdfoutput=1
\documentclass{JHEP3}
\usepackage[utf8]{inputenc}
\usepackage{amstext}
\usepackage{graphicx}
\usepackage{amssymb}
\usepackage{esint}

\newcommand{\rL}{\rho_{\Lambda}}
\newcommand{\rLi}{\rho_{\Lambda}^{i}}
\newcommand{\rLe}{\rho_{\Lambda{\rm eff}}}
\newcommand{\pLe}{p_{\Lambda{\rm eff}}}
\newcommand{\rind}{\rho_{\rm ind}}
\newcommand{\rF}{\rho_{F}}
\newcommand{\pF}{p_{F}}
\newcommand{\G}{{\cal G}}
\newcommand{\F}{{\cal F}}
\newcommand{\M}{{\cal M}}
\newcommand{\E}{{\cal E}}
\newcommand{\rLo}{\rho_{\Lambda}^0}
\newcommand{\CC}{\Lambda}
\newcommand{\rM}{\rho_m}
\newcommand{\rR}{\rho_r}
\newcommand{\weff}{\omega_{\rm eff}}
\newcommand{\wF}{\omega_{F}}
\newcommand{\rLV}{\rho_{\CC\rm vac}}
\newcommand{\rLVo}{\rho_{\CC\rm vac\,0}}
\newcommand{\BCC}{\Lambda_{\rm vac}}
\newcommand{\VP}{\langle V(\varphi)\rangle}
\newcommand{\rV}{\rho_{\rm vac}}
\newcommand{\rLI}{\rho_{\CC\rm ind}}
\newcommand{\EP}{V_{\rm eff}}
\newcommand{\tEP}{{V}_{\rm scal}}
\newcommand{\rLP}{\rho_{\CC{\rm ph}}}
\newcommand{\EPR}{V_{\rm eff}}
\newcommand{\VEPo}{\langle V_{\rm eff\,0}(\varphi)\rangle}
\newcommand{\ZPE}{V_{\rm ZPE}}
\newcommand{\ZPEo}{V_{\rm ZPE\,0}}
\newcommand{\Hp}{\cal H}
\newcommand{\rX}{\rho_X}
\newcommand{\pX}{p_X}
\newcommand{\wX}{\omega_X}

\newcommand{\pL}{p_{\CC}}
\newcommand{\newtext}[1]{\text{#1}}
\newcommand{\revisedtext}[1]{\text{#1}}
\newcommand{\newnewtext}[1]{\text{#1}}
\newcommand{\newfinal}[1]{\text{#1}}
\newcommand{\newnewfinal}[1]{\text{#1}}
\title{Dynamically avoiding fine-tuning the cosmological constant: the ``Relaxed Universe''}

\author{%
Florian Bauer$^a$, Joan Solà$^a$, Hrvoje Štefancic$^b$\\
$^{a}$ High Energy Physics Group, Dept.\ ECM, and Institut
de Ciències del Cosmos\\
Univ.\ de Barcelona, Av.\ Diagonal 647, E-08028 Barcelona,
Catalonia, Spain\\
$^{b}$ Theoretical Physics Division, Rudjer Boškovic Institute, PO Box 180, HR-10002 Zagreb, Croatia\\
E-Mail: \email{fbauer@ecm.ub.es}, \email{sola@ecm.ub.es}, \email{shrvoje@thphys.irb.hr}
}

\abstract{%
We demonstrate that there exists a large class of $\F(R,\G)$ action
functionals of the scalar curvature and of the Gau\ss-Bonnet
invariant which are able to relax \textit{dynamically} a large
cosmological constant (CC), whatever it be its starting value in the
early universe. Hence, it is possible to understand, \textit{without
fine-tuning}, the very small current value $\Lambda_0\sim H_0^2$ of
the CC as compared to its theoretically expected large value in
quantum field theory and string theory. In our framework, this
relaxation appears as a pure gravitational effect,
\newnewfinal{where} no \textit{ad hoc} scalar fields are needed. The
action involves a positive power of a characteristic mass parameter,
$\M$, whose value can be, interestingly enough, of the order of a
typical particle physics mass of the Standard Model of the strong
and electroweak interactions or extensions thereof, including the
neutrino mass. The model universe emerging from this scenario (the
``Relaxed Universe'') falls within the class of the so-called
$\CC$XCDM models of the cosmic evolution. \newfinal{Therefore},
there is a ``cosmon'' entity $X$ (\newfinal{represented} by an
effective object, not a field), which in this case is generated by
the effective functional $\F(R,\G)$ and is responsible for the
dynamical adjustment of the cosmological constant. This model
universe successfully mimics the essential past epochs of the
standard (or ``concordance'') cosmological model ($\Lambda$CDM).
\newfinal{Furthermore},
\newtext{it provides interesting clues to the coincidence problem} and it may
even connect naturally with primordial inflation. }

\preprint{}

\begin{document}

\section{Introduction}

The status of cosmology as a quantitative subdiscipline of physics
is largely based on the high-quality observations made during the
last two decades. The availability of these data enabled
cosmologists to demonstrate that cosmological models are able to
explain not only qualitative features of the universe and its
dynamics, but also quantitatively reproduce the details of the many
observed cosmic phenomena\,\cite{cosmdata}. Yet, the discovery of
the accelerated expansion of the universe at the present epoch
\cite{SNIa} revealed a gap in our understanding of the universe. The
search for the underlying mechanism producing the accelerated
expansion revived an old problem of theoretical physics: the
cosmological constant problem\,\cite{weinberg89,CCproblem}, which
manifests itself as the double conundrum  on the tiny observed value
of the cosmological constant (CC) in Einstein's equations -- the
``old CC problem''\,\newtext{\cite{Zeldovich67}} -- and the puzzling
fact that this value is so close to the current matter density --
the ``cosmic coincidence problem''\,\newnewfinal{\cite{Steinhardt}}.
Even though the simplest candidate for the acceleration mechanism is
the existence of a small positive cosmological constant, $\CC$, now
incorporated into the benchmark model of modern cosmology, or
``concordance'' $\Lambda$CDM model, there is quite a list of
challenging cosmological problems of various sorts
\newnewfinal{afflicting} the standard model of cosmology, see e.g.\
\cite{Perivolaropoulos08a}.

\newnewfinal{Above all these problems} there is a problem of greatest
importance, one which is by far the most troublesome one from the
point of view of Fundamental Physics, to wit: the size of the
cosmological constant energy density required to describe the
cosmological data is in drastic disaccord with the value predicted
by any known fundamental physical theory based on quantum field
theory (QFT)
\newtext{or} string theory.  Indeed, from a theoretical point of
view, the cosmological constant energy density,
\newtext{$\rL=\CC/(8\pi G_N)$} ($G_N$ being Newton's constant), is a
single number, but of remarkable composition as it involves all
sources of vacuum energy: e.g.\ the zero-point energy density of a
quantum field of mass $m$ produces a contribution to the CC energy
density of the order of $m^4$, where bosonic fields contribute
positively and fermionic fields negatively; phase transitions in the
early universe also leave an imprint on the value of $\rL$ in the
form of vacuum energy associated to spontaneous symmetry breaking
(e.g.\ the energy density of the electroweak vacuum); or the
non-perturbative condensates (as e.g.\ the quark and gluon
condensates of the QCD vacuum) etc. All these contributions, which
differ in size and sign, have a common characteristic: they are all
much larger in absolute value than the experimentally determined
vacuum energy density, $\rLo\simeq (2.3\times
10^{-3}\,\text{eV})^4$~\cite{cosmdata}. There is a (remote)
possibility that these manyfold effects might cancel among
themselves to produce the observed value of the CC. This, however,
requires extreme fine-tuning which, in perturbative QFT, needs to be
iterated to all orders of the perturbative expansion. Such
explanation, therefore, although technically possible, is utterly
unconvincing.

If we discard the explanations based on fine-tuning, \newtext{and we
take seriously the idea} that all the contributions to the vacuum
energy do have an impact on the final CC value, then we are
unavoidably left with a value of $\rL$ many orders of magnitude
larger than $\rLo$. Since we have no experimental verification of
the dynamics in the UV limit, the size and sign of the ``initial''
CC energy density -- which we may call $\rLi$ -- are not precisely
known, except the fact that its absolute value must be really huge
when measured in units of the observed CC energy density; i.e.\ the
number $| \rLi |/\rLo$ \newtext{must be} enormous.

To reconcile the anomalously large size of this ratio with the
behavior of the observed universe, the value of $\rLi$ has to be
neutralized in some reasonable manner. One possibility would be the
existence of a mildly broken symmetry requiring the CC to be very
small (zero if the symmetry would be exact). Such a symmetry,
however, is not presently known. Supersymmetry (SUSY)\,\cite{SUSY},
for instance,
\newtext{despite} the first natural hopes\,\cite{WZ74}, cannot cure the CC
problem in any obvious way, because we know that SUSY  must be
broken at the electroweak scale or higher, which means that the SUSY
particles must have masses larger than their conventional
counterparts, and as a result the ``residual'' vacuum energy left in
the universe must again be of the order (or even larger) than that
of the Standard Model (SM) of particle physics, i.e.\ of order
$M_W^4$, where $M_W\sim 100$ GeV is the scale of the weak gauge
boson masses\,\footnote{\newtext{In certain} local realizations of
SUSY, such as in Supergravity theories emerging from $E_8\times E_8$
superstring theory with gluino condensation, one can have vanishing
vacuum energy even if SUSY is broken\,\cite{Dine85}. However, this
does not explain the small value of the CC, specially after one
realizes of the vast ``landscape'' of possible string
vacua\,\cite{Susskind03}, suggesting that there is no particular
reason for a given vacuum choice in this framework. }. Therefore, in
general we expect, roughly, $|\rLi |/\rLo\sim 10^{55}$, which is an
appallingly large number!
\newfinal{Obviously}, a theoretical breakthrough is demanded to
account for this situation.

From a physical point of view, it is especially \newtext{appealing}
to have a dynamical adjustment of the CC value, \newtext{because we
can then avoid} resorting to too preposterous a fine-tuning of the
parameters -- \newfinal{at least 55 digits} in the SM case. This
idea was originally pursued in terms of dynamical scalar
fields\,\cite{OldScalar}, but it was later shown by S. Weinberg that
it is generally obstructed by a ``no-go
theorem''\,\cite{weinberg89}. Subsequently, the more modest idea of
quintessence\,\cite{Quintessence} was proposed without attempting to
explain the smallness of the CC, but only to cope with some aspects
of the cosmic coincidence problem.

In this paper, we wish to face a possible dynamical solution that
can escape the no-go theorem and that is based on a different
concept, \newtext{one that makes no use of scalar fields}.
\newtext{Specifically}, we refer to the concept of ``dynamical effective $\CC$'' or
``effective vacuum energy'' $\rLe$, in which, rather than replacing
$\CC$ by a collection of \textit{ad hoc} ersatz fields, we stick to
the idea that the CC term in Einstein's equations is still a ``true
cosmological term'', although we \newtext{permit} that ``the
observable CC at each epoch'' can be an effective quantity evolving
with the expansion of the universe: $\rLe=\rLe(H)$, where
\newtext{$H$ is} the expansion rate or Hubble function. This
general dynamical $\CC$ approach has been recently re-emphasized in
\cite{ShapSol09,BFLWard09}, and it was explored with different
particular implementations in the past from the point of view of QFT
in curved-space time\,\cite{oldCCstuff1,oldCCstuff2}, including some
recent applications to structure formation\,\cite{newCCstuff} -- see
e.g.\ the review\,\cite{ShapSol0608} and references therein.

The variable CC approach was also tried long ago on purely
phenomenological terms, cf.\,\cite{oldvarCC1,overduin98}.
Furthermore, the idea of a slowly running CC is in general
compatible with the experimental data\,\cite{BPS09a,CCfit}, and has
been even exploited as an alternative description of the notion of
dynamical dark energy (DE) for a possible viable solution of the
aforementioned cosmic coincidence
problem\,\cite{LXCDM,LXCDM2,LXCDMmore}. But, most important of all,
it may also be advantageously utilized as a powerful mechanism
capable of solving (or highly alleviating) the big or ``old'' CC
problem\,\cite{weinberg89}, viz.\, the absolutely formidable task of
trying to explain the value itself of the current vacuum energy
$\rLo$ -- not just its time evolution in the vicinity of it -- on
the face of its enormous input value $\rLi$ left over in the early
times.

Can we explain the measured value of $\CC$ and at the same time
insure that it remains safely small throughout the entire history of
the universe? In this paper, we shall present a thorough attempt of
this sort, actually one which substantially improves previous recent
attempts along these lines\,\cite{Stefancic08,BSS09a}. From now on,
we shall call this new process of dynamical neutralization of the
``true'' cosmological constant: the ``relaxation mechanism'' of the
CC.

Relaxation offers indeed an entirely different perspective on the
old CC problem. Many CC adjustment mechanisms, partly discussed
above, function unilaterally, as e.g.\ when using scalar fields. In
other words, in these cases the dynamics of the relaxation mechanism
works ``intrinsically'' to neutralize the real value of the CC. A
new quality could be added to the relaxation mechanism if it becomes
bilateral. Namely, apart from the mechanism itself for neutralizing
the value of the CC, we could also envisage that the expansion of
the universe, fueled by the big CC value $\rLi$, enables the very
action of the relaxation mechanism. The idea is that the relaxation
mechanism neutralizes the value of $\rLi$, but at the same time the
existence of this value sets the relaxation mechanism to motion.
Since the expansion of the universe should be responsible for the
triggering of the relaxation dynamics, it is reasonable to assume
that such mechanism should be closely related to the expansion
itself, or more precisely, to the interaction governing the
expansion, i.e.\ {\em gravity}.

Despite the appeal of such a feedback, a number of important
questions have to be answered: i) Is it possible that the observed
value of the CC (or effective vacuum energy) is small exactly {\em
because} the ``real'' value of the CC is large?  ii) Is there a way
to dynamically counterbalance the effects of a large CC in all
epochs of the cosmic evolution? iii) Can the feedback between the
relaxation mechanism and expansion be formulated in terms of an
action functional? In this paper, we present a modified gravity
theory based on an action functional $\F(R,\G)$ of the Ricci scalar,
$R$, and the Gauss-Bonnet invariant, $\G$, in which \textit{all} the
above questions have an \textit{affirmative} answer.

A simple model exploring the viability of
\newnewtext{a mechanism cancelling dynamically} the effects of a large CC
was presented in \cite{Stefancic08}. Previously, it had been shown
that the inhomogeneous equation of state (EOS) of a cosmic fluid can
be interpreted as an effective description of modified gravity
theories or even braneworld models \cite{Odin}. In a cosmological
model containing only two components, a large cosmological constant
of an arbitrary sign and a component with an inhomogeneous EOS of
the form $p=w\rho - \beta H^{-\alpha}$, for $\alpha > 0$ the cosmic
expansion asymptotically tends to de Sitter phase with a small
positive effective CC \cite{Stefancic08}. Essentially, the effect of
a very large $\Lambda$ is counterbalanced by the terms of the form
$1/H^{2n}$ when $H$ is sufficiently small. In this cosmological
model, at late times the universe expands as if there existed a
small positive CC even though the actual CC is very large and it can
even be negative. As there is no fine-tuning in this model, such a
dynamical outcome can be rightfully identified as a viable approach
for solving the cosmological constant problem. One specific feature
of the model deserves a special emphasis. Within this relaxation
mechanism, the size of the CC is not only a problem, but also a part
of the solution. To illustrate this fact let us consider  the
asymptotic value of the \newfinal{Hubble parameter}
$H_{\mathrm{asym}} \sim \sqrt{\Lambda_\text{eff}}$ for $\alpha=2$.
For this value of $\alpha$ we have $H_{\mathrm{asym}} \sim
1/|\rLi|$. The bigger $\rLi$ is, the smaller is the effective
observed $\Lambda_\text{eff}$. For some other recent work on
relaxation mechanisms, see also \cite{Barr:2006mp} and
\cite{Batra:2008cc}. For recent alternative ideas on the CC problem,
see\,\cite{Demir09,Maggiore10}.

The mechanism based on the inhomogeneous EOS, despite exhibiting CC
relaxation properties, is incomplete since it does not contain
matter or reproduce the observed sequence of epochs in the history
of the universe. The idea of the CC relaxation mechanism has to be
embedded into a realistic cosmological model. Such a realistic
implementation was made in \cite{BSS09a} in the formalism of
variable $\Lambda$ interacting with a dark matter component, i.e.\
within the context of the so-called $\Lambda$XCDM
model\,\cite{LXCDM}. The variable CC contains contributions in the
form of functions of curvature invariants such as $R^2$, $R^{ab}
R_{ab}$ and $R^{abcd} R_{abcd}$. The model successfully produces the
sequence of radiation, matter and de Sitter phases in the expansion
of the universe. During the de Sitter phase, the action of a large
CC is counterbalanced by the term $1/H^{2n}$ when $H$ is
sufficiently small, similarly as with having inhomogeneous EOS.
However, during the matter and radiation dominated phases, the
effect of a large $\Lambda$ is equilibrated by the terms $ \sim
1/(q-1/2)$ and $ \sim 1/(q-1)$, respectively, where $q$ is the
deceleration parameter. Within this model one can easily produce an
expansion history of the universe close to that of the $\Lambda$CDM
model despite the persistence of the huge primeval CC, which is
permanently subdued by the dynamical mechanism. Besides, a dedicated
analysis of the growth of perturbations shows that the formation of
structures \newtext{in this kind of models} is also comparable to
the $\Lambda$CDM \cite{Bauer2009}.

The model introduced and elaborated in~\cite{BSS09a} is thus
phenomenologically very successful, but it lacks the action
principle formulation. To fill this remaining gap, we proposed
recently a specific model universe~\cite{BSS09b} based on an action
functional for modified gravity. In the present paper, we extend and
generalize this approach while maintaining the phenomenologically
pleasing features discussed above. Specifically, we discuss here a
modified gravity theory (``The Relaxed Universe'') based on a class
of $\F(R,{\cal G})$ action functionals \newfinal{capable} of
relaxing the highly ``stressed'' primeval state of our universe,
namely a state which is beset by a large vacuum energy which
prevents the startup of the normal thermal history. \newfinal{The
relaxation mechanism} allows not only to unblock this situation, it
also operates very actively during \textit{all} the subsequent
epochs of the cosmic expansion without fine-tuning the parameters of
the theory at any time. Furthermore, it involves a characteristic
mass scale, $\M$, whose value is of the order of a typical particle
physics scale whether of the SM or of a Grand Unified Theory (GUT),
and therefore avoids introducing extremely tiny masses
(\newtext{$m_{\phi}\sim H_0\sim 10^{-33}$ eV}) plaguing most
proposals, as e.g.\ quintessence~\cite{Quintessence}.

It should finally be emphasized that the cosmological model
introduced here \newnewtext{goes} beyond the idea of modified
gravity with late time cosmic
acceleration\,\cite{SotiriouFaraoni08,Polarski}. Indeed, let us
stress once more that the proposed gravity modifications inherent to
our mechanism are crucial during the \textit{entire cosmic history},
and not just in the late universe. The reason is that they primarily
counterbalance the effect of a huge CC during \textit{all}
accessible epochs (radiation and matter), and without any
fine-tuning. The late time acceleration phenomenon appears here only
as a special, but certainly very important effect, either in
quintessence-like, de Sitter or phantom-like disguise.

The paper is organized as follows. In the next section, we sketch
the old CC problem and its relation with the notion of fine-tuning.
In section \ref{sec:Einstein-equations} we define and solve the
$\F(R,\G)$-cosmology. In section \ref{sec:CC-relaxation} we discuss
the working principle of the relaxation mechanism. The analysis of
some concrete implementations of the model is performed in section
\ref{sec:numerical}, including detailed numerical results.
\newtext{In the last section} we deliver our conclusions.
\newtext{Finally}, in appendix~\ref{sec:FofRSTG-action} we provide information on our
notation and conventions, in appendix~\ref{sec:finetuning} we extend
the discussion of the severity of the CC fine-tuning problem in QFT,
and in appendix~\ref{sec:SolarSystem} we \revisedtext{briefly}
discuss the relaxation mechanism in the solar system.

\section{The old CC problem as a fine-tuning problem}\label{CCfinetuning}

Before we present our relaxation mechanism, let us summarize the old
CC problem and discuss why the fine-tuning problem unavoidably
appears if one does not modify some aspects of the theory of gravity
in interaction with matter. We wish to illustrate the problem within
the context of the standard model (SM) of particle physics, and more
specifically within the Glashow-Weinberg-Salam model of electroweak
interactions. This is the most successful QFT we have at present
(together with the QCD theory of strong interactions), both
theoretically and phenomenologically, and therefore it is the ideal
scenario where to formulate the origin of the problem. As is
well-known, the unification of weak and electromagnetic interactions
into a renormalizable theory requires to use the principle of local
gauge symmetry in combination with the phenomenon of spontaneous
symmetry breaking (SSB). \newtext{It is} indeed the only known way
to generate all the particle masses by preserving the underlying
gauge symmetry. \newfinal{In the SM}, one must introduce a
fundamental complex doublet of scalar fields. However, in order to
simplify the discussion, let us just consider a field theory with a
real single scalar field $\varphi$, as this does not alter at all
the nature of the problem under discussion. To trigger SSB, one must
introduce a potential for the field $\varphi$, which in
renormalizable QFT takes the form (the
\newtext{tree-level} Higgs potential):
\begin{equation}\label{Poten}
V(\varphi)=\frac12\,m^2\,\varphi^2+\frac{1}{4!}\,\lambda\,\varphi^4\,,\
\ \ \ \ (\lambda>0)\,.
\end{equation}
Since we are dealing with a problem related with the CC, we must
inexcusably consider the influence of gravity. To this effect, we
shall conduct our investigation of the CC problem within the
semiclassical context, i.e.\ from the point of view of quantum field
theory (QFT) in curved space-time\,\newtext{\cite{Parker09}}. It
means that we address the CC problem in a framework where gravity is
an external gravitational field and we quantize matter fields
only\,\cite{ShapSol09,Shapiro:2008sf}. The potential in equation
(\ref{Poten}) is given at the moment only at the classical level,
but it will eventually acquire quantum effects generated by the
matter fields themselves. In this context, we need to study what
impact the presence of such potential may have on Einstein's
equations both at the classical and at the quantum level.

Einstein's field equations for the classical metric in vacuo are
derived from the Einstein-Hilbert (EH) action with a cosmological
term  $\BCC$ (\newtext{hereafter} the CC \textit{vacuum term}). The
EH action in vacuo reads:
\begin{eqnarray}
S_{EH} = \frac{1}{16\pi\,G_N}\,\int d^4 x\sqrt{|g|}\, \left(\,R -
2\,\BCC\,\right)=\,\int d^4 x\sqrt{|g|}\,\left(\frac{1}{16\pi\,G_N}
\,R-\rLV\right)\,. \label{EH}
\end{eqnarray}
(See the Appendix~\ref{sec:FofRSTG-action} for our notation and sign
conventions). Here we have defined $\rLV$, the energy density
associated to the CC vacuum term:
\begin{equation}\label{rLV}
\rLV=\frac{\BCC}{8\pi\,G_N}\,.
\end{equation}
The classical action including the scalar field $\varphi$ with its
potential (\ref{Poten}) is
\begin{eqnarray}
S = S_{EH} +\int
d^4x\,\sqrt{|g|}\,\left[\frac12\,g^{ab}\,\partial_{a}\varphi\,\partial_{b}\varphi-V(\varphi)\right]\,.
\label{Stotal}
\end{eqnarray}
Due to the usual interpretation of Einstein's equations as an
equality between geometry and a matter-energy source, it is
convenient to place the $\rLV$ term as a part of the matter action,
$S[\varphi]$. Then the total action (\ref{Stotal}) can be
reorganized as
\begin{eqnarray}
S = \,\frac{1}{16\pi\,G_N}\,\int d^4 x\sqrt{|g|}\,R+S[\varphi]\,,
\label{Stotal2}
\end{eqnarray}
with
\begin{eqnarray}\label{Sphi}
    && S[\varphi]=\int
    d^4x\,\sqrt{|g|}\,\left[\frac12\,g^{ab}\,\partial_{a}\varphi\,\partial_{b}\varphi-\rLV-{V}(\varphi)\right]
    \equiv \int
    d^4x\,\sqrt{|g|}\ \mathcal{L}_{\varphi}\,,
\end{eqnarray}
where $ \mathcal{L}_{\varphi}$ is the matter Lagrangian for
$\varphi$. For the moment, we will treat the matter fields contained
in $\mathcal{L}_{\varphi}$ as classical fields, and in particular
the potential $V$ is supposed to take the classical form
(\ref{Poten}) with no quantum corrections. If we compute the
energy-momentum tensor of the scalar field $\varphi$ in the presence
of the vacuum term $\rLV$, let us call it
$\tilde{T}_{ab}^{\varphi}$, we obtain
\begin{eqnarray}\label{Tmunu}
    &&\tilde{T}_{ab}^{\varphi}={2\over\sqrt{|g|}}\,{\delta\,S[\varphi]\over\delta\,g^{ab}}
=2\,{\partial\mathcal{L}_{\varphi}\over\partial\,g^{ab}}-
g_{ab}\,\mathcal{L}_{\varphi} = g_{ab}\,\rLV+{T}_{ab}^{\varphi}\,,
\end{eqnarray}
where we have used $\partial\sqrt{|g|}/\partial  g^{ab} =
-(1/2)\sqrt{|g|}\,g_{ab}$. Here
\begin{eqnarray}\label{Tmunu2}
    &&{T}_{ab}^{\varphi} = \left[\,\partial_{a}\varphi\,\partial_{b}\varphi-\frac12\,g_{ab}\
\partial_{c}\varphi\,\partial^{c}\varphi
\right]+g_{ab}\,V(\varphi)
\end{eqnarray}
is the ordinary energy-momentum tensor of the scalar field
$\varphi$.

In the vacuum (i.e.\ in the ground state of $\varphi$) there is no
kinetic energy, so that the first term on the \textit{r.h.s} of
(\ref{Tmunu2}) does not contribute \newtext{in that state}. Only the
potential may take a non-vanishing vacuum
\newtext{expectation} value, which we may call $\VP$. Thus, the
ground state value of (\ref{Tmunu}) is
\begin{equation}\label{vacTmunu}
    \langle \tilde{T}_{ab}^{\varphi}\rangle=g_{ab}\,\rLV+\langle
    {T}_{ab}^{\varphi}\rangle=g_{ab}\,(\rLV+\VP)\equiv\rV^{\rm
    cl}\ g_{ab}\,,
\end{equation}
where $\rV^{\rm cl}$ is the \textit{classical vacuum energy} in the
presence of the field $\varphi$.

If $m^2>0$ in equation (\ref{Poten}), then $\langle\varphi\rangle=0\
\Rightarrow$ $\VP=0$ and the classical vacuum energy is just the
original $\rLV$ term,
\begin{equation}\label{TmunuNoSSB}
    \langle \tilde{T}_{ab}^{\varphi}\rangle=g_{ab}\,\rLV\,.
\end{equation}
This result also applies in the free field theory case.  However, if
the phenomenon of SSB is active, which precisely occurs when
$m^2<0$, we have a non-trivial ground-state value for $\varphi$:
\begin{equation}
\langle\varphi\rangle =\sqrt{\frac{-6\,m^{2}}{\lambda}}\,.
\label{5N}
\end{equation}
In this case, there is an \textit{induced part} of the vacuum energy
at the classical level owing to the electroweak phase transition
generated by the Higgs potential. This transition induces a
non-vanishing contribution to the cosmological term which is usually
called the ``induced CC''. At the classical level, \newtext{it is}
given by
\begin{equation}
\rLI\equiv\VP=-\frac{3\,m^{4}}{2\lambda}=
-\frac18\,M_{\Hp}^2\,\langle\varphi\rangle^2=
-\frac{1}{8\sqrt{2}}\,M_{\Hp}^2\,M_F^2\,, \label{eq:Higgstree}
\end{equation}
In the last two equalities, we have used the physical
Higgs mass squared $M_{\Hp}^2=-2m^2>0$. Indeed, if we redefine the Higgs field as
$\Hp=\varphi\,- \langle\varphi\rangle$, then its value at the minimum will
obviously be zero. \newtext{This is} the standard position for the
ground state of the field before doing perturbation theory. The
physical mass is just determined by the oscillations of $\Hp$ around
this minimum, i.e.\ it follows from the second derivative of $V$ at
$\varphi=\langle\varphi\rangle$. We have also introduced the so-called Fermi's
scale $M_F\equiv G_F^{-1/2}\simeq 290\,\text{GeV}$, which is defined
from Fermi's constant obtained from muon decay, $G_F\simeq
\newtext{1.166}\times 10^{-5}\,\text{GeV}^{-2}$.

In view of the previous SSB contribution, it is clear that we must
replace ${T}_{ab}^{\varphi}\to\tilde{T}_{ab}^{\varphi}$ in the
expression of Einstein's equations in vacuo. Furthermore, in the
presence of incoherent matter contributions (e.g.\ from dust and
radiation) described by a perfect fluid we have the additional term
$T_{ab}=(\rho+p)u_{a}u_{b}-p\, g_{ab}$. Therefore, Einstein's
equations in terms of coherent and incoherent contributions of
matter, plus the vacuum energy of the fields, finally read
\begin{equation}\label{EEvac}
R_{ab}-\frac{1}{2}g_{ab}R=-8\pi\,G_N\,\left(\langle\tilde{T}_{ab}^{\varphi}\rangle+
T_{ab}\right)=-8\pi\,G_N\,\left[g_{ab}\left(\rLV+\rLI\right)+T_{ab}\right]\,.
\end{equation}
We conclude that the ``physical value'' of the CC, at this stage, is
not just the original term  $\rLV$, but
\begin{equation}\label{rLphclass}
\rLP=\rLV+\rLI\,,
\end{equation}
where the induced part is given by (\ref{eq:Higgstree}). However, it
is pretty obvious that a severe fine tuning problem is conjured in
equation (\ref{rLphclass}) when we compare theory and experiment.
Indeed, the lowest order contribution from the Higgs potential, as
given by equation (\ref{eq:Higgstree}), is already much larger than
the observational value of the CC. Using the LEP lower bound on the
Higgs mass ($M_{\Hp}\gtrsim 114\,\text{GeV}$), equation
(\ref{eq:Higgstree}) yields $\rLI\simeq -10^8\,\text{GeV}^4$.
Roughly speaking, the VEV of the Higgs potential is (naturally) in
the ballpark of the fourth power of the electroweak VEV, i.e.\
$\rLI\sim v^4$, where $v={\cal O}(100)$ GeV. Thus, being the CC
observed value of order $\rLo\sim 10^{-47}$ GeV$^4$, the electroweak
vacuum energy density is predicted to be $55$ orders of magnitude
larger than $\rLo$!

Suppose that the induced result would exactly be $\rLI=
-10^8\,\text{GeV}^4$ and that the vacuum density would exactly be
$\rLo=+10^{-47}\,\text{GeV}^4$. In such case one would have to
choose the vacuum term $\rLV$ in equation (\ref{rLphclass}) with a
precision of $55$ decimal places in order to fulfill the equation
\begin{equation}\label{finetuningclassical}
10^{-47}\,\text{GeV}^4=\rLV+\rLI=\rLV-10^8\,\text{GeV}^4\,.
\end{equation}
This is of course the famous fine-tuning problem. This problem is in
no way privative of the cosmological constant approach to the DE,
but it is virtually present in \textit{any} known model of the DE,
in particular also in the quintessence
approach\,\cite{Quintessence}. Indeed, the quintessence scalar field
potential $V(\varphi)$ is supposed to precisely match the value of
the measured DE density at present starting from a high energy
scale, usually some GUT scale $\varphi=M_X\sim 10^{16}$ GeV. In
order to achieve this, an ugly fine-tuning of its initial value $\langle V\rangle\sim 10^{64}$ GeV$^4$ is unavoidable. Therefore, the quintessence
approaches, apart from introducing extremely unnatural small mass
scales of the order of the Hubble function at present (hence masses
as small as $m_{\varphi}\sim 10^{-33}$ eV), are plagued with
fine-tuning problems in no lesser degree than the original CC
problem itself.

However, this is not quite the end of the story yet. In QFT the
induced value of the vacuum energy is much more complicated than
just the simple result (\ref{eq:Higgstree}), and the fine-tuning
problem is much more cumbersome than the one expressed in equation
(\ref{finetuningclassical}), see the Appendix~\ref{sec:finetuning}
for a more detailed exposition. At the end of the day, we really
need some mechanism that is able to concoct the tuning
``dynamically'', i.e.\ automatically, and without requiring the
\newtext{intervention} of some carefully ``designed'' counterterm encoding the
aforementioned fabulous numerical precision. In the rest of the
paper, we will try to convince the reader that such mechanism to
avoid fine-tuning does exist.

\section{$\F(R,\G)$-cosmology}\label{sec:Einstein-equations}

As already mentioned, in Ref.~\cite{BSS09a} we investigated a
powerful mechanism for relaxing dynamically the vacuum energy or
effective cosmological constant. For all of its virtues, it is
nevertheless based on introducing a direct modification of the
gravitational part at the level of the field equations, and hence
without any obvious connection with an action principle. A deeper
step in the theoretical construction process would be to implement
it from an action functional. For this reason, we investigate here
the CC relaxation mechanism within the context of the $\F(R,\G)$
modified gravity setup and in the metric formalism, where~$\F$
defines a functional of the Ricci scalar $R$ and the Gau\ss-Bonnet
invariant~$\G$,
\begin{equation}\label{GB}
{\cal G}\equiv R^2-4 R_{ab}R^{ab}+R_{abcd}R^{abcd}\,.
\end{equation}
(For more details on notation, see once more the
Appendix~\ref{sec:FofRSTG-action}). The reason why the higher
curvature term is $\G$ rather than the individual higher derivative
components \newtext{that define it} will become clear later.

\subsection{\newtext{The generic} class of the $\CC$XCDM models}\label{LXCDMmodels}

\newtext{Our} theoretical $\F(R,\G)$ construct is placed within the general class of
the so-called $\Lambda$XCDM models of the cosmological
evolution~\cite{LXCDM,BSS09a}, in which the cosmological term is
supplemented (\textit{not} replaced!) with an effective entity $X$
at the level of the field equations. The resulting cosmological
system is thus characterized by a compound dark energy made out of a
(constant or variable) cosmological term $\rL$ \textit{and} the
contribution $\rho_X$ corresponding to a new entity $X$ (called the
``cosmon''). The total DE density (or
\newtext{``effective CC density''}) reads
\begin{equation}\label{rD}
\rLe=\rho_{\CC}+\rX\,.
\end{equation}
In the simplest formulation of the $\CC$XCDM model, in which
Newton's coupling $G_N$ is constant, this overall DE density is
locally and covariantly conserved\,\cite{LXCDM}:
\begin{equation}\label{conslawDE}
\dot{\rho}_{\Lambda{\rm eff}}+3\,H\,(\rLe+\pLe)=0\,,
\end{equation}
where $H(t)\equiv{\dot{a}}/{a}$ is the expansion rate (overdots
represent time-derivatives with respect to the cosmic time). We can
reexpress this equation in terms of $\weff$, the effective EOS of
the compound DE system:
\begin{equation}\label{eEOSdef}
\dot{\rho}_{\Lambda{\rm eff}}+3\,H\,(1+\weff)\,\rLe=0\,,\ \ \
\weff=\frac{\pLe}{\rLe}=\frac{\pL+\pX}{\rL+\rX}\,,
\end{equation}
where $\pL=-\rL$ is the CC pressure and $\pX$ is the pressure of the
cosmon component. It follows that matter is also covariantly
conserved.

As emphasized in \cite{LXCDM}, the cosmon is \textit{not} to be
viewed in general as a field, but as a complex object emerging from
the full structure of the effective action. Its dynamics is
determined either by suggesting an explicit modification of the
effective action, or by providing a particular evolution law for the
CC term, $\rL=\rL(t)$. In either case $\rX=\rX(t)$ is then
completely determined by the local conservation law
(\ref{conslawDE}). This can be further appreciated by rewriting that
law as follows:
\begin{equation}\label{conslawDE2}
\dot{\rho}_{\Lambda}+\dot{\rho}_X+\, 3\,H\,(1+\wX)\,\rX=0\,,
\end{equation}
where we have defined the effective EOS of the cosmon:
$\wX=\pX/\rX$. The effective EOS of the compound DE system can now
be written as
\begin{equation}\label{eEOS}
\weff=\frac{-\rL+\wX\,\rX}{\rL+\rX}= -1+(1+\wX)\,\frac{\rX}{\rLe}\,.
\end{equation}
In general, both $\wX$ and $\weff$ will be non-trivial functions of
the cosmic time or the cosmological scale factor, $a=a(t)$, or the
redshift: $z=(1-a)/a$. In the original $\CC$XCDM model of
Ref.\,\cite{LXCDM}, one starts from a given evolution law
$\rL=\rL(H)$ for the CC term, specifically one which is motivated by
the presence of quantum corrections, and the EOS parameter $\wX$ for
the cosmon is taken constant. This setup enabled a full analytical
treatment and an explicit determination of the cosmon energy density
as a function of the redshift, $\rX=\rX(z)$. In the present case,
however, $\rL$ is just a constant (given by the initial $\rLi$), and
the dynamics of the cosmon -- represented by the conservation law
(\ref{conslawDE2}) -- will be completely controlled by the influence
of a new gravitational term that modifies the EH action. In fact,
the complete action functional is composed, apart from the
conventional matter part and the EH term, also of the aforementioned
function of the invariants $R$ and $\G$ -- whose contribution to the
total action will, for brevity sake, be referred to as the
``$\F(R,\G)$-functional''. This functional determines the dynamics
of $X$. Notice that the apparent simplification produced by the fact
that $\rL$ is now just a constant is compensated by the fact that
the cosmon EOS will be a complicated function of time or redshift:
$\wX=\wX(z)$. As a result, a complete analytical treatment will not
be possible. The reward, however, will be a powerful formulation of
the $\CC$XCDM model in which the cosmon will be able to efficiently
deal (as its name intends to suggest) with the old CC problem, and
specifically with the toughest aspect of it: the fine-tuning
problem.

Let us note that the definition of the cosmon $X$ in the wide class
of $\CC$XCDM models is really very general because it introduces the
energy density $\rho_X$ not at the core level of the action, but at
the level of the field equations. In the present case, our $\CC$XCDM
model is constructed \textit{ab initio} at the level of an action
functional containing the piece $\F(R,\G)$. This is of course a
great advantage from the theoretical point of view. However, in
order to recognize what is $\rX$ in the present case, and finally
unveil the corresponding $\rLe$ (i.e.\ the ``effective CC'') that
results from the presence of the $\F$-functional, one has to account
for the field equations of the complete effective action. We do this
in the next section.

\subsection{Action principle and field equations}\label{action}

The complete effective action of our cosmological model
reads
\begin{equation} \mathcal{S}=\int
d^{4}x\,\sqrt{|g|}\left[\frac{1}{16\pi G_{N}}R-\rLi-
\F(R,\G)+\mathcal{L}_{\phi}\right]\,,\label{eq:CC-Relax-action}\end{equation}
where ${\cal L}_{\phi}$ stands for the Lagrangian of the matter
fields. Clearly, our functional constitutes an extension of the EH
action with CC in which we have added the $\F(R,\G)$ part, i.e.\ a
generalization of equations (\ref{Stotal2})-(\ref{Sphi}). We take
$\F$ within the class of functions of the form
\begin{equation}\label{eq:FFP}
\F(R,\G)=\beta\, F(R,\G)+A(R)\,,
\end{equation}
in which $\beta$ is a parameter, $F(R,\G)$ a non-polynomial function
(see below) and $A(R)=a_2 R^2+a_3 R^3...$ is a (low order)
polynomial of $R$. The latter has neither linear nor independent
term, the reason being that these terms can already be included as
part of the EH action with CC. Notice that we do not use $\G$ in the
structure of the polynomial $A$ because the first term would just be
$\G$, which is a topological invariant in four dimensions. As for
$F(R,\G)$, we take it as a function of negative mass dimension. We
will usually call it the ``$F$-term''. The canonical implementation
of it \newtext{is} a rational function of $R$ and $\G$ vanishing as
$R$ and $\G$ are sufficiently large, \newtext{say}
\begin{equation} F(R,\G)=\frac{1}{B(R,\G)}\,,\label{eq:FRG-1B}\end{equation}
with $B(R,\G)$ a polynomial in $R$ and $\G$. Within this canonical
ansatz (which we will adopt for most of our considerations), the
complete functional (\ref{eq:FFP}) reads
\begin{equation}\label{eq:FFP-1B}
\F(R,\G)=\frac{\beta}{B(R,\G)}+A(R)\,.
\end{equation}
Since, in the FLRW metric, $F(R,\G)$ defines a functional $\Phi(H)$
of the Hubble function or expansion rate $H(t)\equiv{\dot{a}}/{a}$
and its time-derivatives, we can express the required condition as
\begin{equation}\label{eq:limF}
F(R,\G)=\Phi(H)\to 0\, \ \ \ \text{for}\ \ \ H\gtrsim H_{\rm
rad}^*\,,
\end{equation}
where $H_{\rm rad}^*$ is the Hubble rate at the time when the very
early radiation period sets in, hence after both inflation and
reheating have already taken place. We shall explain the motivation
for this condition later on. Since $F(R,\G)$ dies off with
curvature, the parameter~$\beta$ in equation (\ref{eq:FFP}) must
have a positive dimension of energy, namely
\begin{equation}\label{eq:betaM}
\beta\equiv \M^N\,, \ \ \ \ \text{with}\ \ N>0\,,
\end{equation}
where $N=n_B+4$ is an integer in which \newtext{$n_B$ is equal} to
the mass dimension of the $B(R,\G)$ polynomial, and $\M$ represents
some characteristic cosmic mass scale associated to our
$\F$-functional. For instance, if $B(R,\G)$ is a polynomial
involving just quadratic terms in $R$ and linear in $\G$, we would
have $N=8$.  Notice that if the polynomial $A(R)$ reduces to the
monomial $a_2 R^2$, then $a_2$ is a dimensionless coefficient. In
this case, the only dimensionful scales carried by $\F$ are $\beta$
and those that might involve $B(R,\G)$ in the form of monomials of
$R$ (respect.\ $\G$) of order $3$ (respect. 2) or above. The mass
scale ${\cal M}$ should have some physical significance, and
therefore it will be interesting to check which are the typical
values allowed for $\M$ (depending on the choice of $F$) in order to
implement realistically the relaxation mechanism in our Universe.
Generalizations of $F=1/B$ offer more possibilities for the mass
scale~$\M$ and will be discussed later on in
Sec.~\ref{sec:GeneralModels}.

Another essential ingredient of the effective action
(\ref{eq:CC-Relax-action}) is of course the ``initial'' cosmological
constant term
\begin{equation}\label{eq:CCterm}
\Lambda^{i}=8\pi G_{N}\,\rLi\,,
\end{equation}
in which $\rLi$ stands for all possible vacuum energy density
contributions (of arbitrary size) pertinent to ``initial'' phase
transitions in the early universe, e.g.\ the GUT phase transitions,
the electroweak transition, the QCD quark-gluon transition, and in
general all vacuum energy density contributions associated to the
matter fields (bosonic or fermionic) of the Lagrangian
$\mathcal{L}_{\phi}$. For instance, $\rLi$ embodies the important
electroweak contribution from the SM Higgs potential, i.e.\
$|\rho_{\rm EW}|\sim M_F^2\,M_{\Hp}^2\sim 10^8\,\text{GeV}^4$. In
addition, $\rLi$ integrates the QCD contribution, which is of order
$\rho_{\rm QCD}\sim \Lambda_{QCD}^4$, with $\Lambda_{QCD}\sim
0.1\,\text{GeV}$. Taken alone, any of these SM contributions from
particle physics is enormous as compared to the current value of the
CC density, $\rLo\sim 10^{-47}\,\text{GeV}^4$. On the other hand,
$\rLi$ may contain much larger contributions; in general it will be
dominated by the maximum effect prevailing in the early universe,
which should be determined by the strongest GUT phase transition,
say $\rLi\sim M_X^4$, with $M_X\sim 10^{16}\,\text{GeV}$. At the
same time, $\rLi$ includes the vacuum term $\rLV$, too, if only for
renormalizability reasons. This term is completely free and, in the
traditional approach explained in section~\ref{CCfinetuning}
and Appendix~\ref{sec:finetuning}, it can be used (after renormalization) to fine-tune all the other contributions.

For all of its non-trivial composition, $\rLi$ is \textit{not} yet
the physical CC of the $\F(R,\G)$-cosmology, even though it contains
all the ingredients of the traditional approach, i.e.\ all the terms
on the \textit{r.h.s.}\ of equation (\ref{rLphclass}). If it were,
the tuning that we ought to apply to $\rLV$ to compensate for the
energy released during the GUT phase transition would make the
$55$-digits-electroweak-tuning described in section
\ref{CCfinetuning} pale in comparison, and hence would further
increase the severity of the CC problem to the utmost level! But
this is not what we shall assume here, so $\rLV$ will be taken as
any other contribution, it is not important which one. What is
important is that we will \textit{not} take a special (fine-tuned)
value for it (in contrast to the traditional approach, see Appendix~\ref{sec:finetuning}, \newtext{and we will need not do it at any stage}.

How to cut off from the root the unending escalade of fine-tunings
plaguing the \newtext{traditional} approach and still render a
sound value for the physical CC? A first key appears when we compute
the ``effective Einstein's equations'' of the $\F(R,\G)$-cosmology.
They emerge from functionally differentiating
(\ref{eq:CC-Relax-action}) with respect to the metric, with the
result (cf.\ Appendix~\ref{sec:FofRSTG-action} for details):
\begin{equation}
0=2\frac{\delta\mathcal{S}}{\delta g^{ab}}=\int
d^{4}x\,\sqrt{|g|}\left[\frac{1}{8\pi G_{N}}G_{ab}+ g_{ab}\,\rLi+2
E_{ab}+T_{ab}\right],\label{eq:Mod-Einstein-Eqs0}\end{equation}
where the surface terms have been omitted. Therefore, instead of
(\ref{EEvac}) we now have
\begin{equation}\label{eq:Mod-Einstein-Eqs}
G_{\,\,b}^{a}= -8\pi G_{N}\left[\rLi\,\delta_{\,\,b}^{a}+2
E_{\,\,b}^{a}+T_{\,\,b}^{a}\right]\,.
\end{equation}
Apart from the Einstein
tensor~$G_{\,\,b}^{a}=R_{\,\,b}^{a}-({1}/{2})\,\delta_{\,\,b}^{a}\,R$,
we have the constant vacuum energy $\rLi$ and the energy-momentum
tensor~$T_{ab}$ of matter. \newtext{Notice} that we have already
absorbed in $\rLi$ the vacuum effects (e.g.\ phase transitions and
\newtext{quantum effects}) associated to the coherent matter contributions from the
fields $\phi$ in the Lagrangian. \newnewtext{Therefore},  ~$T_{ab}$
in equation (\ref{eq:Mod-Einstein-Eqs0}) involves only the
incoherent matter contributions. Finally, there is a new (``extra'')
gravitational tensor $E_{\,\,b}^{a}$ coming solely from the
$\F(R,\G)$-term in the action~(\ref{eq:CC-Relax-action}), which we
have also placed on the \textit{r.h.s.}\ of the field equations in
order not to distort the standard Einstein part, usually positioned
on the \textit{l.h.s.}\

On a spatially flat FLRW background with line element
$ds^{2}=dt^{2}-a^{2}(t)d\vec{x}^{2}$, scale factor~$a(t)$ and
expansion rate $H(t)\equiv{\dot{a}}/{a}$, the tensor components of
the various gravitational parts in (\ref{eq:Mod-Einstein-Eqs}) are
given by
\begin{eqnarray}
G_{\,\,0}^{0} & = & -3H^{2}\nonumber \\
G_{\,\, j}^{i} & = & -\delta_{\,\, j}^{i}(2\dot{H}+3H^{2})\,,
\label{eq:Gij}\end{eqnarray} and
\begin{eqnarray}
E_{\,\,0}^{0} & = & \frac{1}{2}\left[
\F(R,\G) -6(\dot{H}+H^{2})\F^{R} +6H\dot{\F}^{R} -24H^{2}(\dot{H}+H^{2})\F^{\G} +24H^{3}\dot{\F}^{\G}
   \right]\label{eq:E00}\\
E_{\,\, j}^{i} & = & \frac{1}{2}\,\delta_{\,\,j}^{i}\left[
\F(R,\G) -2(\dot{H}+3H^{2})\F^{R} +4H\dot{\F}^{R} +2\ddot{\F}^{R}\right.\nonumber \\
 & & \left. -24H^{2}(\dot{H}+H^{2})\F^{\G} +16H(\dot{H}+H^{2})\dot{\F}^{\G} +8H^{2}\ddot{\F}^{\G} \right],\label{eq:Eij}
\end{eqnarray}
where $\F^{Y}\equiv\partial F/\partial Y$ are the partial
derivatives of $\F$ with respect to~$Y=R,\G$.
%
%
%
Let us remark that in all these expressions the time derivative of
the expansion rate can be reexpressed as $\dot{H}=-H^{2}(q+1)$,
where
\begin{equation}\label{eq:q}
q=-\frac{\ddot{a}\,a}{\dot{a}^2}
\end{equation}
is the deceleration parameter. This quantity will be very important
in our discussions, as we shall see immediately. For one thing the
two fundamental curvature invariants on which our action functional
$\F$ depends can just be expressed in the FLRW metric in terms of
$H$ and $q$ as follows:
\begin{equation}\label{eq:invariants}
R=6H^{2}(1-q)\,,\ \ \ \ \ \ \ \ \ \ \G=-24H^{4}q\,.
\end{equation}
Notice from equations (\ref{eq:Mod-Einstein-Eqs}) and (\ref{eq:E00})
that the effective CC density in our $\F(R,\G)$-cosmology, and
therefore the quantity playing the role of DE in our framework, is
\textit{not just} the parameter $\rLi$, but the full expression
\begin{equation}\label{eq:rLe}
\rLe(H)=\rLi+\rind(H)\,,
\end{equation}
in which
\begin{equation}\label{eq:rF}
\rind= 2 E_{\,\,0}^{0}
\end{equation}
constitutes that part of the effective CC which is genuinely induced
by the $\F$-functional. \newtext{The expression} (\ref{eq:rF}) can
be called the (gravitationally) ``induced dark
energy''\,\footnote{The name seems appropriate as long as the
gravitational functional induced by $\F(R,\G)$ is treated as a part
of the full energy momentum tensor. It would be misleading to call
$\rind$ \newtext{just} ``induced CC'', because as we have seen in
section \ref{CCfinetuning} (cf.\ also Appendix~\ref{sec:finetuning}) this name is
usually reserved for the classical and quantum contributions to the
CC emerging from the matter part (e.g.\ the quantum corrected VEV of
the Higgs potential), which we have already absorbed in $\rLi$ from
the very beginning. }. It adds up to the original cosmological
constant $\rLi$ to produce the quantity $\rLe$ or ``effective vacuum
energy density''.
\newtext{The induced DE} is obviously dynamical, and with it the
total effective DE density too. Therefore, $\rLe$ defined by
(\ref{eq:rLe}) runs with the expansion of the universe.
\newtext{Because of} (\ref{eq:invariants}), $\rLe$ is a function of
the expansion rate $H$, the deceleration parameter $q$ and its first
time derivative: $\rLe=\rLe(H,q,\dot{q})$, but for simplicity we
shall indicate it sometimes simply as $\rLe=\rLe(H)$ -- as we shall
do also with other cosmological quantities. Most important, in the
context of the $\F(R,\G)$-cosmology, the sum (\ref{eq:rLe}) is the
very observable quantity that should be accessible to observation,
as it is this quantity that clearly takes the role of the CC in the
effective Einstein's equations emerging from the action functional
(\ref{eq:CC-Relax-action}).  Put another way, $\rLe$ is the truly
``observable DE density'' of the $\F(R,\G)$-cosmology.
\newtext{Notice} that we cannot disentangle observationally the two
terms in the sum (\ref{eq:rLe}), and therefore it does not matter if
$\rLi$ is very large provided the induced term $\rind$ is also
large, but with \newtext{opposite sign}, such that the sum
(\ref{eq:rLe})
\newtext{leaves a small} remainder. Obviously, for this cosmology to
be realistic, we expect that \newtext{this remainder}
\newtext{is small enough}
and \newtext{moreover} runs only mildly with $H$,
such that it can mimic approximately the $\CC$CDM concordance model.
But, at the same time,
\newtext{and in order to avoid} the fine-tuning problem, there must
be a non-trivial \textit{dynamical interplay between the two terms}
in (\ref{eq:rLe}), leaving just a mild running residue
\newtext{\textit{at all times} of the cosmological evolution \textit{after inflation}},
\newtext{\textit{not just now}}. Let us remark that models with mildly
running cosmological term provide a global fit to LSS and CMB data
perfectly comparable to the $\CC$CDM model\,\cite{BPS09a}.

\subsection{Searching for the class of $\F$ functionals}
\label{searchingF}

We will argue that for a realistic approach of the new
$\F(R,\G)$-cosmology, we expect the following two conditions to
occur:

\begin{enumerate}

\item[i)] The effective cosmological term $\rLe=\rLe(H)$ \newtext{in equation}
(\ref{eq:rLe}) must essentially coincide with the enormous value of
the vacuum energy density in the vicinity of the post-inflationary
time, when the seeds of a huge permanent vacuum energy are first
sowed -- and need to be removed. This value is what defines our big
``initial'' cosmological constant $\rLi$, and corresponds to an
epoch characterized by an expansion rate $H\lesssim M_X$. The
function (\ref{eq:FFP}) must, therefore, allow for the following
behavior of the quantity (\ref{eq:rLe}):
\begin{equation}\label{eq:RG1}
\rLe(H)\simeq \rLi\lesssim M_X^4  \ \ \ \ \ (H_{\rm rad}^*<
H\lesssim M_X)\,.
\end{equation}
That this relation can be amply fulfilled in our framework, can be
argued as follows. To start with, remember that we have imposed the
condition (\ref{eq:limF}) on the $F$-term. For $H\sim M_X$ such
condition should presumably be satisfied (because $H_{\rm rad}^*\ll
M_X$),  and hence in this range we have
\begin{equation}\label{eq:RG2}
\F(R,\G)\to A(R)=a_2 R^2+... \ \ \ \ \
\end{equation}
where we have taken the simplest \newtext{non-trivial} possibility
for the polynomial $A(R)$ in (\ref{eq:FFP}). Despite that this term
grows with $R$, its value near the startup of the radiation epoch is
not sufficiently large yet as to distort the goodness of the
condition (\ref{eq:RG1}). For instance, for a typical GUT phase
transition with $M_X\sim 10^{16}\,\text{GeV}$ we have $R\sim H^2\sim
8\pi\,\rho_X/M_P^2\sim 8\pi\,M_X^4/M_P^2$, and thus
\begin{equation}\label{eq:ratioRrLi}
\frac{R^2}{\rLi}\sim 64\pi^2\,\frac{M_X^4}{M_P^4}\sim 10^{-9}\,,
\end{equation}
where $M_P\equiv G_N^{-1/2}\sim 10^{19}\,\text{GeV}$ is the Planck
mass. \newtext{Clearly}, the condition (\ref{eq:ratioRrLi}) would
not be so easily satisfied if the vacuum energy left after
primordial inflation would be very close to $M_P^4$. However, it
does not seem realistic (not even necessary) to try to extrapolate
cosmology up to this point, so we avoid this speculative situation
which would probably require a deeper knowledge of the space-time
structure at the level of Quantum Gravity rather than QFT in curved
space-time (as we are dealing with in our approach). In short, the
relaxation mechanism should efficiently wash out the large
contributions to the vacuum energy only \textit{after} inflation has
fully accomplished its role, and more specifically after the
reheating processes have been able to ``restore'' the ``initial''
relativistic matter content of the universe in the form of the
so-called ``radiation epoch''; but of course not before, since
otherwise inflation itself could not have occurred. It follows that
for the practical study of the relaxation mechanism we can simplify
the induced DE (\ref{eq:rF}) to the reduced form
\begin{equation}\label{eq:rF-relax}
\rind \to \rF \equiv \rind\Big|_{A(R)=0}\ \ \ \  (\text{for}\ H<M_X)\,,
\end{equation}
where the \newtext{notation $\rF$} reminds us that this part is
totally attributed to the $F$-term with no contributions from the
polynomial $A(R)$ in equation (\ref{eq:FFP}).

\item[ii)] At the same time, the effective quantity $\rLe(H)$  must
not disturb the standard thermal history of the universe, and
\newtext{should} reach the present epoch with a value
\newtext{$\rLe(H=H_0)$} very close to $\rLo\sim 10^{-47}\,\text{GeV}^4$. In
view of the fact that $\rLo\ll|\rLi|$, it means that we need a huge
\textit{dynamical} cancelation between the two terms on the
\textit{r.h.s.}\ of equation (\ref{eq:rLe}) during both the
radiation and matter epochs: $\rind \simeq -\rLi$. On the face of
(\ref{eq:rF-relax}), in practice this means that we must have
\begin{equation}\label{eq:cancellation}
 \rF \simeq -\rLi\, \ \ \ \ \ (\forall \ H\lesssim
H_{\rm rad}^*\ll M_X\ \ \text{until the present})\,.
\end{equation}
This is of course the most delicate point of our construction and
relies significantly on a suitable choice of the $F$-term. For
example, the choice in equation (\ref{eq:FRG-1B}) is  convenient
because it makes allowance for the requirement
\begin{equation}\label{eq:BF}
B(R,\G)\to 0  \ \ \ \ \ (\forall \ H\lesssim H_{\rm rad}^*\ll M_X\ \
\text{until the present})\,,
\end{equation}
as a starting point to fulfill the relation (\ref{eq:cancellation}).
Indeed, let us note that the condition (\ref{eq:BF}) insures that
$F(R,\G)$ \textit{and} its derivatives $F^{Y}\,(Y=R,\G)$ become
arbitrarily large; \newnewtext{in fact as large as $\rLi$ might be},
but not infinite because $\rLi$ is anyway finite and hence the point
$B=0$ is actually never reached.
How to make the requirement (\ref{eq:BF}) natural (without
fine-tuning) such that the relation (\ref{eq:cancellation}) is
fulfilled for arbitrarily large $\rLi$, \newtext{is something} that
we will discuss in much of the remainder of this paper.
\end{enumerate}
Some further comments are now in order. The aforementioned
conditions are interesting in that they not only define the range of
the history of the early universe in which the relaxation mechanism
of the vacuum energy must operate, they also make clear that the
inflationary scenario \newtext{can be} preserved. This setup might
actually prepare the ground for triggering primordial inflation
itself through an $R^2$-term (or a higher order polynomial) in the
functional~$A(R)$ in equation (\ref{eq:FFP-1B}). Indeed, equation
(\ref{eq:RG2}) is fulfilled for $H> M_X$, so that $R^2$ becomes
dominant in the far UV regime.
\newtext{This is consistent} with the fact that the renormalizable
quantum theory of matter fields on a curved background must
necessarily include the action of vacuum, which contains the higher
order $R^2$-curvature terms\,\cite{Parker09}.
\newtext{While these terms} are irrelevant for scales of order $M_X$ or
below -- cf.\ equation (\ref{eq:ratioRrLi}) -- they nevertheless
become dominant near the Planck scale, where $R\sim M_P^2$, and in
fact they then furnish the driving force for $R^2$-inflation. In
other words, the $\F(R,\G)$-cosmology provides also a possible
natural connection with Starobinsky's
inflation\,\cite{Starobinsky80} and the more recent developments on
anomaly-induced inflation\,\cite{AII1,AII2,Fossil07}.

Finally, a possible \newtext{additional} bonus of the polynomial
term in (\ref{eq:FFP}) is that it may provide an escape \newtext{to}
some instability issues discovered in extended gravity theories in
the metric formalism\,\cite{SotiriouFaraoni08}, as it is known e.g.\
for $\F(R)$ and $\F(R,R_{ab}^2, R_{abcd}^2)$
theories\,\cite{Carroll04}. There are also potential difficulties as
far as concerns the astrophysical implications on the solar system
measurements\,\cite{SotiriouFaraoni08}. In our case, also the
functional defined by $\F(R,\G)$ \newtext{may not be} completely
free from typical modified gravity problems~\cite{Hindawi:1995cu}.
Nevertheless, it behaves better than the general $\F(R,R_{ab}^2,
R_{abcd}^2)$-functionals \newtext{and this should suffice} for the
illustration purposes of our paper. For example, a remarkable
feature of the $\F(R,\G)$ functionals is that they avoid the
Ostrogradski instability, i.e.\ they do not involve vacuum states of
negative energy, and are thought to have a \newnewfinal{reasonable
behavior} in the solar system limit~\cite{Nojiri:2005jg,Comelli05}.
At the same time, Gau\ss-Bonnet models are automatically free from
graviton ghosts and other
singularities\,\cite{Comelli05,BambaOdintsov}.

\newtext{From the foregoing considerations}, it became clear that the polynomial contribution
$A(R)$ is relevant for restoring renormalizability at high energies,
for bridging the relaxation framework with inflation and maybe also
to help curing stability issues, but the polynomial term is not
indispensable for the relaxation mechanism itself. Therefore, we
shall hereafter substitute the full $\F(R,\G)$ functional
(\ref{eq:FFP}) for its reduced part $\beta\,F(R,\G)$, where the
$F$-term is given by (\ref{eq:FRG-1B}).

\subsection{Effective equation of state}\label{EOS}

The matter sector can be adequately described by a perfect fluid
stress tensor $T_{ab}=(\rho+p)u_{a}u_{b}-p\, g_{ab}$ with
four-velocity~$u_{a}$, energy density~$\rho$ and (isotropic)
pressure~$p$, respectively. For a fluid at rest
($u^{a}=\delta_{0}^{a}$) this means
\begin{equation}\label{Tmn}
T_{\,\,0}^{0}=\rho,\,\,\, T_{\,\, j}^{i}=-p\,\delta_{\,\, j}^{i}\,.
\end{equation}
Using a fluid description also for the vacuum energy, we define the
(effective) pressures corresponding to the induced part
(\ref{eq:rF}) \newtext{through} $p_F=-(T_F)_{i}^{i}/3$, where from
(\ref{eq:Mod-Einstein-Eqs}) we have $(T_F)_{j}^{i}=2\,E_{j}^{i}$.
Thus
\begin{equation}
 p_{F}=-\frac23\,E_{\,\, i}^{i}\,,\ \ \ \ \
\pLe=p_F - \rLi\,, \label{eq:rho-pres_F}\end{equation}
where the last expression is the corresponding pressure for the full
effective vacuum energy (\ref{eq:rLe}). Since the Einstein
tensor~$G_{ab}$ and the extra gravitational tensor~$E_{ab}$ are both
covariantly conserved, it follows that the $\F(R,\G)$-cosmology
automatically conserves matter, too, i.e.\ $\nabla^a T_{ab}=0$.
Therefore, the Bianchi identity on the FLRW background leads to the
local covariant conservation laws
\begin{equation}
\dot{\rho_n}+3H(\rho_n+p_n)=0\label{eq:Bianchi}\,,\end{equation}
which are valid for all the individual components ($n=$ radiation,
matter and DE) with energy density~$\rho_n$ and pressure~$p_n$. In
particular, the conserved matter ($p_m=0$) and radiation
($p_r=\rR/3$) components integrate immediately and yield the usual
expressions $\rho_{m}=\rho_{m}^{0}a^{-3}$ and
$\rho_{r}=\rho_{r}^{0}a^{-4}$, whereas the effective CC density is
related to the corresponding pressure through equation
(\ref{conslawDE}), \newtext{which in the present case is equivalent}
to
\begin{equation}\label{localCCF}
\dot{\rho}_{F}+3H(\rF+\pF)=\dot{\rho}_{F}+3H\,\left(1+\wF(t)\right)\rF=0\,,
\end{equation}
because the constant part of the effective CC satisfies
$p_{\CC}^i=-\rho_{\CC}^i$ and therefore cancels out in
(\ref{localCCF}).
\newtext{In the previous} equation, we have defined
\begin{equation}\label{eq:omegaeffF}
\wF(t)=\frac{\pF(t)}{\rF(t)}\,,
\end{equation}
which plays the role of the EOS for the $F$-term.
\newtext{As it is obvious}, the conservation law (\ref{localCCF}) is
a particular case of the general $\CC$XCDM one (\ref{conslawDE2})
for $\rL=\rLi=\text{const}$. The role of the cosmon density $\rX$ is thus
played by $\rF$. The local conservation law (\ref{localCCF}) cannot
be directly integrated because the EOS parameter
(\ref{eq:omegaeffF}) is actually a non-trivial function of the
cosmological evolution. The parallelism with the $\CC$XCDM model can
be made even more manifest if we define the overall EOS of the
compound DE system formed by the constant $\rLi$ and the induced
$\rF$ component as $\pLe=\weff\ \rLe$. It is easy to see that
$\weff$ is related with (\ref{eq:omegaeffF}) as follows (for
convenience we trade time for redshift):
\begin{equation}\label{eq:omegaeff}
\weff(z)=-1+\left(1+\wF(z)\right)\,\frac{\rF(z)}{\rLi+\rF(z)}\,,
\end{equation}
which is again a particular case of the general $\CC$XCDM effective
EOS defined in equation (\ref{eEOS}), with the correspondence
$\rX\to\rF$, $\wX\to\wF$ and $\rLi+\rF(t)=\rLe$.

\newtext{The expressions for $\rF$ and $\pF$} computed in section \ref{action} are
not just opposite in sign. Thus, we do not expect $\wF=-1$, and
therefore $\weff\neq -1$ either. Both are non-trivial functions of
time or redshift: $\wF=\wF(z)$ and $\weff=\weff(z)$. Models with
variable cosmological parameters indeed usually exhibit this
feature\,\cite{SS12}. In our model universe, we see from
(\ref{eq:rF}) and (\ref{eq:rho-pres_F}) that the departure from a
strict cosmological constant behavior is caused by
\newtext{the fact} that $E_{\,\,0}^{0} \neq (1/3)E_{\,\,i}^{i}$ in
equations (\ref{eq:E00}) and (\ref{eq:Eij}). \newtext{This relation}
would hold only if all derivatives $F^{Y}\equiv\partial F/\partial
Y=0$ \ ($Y=R,\G$), but this is impossible for the typically needed
structure for $F$, see equation (\ref{eq:FRG-1B}). The $\weff\neq
-1$ feature, therefore, will actually persist for the entire
cosmological history,
\newtext{but we expect} that the departure from $-1$ will not be very important
near our time because we are currently observing a predominance of
the ``DE epoch'' ($\rM,\rR\ll\rLe$), in which the DE behaves
essentially as a CC term. \newtext{So we should ensure} that
$\weff(z)\simeq -1$ for redshift $z\simeq 0$.

\newtext{Well within the original spirit} of the class of $\CC$XCDM models,
the local covariant conservation law satisfied by the cosmon --
equation~(\ref{localCCF}) -- plays a fundamental role to elucidate
the dynamics of the model. Thanks to this conservation law (which
acts as a first integral of our dynamical system), there is no need
to use the complicated expression (\ref{eq:Eij}) -- which leads to a
differential equation of one order higher than (\ref{eq:E00}).
Therefore, it is not necessary to use (\ref{eq:Eij}) to find $\pLe$
through (\ref{eq:rho-pres_F}). In practice, the effective EOS can be
determined with the help of the local covariant conservation law
\,(\ref{conslawDE}) or just (\ref{localCCF}). For this, $\rLe$ is to
be determined first.

\newtext{But how to find} explicitly the effective vacuum energy $\rLe$?
To this end let us consider the $_{\,\,0}^{0}$-component of the
Einstein equation~(\ref{eq:Mod-Einstein-Eqs}):
\begin{equation}
3H^{2} = 8\pi G_{N}(\rho_{m}+\rho_{r}+\rLe)\,.\label{eq:Einstein-H2}
\end{equation}
This is the generalized Friedmann's equation for the
$\F(R,\G)$-cosmology, in which the role of the CC is played by
$\rLe$. Such equation provides the clue for integrating the field
equations and determine all relevant energy densities. Substituting
equation (\ref{eq:rF}) in the formula (\ref{eq:rLe}) for the
effective vacuum energy $\rLe$, one obtains an expression that
depends on $H$, $q$ and $\dot{q}$. In practice, for any function
$f$, we can trade the time evolution of the resulting expression for
the scale factor dependence through $\dot{f}=aH(a)\,f'(a)$. In
particular, $\dot{H}(a)=aH(a)H'(a)$ and moreover from
$q(a)=-1-aH'(a)/H(a)$ we have
$\dot{q}(a)=-aH'(a)-a^2[H''(a)-H'^2(a)/H(a)]$. Proceeding in this
way with all the terms, we arrive at the form
$\rLe=\rLe(a,H(a),H'(a),H''(a))$. Therefore, Friedmann's equation
(\ref{eq:Einstein-H2}) can be finally cast as a second order
differential equation for $H(a)$ that can be solved numerically.
Once $H(a)$ is known, the effective vacuum energy $\rLe=\rLe(a)$ is
also known, and can be plugged in the local conservation
law~(\ref{conslawDE}) to determine $\pLe=\pLe(a)$, and from here the
desired EOS $\weff(a)=\pLe(a)/\rLe(a)$ ensues.  We shall follow this
procedure in practice.

Furthermore, the $_{\,\, j}^{i}$-component \newfinal{contains}
information on the acceleration, and can be expressed in terms of
the deceleration parameter (\ref{eq:q}) as follows:
\begin{equation}
3\,H^{2}\,q = 4\pi
G_{N}\left[2\,\rho_{r}+\rho_m+(1+3\weff)\,\rLe\right]\,.\label{eq:Einstein-q}
\end{equation}
Using (\ref{eq:Einstein-H2}), we can recast this expression such
that its \textit{r.h.s.} contains just the sum of pressure
components:
\begin{equation}
H^{2}(q-1/2) = 4\pi G_{N}(p_{r}+\pLe)=4\pi
G_{N}\left[\rR/3+\weff\,\rLe\right]\,.\label{eq:Einstein-q-1-2}
\end{equation}
Similarly, we obtain
\begin{equation}
3\,H^{2}(1-q)  =  4\pi G_{N}\left[\rM+\rLe-3\pLe\right]=4\pi
G_{N}\,\left(\rM+(1-3\weff)\,\rLe\right)\,.\label{eq:Einstein-q-1-3}
\end{equation}
The last two equations are convenient forms of the dynamical
equation for the acceleration, \newnewfinal{they are written} in
terms of the deceleration parameter (\ref{eq:q}) and will be used
later on. By ignoring $\rLe$ for the moment it is easy to see from
(\ref{eq:Einstein-q-1-3}) that the radiation epoch (in which $\rM$
can be neglected) is characterized by $q=1$, whereas from
(\ref{eq:Einstein-q-1-2}) it is transparent that the matter epoch
(in which $\rR$ can be neglected) is characterized by $q=1/2$. In
Secs.~\ref{sec:EOS-matter-era} and~\ref{sec:EOS-radiation-era} we
will show that these observations still hold when taking into
account $\rLe$.

\newfinal{Let us note} that the above formulae just \newnewfinal{follow} the normal pattern
of equations characterizing a cosmological medium which is composed
of several fluids. If these fluids have  EOS parameters
$\omega_n=p_n/\rho_n$ and density parameters
$\Omega_n(a)=\rho_n(a)/\rho_c(a)$ normalized with respect to the
critical density $\rho_c(a)=3H^2(a)/(8\pi G_N)$, one can easily show
that the deceleration parameter (\ref{eq:q}) can be expressed as
\begin{equation}\label{qomegas}
q=\sum_n\,(1+3\omega_n)\,\frac{\Omega_n}{2}\,; \ \ \ \ \ \ \ \
\sum_n\,\Omega_n=1\,.
\end{equation}
Clearly $q=(1,1/2,-1)$ for radiation ($\omega_R=1/3$), matter
($\omega_m=0$) and standard vacuum energy ($\omega_{\CC}=-1$)
dominated epochs respectively, which correspond to having the
dominant density parameter in each epoch $\Omega_n=1$ and all the
others zero. In the present epoch, we have a mixture of matter and
DE in which the latter behaves very approximately as vacuum energy,
therefore the current value of the deceleration parameter in the
$\CC$CDM model is generally expressed as
$q_0=\Omega_m^0/2-\Omega_{\Lambda}^0$. The only note of caution is
that, within the framework under consideration, the effective CC
term $\rLe$ does not behave as standard vacuum energy because it has
a non-trivial EOS $\weff$, which is not equal to $-1$, and in
general is a complicated function of time $\weff=\weff(t)$ or of the
redshift --  see (\ref{eq:omegaeff}). With this only proviso,
equation (\ref{eq:Einstein-q}) is easily seen to follow from the
general one (\ref{qomegas}) accounting for a mixture of fluids.
\newfinal{In this way}, defining $\Omega_{\Lambda {\rm
eff}}(a)=\rLe(a)/\rho_c(a)$ also for the effective vacuum fluid of
our model, the value of the deceleration parameter at present is to
be written as
\begin{equation}\label{q0relax}
q_0=\Omega_m^0/2+(1+3\weff^0)\Omega_{\Lambda {\rm
eff}}^0/2=\frac12\left(1+3\weff^0\,\Omega_{\Lambda {\rm
eff}}^0\right)\,,
\end{equation}
where the second equality is valid only if the universe is spatially
flat. Here $\Omega_{\Lambda {\rm eff}}^0$ and $\weff^0$ are the
current values of these quantities. Equation (\ref{q0relax}) is
obviously consistent with (\ref{eq:Einstein-q}) when the radiation
contribution is neglected. In sections~\ref{sec:EOS-matter-era}
and~\ref{sec:EOS-radiation-era} we will discuss the relation of the
non-trivial EOS $\weff$ of the effective vacuum energy $\rLe$ with
the EOS of matter and radiation in the various epochs.

\subsection{Evading a ``no-go theorem''}

Before further exploring our $\F(R,\G)$-cosmology, let us briefly
comment why it has a chance to evade Weinberg's ``no-go theorem''
for dynamical adjustment mechanisms of the cosmological
term~\cite{weinberg89}. The theorem is formulated for a system of
scalar fields $\varphi_j$ non-minimally coupled to gravity, and the
basic claim is that it is impossible to find a stable vacuum state
for this system that coincides with flat space-time in the gravity
sector, unless fine-tuning is used. In a very simplified form, the
proof is based on studying the consistency of the combined set of
equations defining the existence of the necessary extremum for
constant scalar fields and metric, namely the set of derivatives of
the matter Lagrangian with respect to all the fields equated to
zero:
\begin{equation}\label{nogo1}
\frac{\partial\mathcal{L}_{\phi}}{\partial\varphi_j}=0\,,\ \ \ \ \ \
\ \ \frac{\partial\mathcal{L}_{\phi}}{\partial g_{ab}}=0\,.
\end{equation}
The solutions of this system must be compatible with the solution of
the trace $T\equiv T_{\,\,a}^a$ of the energy-momentum tensor being
zero at the same point in (constant) field space, i.e.\
\begin{equation}\label{nogo2}
T=2\,g_{ab}\,\frac{\partial\mathcal{L}_{\phi}}{\partial
g_{ab}}-4\,\mathcal{L}_{\phi}=0\,.
\end{equation}
What are the chances for this possibility? A good start would be
that the two expressions on the respective \textit{l.h.s.}\ of the
system (\ref{nogo1}) would be proportional, or related by a linear
transformation. Then, a ground state solution in $\varphi_j$-space
would automatically be compatible with a solution of constant metric
$g_{ab}$, which we may suggestively call $\eta_{ab}$ (Minkowski).
Besides, this would also imply that the first term on the
\textit{r.h.s.}\ of (\ref{nogo2}) would vanish. Unfortunately, this
does not guarantee yet that the energy-momentum tensor vanishes
unless the second term, viz.\ $-4\mathcal{L}_{\phi}=+ 4\,V_{\rm
eff}(\varphi_j)$, also vanishes. But this could not happen unless we
would fine-tune to zero the value of the ground state of the
effective potential of the scalar fields, quite in the same
contrived way as discussed in section \ref{CCfinetuning} -- except
that here we would have exactly zero on the \textit{l.h.s.}\ of
equation (\ref{finetuningclassical}). Therefore, in general there is
a ``no-go'' conclusion about the possibility of having $\eta_{ab}$
as the metric solution just at the point of field space where it is
localized the ground state of the $\varphi_j$ fields. In the old
days, it was expected that if this ground state value is zero, then
there would be some hope that some symmetry or dynamical mechanism
would help reaching it without fine-tuning. But it does not seem to
be the case, as Weinberg's no-go theorem claims\,\cite{weinberg89}.

The root of the problem lies on the fact that the above system of
equations is over-constrained. If we would, instead, not require to
have a constant solution $\eta_{ab}$ in the current vacuum state,
the problem should not arise, because then the metric at any time --
and in particular in the present universe -- could be a dynamical
one, typically the FLRW metric, with $g_{ab}$ a function of the
scale factor $a=a(t)$. In this case, the second equation in
(\ref{nogo1}) would not hold. If, in addition, we do not require
that the expression for the trace of the energy-momentum tensor
vanishes at the ground state for matter fields -- and in particular
neither at a point where $g_{ab}$ is constant -- the system becomes
\newtext{less and less} constrained and we should not expect any impediment
to demand that $T$ carries, at the present time, some non-vanishing,
even if small, energy density compatible with the curved space-time
metric of our current epoch. In short, by loosing the constraints
(by allowing dynamical metric -- hence space-time curvature -- and
non-zero vacuum energy at any time) the no-go conclusion disappears.
This is exactly our situation. The ``only'' final difficulty lies in
achieving the right non-vanishing value of the current $T$, namely
one which is small enough for particle physics standards. Here is
precisely where the full power of the relaxation mechanism enters.
The rest of the paper is devoted to explain why and how this is
possible.

\section{Dynamical relaxation of the vacuum energy}\label{sec:CC-relaxation}

In Einstein's General Relativity, the theoretically expected large
vacuum energy density $\rho_{\Lambda}^{i}$ which was released at the
early stages of the cosmic evolution would drastically change the
essential features of the standard cosmological paradigm, in
particular it would prevent the well-established thermal history and
all the astounding successes of the Big Bang universe. This problem
can be solved either by extreme fine-tuning or by a dynamical CC
relaxation mechanism, which is the subject of this work.

The big value $\rLi$ should prevail at times prior to the radiation
epoch, in particular during the fast de Sitter expansion that
characterizes the primordial inflationary phase. However, a
``residual'' vacuum energy of respectable (even of comparable) size
is expected to remain in the universe in the vicinity of the
incipient radiation epoch, i.e.\ the epoch that ensues after the
universe loiters for a while in the reheating state, namely that
state which is responsible for ``re-creating'' all the
(relativistic) matter out of the decay of the inflaton or any other
inflationary driving force. In fact, there is no reason to expect
that after inflation the universe will roll down into a vacuum state
of very small energy. The ``residual'' energy left in the reheating
vacuum can be called again $\rLi$, because it could perfectly be of
the same order of magnitude. Let us take into account that nothing
is accessible to us before this time, and much less if we move deep
into the inflationary era. Therefore, the relaxation mechanism must
be operative only after inflation has ceased and the turbulent state
of the universe, caused by the reheating mechanism, has finally
homogenized the fluid and triggered the primeval radiation epoch
within the FLRW metric.

Of course we cannot easily describe the interpolation processes that
made possible the transition from the de Sitter inflationary phase
into the FLRW phase, and much less without a fundamental microscopic
understanding of the very early universe (string theory,
brane-world, M-theory?). This goes beyond our main purpose in this
paper, which is only to demonstrate that a dynamical mechanism to
relax the CC can be explicitly constructed. Thus, we shall just
assume that the transition took place and that, after reheating, the
universe was left in principle with a significant vacuum energy
$\rLi$ of the order of the initial de Sitter one. There is no reason
whatsoever (apart from an unacceptable fine-tuning of the initial
conditions) to expect that a sizeable vacuum energy is not there, so
unless the universe unleashes automatically some countermeasures to
reduce it fast at a minimum level, it may completely ruin the onset
and full development of the standard thermal history of the Big Bang
model, in particular the primordial and very successful
nucleosynthesis of the light elements. For this reason, the
neutralization process of $\rLi$ must be immediately put to work
with utmost efficiency.

The previous description tells us \textit{when} our relaxation
mechanism is supposed to start working. But the next (and highly
non-trivial) question is: \textit{how} does it work? To introduce the
mechanism of relaxation in our modified gravity framework, let us
make some ansatz within the class of the functionals $\F(R,\G)$
defined by equation (\ref{eq:FFP-1B}). Remember that we need to
satisfy some properties described in section \ref{searchingF}. For
definiteness, let us choose for the polynomial in $R$ and $\G$ the
expression
\begin{equation}\label{eq:Bgeneral}
B(R,\G)=b_2\,R^2+c\,\G+b_n\,R^n\,,
\end{equation}
in which $n$ is some integer different from $2$. This simple ansatz
is already sufficient to explain the basic principle, which carries
over to more complicated models like $F=R^{2}/B^{3}$ etc that we
will address briefly later on. It is obvious from the choice
(\ref{eq:Bgeneral}) that in general the corresponding $F$-term will
trivially satisfy the condition (\ref{eq:limF}). But a more
difficult question is whether it can satisfy the dynamical
neutralization condition (\ref{eq:cancellation}) as well, which must
hold for \textit{all} $H$ below $H_{\rm rad}^*$ until the present
epoch. Amazingly enough, achieving this feat is possible by an
appropriate choice of the first two coefficients of the polynomial
(\ref{eq:Bgeneral}), whereas the third coefficient and the power $n$
just control some smoothing properties of the thermal history, as we
shall see below. Notice that if $b_2$ and $c$ are dimensionless,
then $b_n$ is dimensionful, with dimension $4-2n$ of mass, i.e.\
$M^{4-2n}$. At the same time, it is clear from dimensional analysis
that, for the present ansatz, the coefficient $\beta$ in equation
(\ref{eq:betaM}) has mass dimension $N=8$.

\subsection{A toy model}\label{sec:toy-model}

Let us now explain how the relaxation mechanism works, and let us do
it by making use of a simpler (albeit non-trivial) toy-model example
which contains already some of the main ingredients. Relaxation
means that the observed energy density $\rLe$ can be made much
smaller in magnitude than the initial $\rho_{\Lambda}^{i}$. This can
be achieved by making the induced part $\rind$ (\ref{eq:rF}) large
in magnitude, and opposite in sign to $\rLi$, such that the two
terms in equation (\ref{eq:rLe}) conspire to keep the sum
$|\rLi+\rind|\ll |\rLi|$ for the full stretch of the
post-inflationary cosmic evolution. With the current ansatz, the
previous condition is realized when the denominator $B$ is
sufficiently small, $B\rightarrow 0$, but non-zero.

As explained in section \ref{searchingF}, for the study of the
relaxation mechanism we can take the reduced form~$\rF$
(\ref{eq:rF-relax}) of $\rind$. As a warm-up, let us consider the
polynomial (\ref{eq:Bgeneral}) for the particular case where $b_2=1$
and $c=b_n=0$. Then, $B$ takes the simplest possible
\newtext{structure} $B=R^{2}=36\,H^4(1-q)^{2}$ -- see
equation (\ref{eq:invariants}) -- and
\newtext{we have}
\begin{equation}\label{eq:warmup}
\rLe(H)=\rLi+\rF=\rLi+\beta\,\left[\frac{1}{36\,H^4(1-q)^{2}}+\E(H,q,\dot{q})\right]\,,
\end{equation}
where $\E(H,q,\dot{q})$ represents the terms in (\ref{eq:E00})
containing derivatives of $\F$ (i.e., essentially of $F$). Notice
that this toy-model example is similar to the one studied in
Ref.~\cite{BSS09a}, see equation (6) of the latter. The difference,
however, is that here we use $1/R^2$ rather than $1/R$, and that we
have the presence of the function $\E(H,q,\dot{q})$. The latter is a
direct consequence of performing our analysis of the relaxation
mechanism from an action functional rather than imposing the form of
the new terms at the level of the field equations. However, none of
these differences will change the qualitative behavior of the
relaxation mechanism nor the fact that this setup, despite it
contains the first clues to the dynamical relaxation, is still too
simple for making it work realistically.

The dynamical relaxation of $\rLe$ originates from the~$(1-q)$
factor in the denominator of the expression $\rF$ in equation
(\ref{eq:warmup}). The large~$\rLi$ left over in the immediate
post-inflationary period drives the deceleration parameter~$q$ to
larger values until $q\to 1$, which corresponds to radiation-like
expansion. In other words, the very existence of the radiation
period is triggered automatically by the presence of this term,
which can be thought of as a countermeasure launched by the universe
against the presence of the large ``residual'' vacuum energy $\rLi$
at the pre-radiation era. In view of the form of (\ref{eq:warmup}),
we expect that there will be a significant dynamical neutralization
of $\rLi$ during this epoch, for an appropriate sign of the
parameter $\beta$. Although $q$ is driven dynamically to $q\to 1$,
it cannot cross $q=1$ from below since, then,~$\rF$ would dominate
over~$\rLi$ and stop the cosmic deceleration before~$q$ reaches~$1$.

Let us also clarify that the function $\E$ in (\ref{eq:warmup}) is
not just a passive spectator, as it contributes alike to the
neutralization process. The reason is that the terms of $\E$ with
derivatives $F^{R}$ and $\dot{F}^{R}$ furnish contributions to $\rF$
of the form $1/(1-q)^3$ and $1/(1-q)^4$, respectively.
\newtext{Thus}, we end up with a general expression of the form
\begin{equation}\label{eq:warmup2}
\rLe(H)=\rLi+\rF=\rLi+\beta\,\left[\frac{{\cal
N}_2}{(1-q)^{2}}+\frac{{\cal N}_3}{(1-q)^3}+\frac{{\cal
N}_4}{(1-q)^4}\right]\,,
\end{equation}
where the functions ${\cal N}_{i}(H,q,\dot{q})\,(i=2,3,4)$ do not
contain the factor $B$.
\newtext{As we see}, the additional dynamical terms on the
\textit{r.h.s.}\ of equation (\ref{eq:warmup2}) stay on equal
footing as far as their ability to neutralize the
$\rho_{\Lambda}^{i}$ term. As advanced in point ii) of section
\ref{searchingF}, one can show that the validity of the argument is
general for any $F(R,\G)$ of the form (\ref{eq:FRG-1B}). At the same
time, by
\newtext{dimensional reasons} we have ${\cal N}_{i}(H,q,\dot{q})\to 0$ as $H\to\infty$,
and therefore the condition (\ref{eq:limF}) is satisfied. By the
same token, ${\cal N}_{i}(H,q,\dot{q})\to\infty$ as $H\to 0$.

Despite there is a tremendous cancelation in (\ref{eq:warmup2})
between $\rLi$ and the ``$\beta$-terms'', i.e.\
$|\rLi+\rF|\ll|\rLi|$, there is in fact no fine-tuning anywhere. The
compensation is dynamical, and hence automatic, i.e.\ triggered by
the evolution itself of the universe. The point is that $\rLi$ and
$\rF$ want to drive the deceleration parameter $q$ to different
directions. Let us e.g.\ consider a dominant negative vacuum energy
density $\rho_{\Lambda}^{i}<0$ (as it would be e.g.\ the case of the
electroweak energy of the Higgs potential in the SM (see section
\ref{CCfinetuning}). Then, $\rLe\approx\rho_{\Lambda}^{i}<0$ at the
initial stage of the radiation epoch. This big negative vacuum
energy would tend to produce a dramatic deceleration of the
expansion, but at the same time~$q$ is fast driven to $1$ until the
terms in $\rho_{F}$ that increase with inverse powers of $(1-q)$
become sufficiently big to compensate for $\rLi$. Put another way,
$\rF$ acts as a ``dynamical counterterm''. Ultimately, the
``fine-tuning'' between $\rLi$ and $\rF$ is indeed there, but it is
\textit{not} ``man-made'', it is rather dictated dynamically by the
universe itself!

Worth emphasizing is the fact that the relaxation solution is
dynamically stable. To see this, take again the case of a large and
negative $\rLi$. The driving of $H$ (by $\rLi<0)$ to small values
becomes compensated by the large and positive contribution of~$\rF$, which, as we have seen, grows as~$H$ decreases. On the other
hand, any attempt of $H$ at growing inordinately large would be
deactivated automatically by the decreasing $\rF$, which would make
the term~$\rLi<0$ to take over again and render $H$ stable. In
other words, $\rLi$ and $\rF$ monitor each other, and this feedback
results in the complete stabilization of the expansion rate. As
already mentioned, this stabilization is what impedes $q$ ever
reaching the exact value $1$. It just approaches $1$ the exact
amount to get the $\rLi$ term sufficiently counterbalanced.

In this dynamical relaxation process for the CC, tiny changes of the
deceleration~$q$ near $1$ are sufficient to compensate for changes
in the detailed structure of $F$ or for large variations in the
value of $\rLi$. In the latter case, it means that the mechanism
automatically self-adapts to any modification of the initial
conditions setting the value of $\rLi$. For example, it works
equally well if the original vacuum energy is \newtext{of the order}
$\rLi\sim (10^{2}\,\text{GeV})^4$ (as in the SM) or if it is much larger
(e.g.\ in a typical GUT) and very much accurate, say with the precise
value $\rLi=(5.648310279\times 10^{16}\,\text{GeV})^4$ etc. This
self-adapting dynamics is also the reason for the absence of
fine-tuning in our setup.

Note that, at the equilibrium point (in this example $q\approx 1$),
both terms in $\rLe$ are almost equal to each other apart from
opposite signs (for an appropriate sign choice of $\beta$).
Therefore, each term $\rLi$ and $\rF$ could be well approximated as
a cosmological constant. However, their sum is not constant in
general, which is the result of the implicit time-dependence in
$\rF(H(t))$. If mild enough, the running property of $\rLe(H)$ with
the expansion rate can remain almost undetected to us and can
perfectly simulate the $\CC$CDM model. Overall, two  large
approximate CC terms conspire to give a much smaller CC-like term,
the observed one! We shall see explicit numerical examples in
section \ref{sec:numerical}. Besides, there is a corresponding
compensation of the terms $p_{F}$ and $-\rLi$ in the effective
vacuum pressure $\pLe$ in (\ref{eq:rho-pres_F}), and we have already
seen in (\ref{eq:omegaeff}) that the EOS of the effective vacuum
energy density is not constant in general.

Let us point out that the cases $\rLi<0$ and $\rLi>0$ are
qualitatively distinct. If $\rLi$ is large and negative (e.g.\ as in
the electroweak vacuum), then the driving of $q$ to $1$ is enforced
automatically by the $\rF$ term in (\ref{eq:warmup}), just to avoid
that $H^2<0$. In this sense, the $\rLi<0$ vacuum ``brings forth''
the radiation epoch as something inevitable after the primordial
post-inflation period. However, in the alternative situation in
which $\rLi$ is positive and large, in principle nothing prevents
the universe from still continuing in the de Sitter phase, unless
the vacuum energy starts to decay into radiation (e.g.\ by virtue of
some particle physics processes associated to the reheating
mechanism). \newtext{We cannot} describe this decaying mechanism in
our framework, but we must assume it has happened, and hence it
should trigger the formation of ``bubbles'' of the $q=1$ state in
the vacuum. \newtext{From here onwards} the relaxation mechanism
takes its turn an can automatically remove most of the vacuum energy
from this state, thereby transforming it into a a normal heat bath
of relativistic particles. Only after most of the vacuum energy
would be neutralized, the evolution of the relativistic particles
(radiation) in the heat bath could follow the pattern of the
standard FLRW radiation epoch.

In summary, in this toy-model example we have all the essential
ingredients for the relaxation mechanism to work. The latter may not
only trigger the appearance of the radiation epoch (specially if
$\rLi<0$) and protects it from the devastating effects of a large
vacuum energy remnant in the post-inflationary time; quite
remarkably, it also predicts a very small value of the effective
vacuum energy at the present time, which, amazingly enough, is the
most sought-for ``miracle'' needed to solve the big cosmological
constant problem. Indeed, in the current epoch the condition
$q\simeq 1$ has long ceased to hold and the compensation of $\rLi$
by $\rF$ in equation (\ref{eq:warmup}) can only occur because $H$ in
the denominator of $\rF$ has attained a very small value. How small
is this value? The presence of the complicated term $\E$ may obscure
an analytic estimate here, and although we shall present later an
exact numerical solution of a more realistic model, let us now
simplify things momentarily by considering the effect of the
$F$-term only -- i.e.\ imagine that the $\E$-term in
(\ref{eq:warmup}) is absent.
\newtext{This is} tantamount to say that we assume ${\cal N}_3={\cal N}_4=0$
in equation (\ref{eq:warmup2}). Then, since in the current universe
we must have $|\rLi+\rho_{F}(H)|/\rLi\ll 1$, with
$\rF\sim\beta/H^4$, it follows that the value of $H$ that solves
this equation is approximately given by
\begin{equation}\label{Hstar}
H_{*}\sim \left(\frac{\beta}{|\rLi|}\right)^{1/4}\,,
\end{equation}
which is \newtext{well defined} because, in this example, $\rLi<0$
and hence we have to choose $\beta>0$. Furthermore, from this
expression it is patent that the small value of $H_{*}$ at the
present time is just caused by the large magnitude of $\rLi$ at the
early times! The values of $H_{*}$ and $\rLe$ are
\newtext{connected} by equation (\ref{eq:Einstein-H2}), i.e.\ approximately by $3H_{*}^2\simeq
8\pi G_N\,\rLe$ (if we neglect the current matter contribution,
which is anyway smaller than the observed CC). Finally, we can
attain $H_{*}\simeq H_0$ by an appropriate choice of the magnitude
of the parameter $\beta$, or equivalently by the mass scale ${\cal
M}$ in equation (\ref{eq:betaM}), with $N=8$. For instance, taking
$\rLi\sim M_X^4$, with $M_X\sim 10^{16}$ GeV, and using $H_0\sim
10^{-42}$ GeV, we easily find ${\cal M}\sim 10^{-4}$ eV, which is in
the range of light neutrino masses, i.e.\ a reasonable mass scale
for particle physics standards!

\subsection{More realistic cosmological models}\label{sec:Realistic-model}

As we have seen in the previous section, the simple \newtext{choice}
$b_2=1$ and $c=b_n=0$ made for the coefficients of the polynomial
(\ref{eq:Bgeneral}) provides a cosmology endowed of truly remarkable
properties. Unfortunately, that choice is too simpleminded for a
realistic description of our universe. The reason is that while the
cosmos in that scenario goes through a radiation epoch ($q=1$) there
is no possibility to drive it into a subsequent matter epoch
($q=1/2$). Obviously, this is a fundamental shortcoming. In the
following we shall try to amend this difficulty and we will discuss
the evolution of the effective vacuum energy in a more realistic CC
relaxation model in which all relevant epochs are finally included.
This case will be more complicated than the toy-model from the
previous section, but the working principle is the same and for this
reason we have explained it with some detail there. Our starting
point is a model still based on the $F$-term ansatz
(\ref{eq:FRG-1B}) and with a $B$-polynomial of the generic form
(\ref{eq:Bgeneral}). Again, for the analysis of this setup it is
useful to characterize the CC relaxation ($|\rLe|\ll|\rLi|$) with
the condition $B\rightarrow 0$, although $B$ does not strictly
vanish, as we have explained in the toy-model example. From this
condition, we will derive in the next section approximate analytical
results for the evolution of the effective CC term (\ref{eq:rLe}),
which we shall support with numerical simulations.

For a realistic model of this kind, the polynomial $B$ must have
appropriate \newtext{coefficients} $b_2$ and $c$ such that the $R^2$
and $\G$ terms produce a neat $(q-1/2)$ factor, and besides we need
a non-vanishing coefficient $b_n$ to insure that the $R^n$ term will
provide a $(q-1)$ factor as well. Therefore, we introduce the
polynomial
\begin{equation} B(R,\G)=\frac{2}{3}R^{2}+\frac{1}{2}\G+(y\,
R)^{n}=24H^{4}(q-\frac{1}{2})(q-2)+\left[6\,y\,H^{2}(1-q)\right]^{n}\,,\label{eq:B}
\end{equation}
where we have used equation (\ref{eq:invariants}). Comparing with
the example above, it is easy to see that the second term $\sim
H^{2n}(1-q)^{n}$ will relax the effective CC in the radiation era
($q\approx1$). In order for this term to dominate at Hubble
rates~$H$ characteristic of that epoch, we must require $n>2$ as
this insures that the last term of equation (\ref{eq:B}) increases
faster than $H^4$ at high $H$ -- i.e.\ faster than the first
term. The latter, on the other hand, will be responsible for the
relaxation in the matter era~($q\approx{1}/{2}$) for lower values of
$H$, namely for $H<H_{\rm eq}$, where $H_{\rm eq}$ is the Hubble
rate just at the transition time from radiation to matter.
Numerically, $H_{\rm eq}\sim 10^{5}\, H_{0}$, corresponding to a
temperature of $T\sim \text{eV}$. Additionally, the exponent $n>2$
determines the smoothness of the radiation--matter transition.
Finally, the dimensional parameter $y$ fixes the redshift of the
transition. Obviously, it will be of order
\begin{equation}\label{eq:yparam}
y\sim H_{\rm eq}^{(4-2n)/n}\,,
\end{equation}
as this is the point where the two terms on the \text{r.h.s.}\ of
(\ref{eq:B}) will be of the same order. We thus have only two free
parameters in the polynomial (\ref{eq:B}), which if added to the
parameter $\beta$ (or, equivalently, the mass scale $\M$) in
(\ref{eq:betaM}), it makes a total of three free parameters:
\begin{equation}\label{eq:freeparam}
(\beta,n,y)\,.
\end{equation}
With only this small number of parameters the relaxation mechanism
can be made to work in a pretty realistic way, as we shall
demonstrate explicitly in the next sections.

Before closing this section, the following comment is in order. For
$H\gg H_{\rm rad}^*$ (see section \ref{action}), and specially near
the de Sitter phase at $H\sim M_X$, the deceleration parameter
should be forced by the mechanism of primordial inflation to stay
near $q=-1$, and therefore the $F$-term should satisfy the condition
(\ref{eq:limF}) since the polynomial (\ref{eq:B}) becomes
numerically large at high $H$ when it is away from the region where
$q=1$. Notwithstanding, we must admit that we do not have at present
a precise control of the interpolation regime between the
inflationary period and the onset of the standard FLRW cosmological
evolution. This is actually a general problem plaguing all
inflationary models. Therefore, we are not supposed to describe at
this stage the corresponding evolution of the relaxation
$\F$-functional from one period to the other.  In particular, the
functional form of the $F$-term in the general structure of $\F$
could change during inflation; in fact, its ultimate origin goes
beyond the scope of this investigation. But irrespective of the
details of the underlying fundamental theory of the $\F$-functional,
we expect that the condition (\ref{eq:limF}) should be satisfied in
order to preserve the mechanism of inflation prior to the startup of
the FLRW cosmology. At energies close to $H\sim M_X$ the behavior of
the complete $\F$-functional (\ref{eq:FFP}) should not interfere
with this fact, and it \newtext{is} thus reasonable that it takes
the polynomial form (\ref{eq:RG2}). After all, this UV form of the
effective action (if no powers higher than $R^2$ are involved) is
the one that is expected for the standard renormalizable effective
action of QFT in curved space-time\,\cite{Parker09,Shapiro:2008sf}.
At the same time, it would naturally provide Starobinsky's type
mechanism of primordial inflation\,\cite{Starobinsky80} and modified
formulations thereof\,\cite{AII1,AII2,Fossil07}. This is a most
natural expectation in a framework where the main job of solving the
cosmological constant problem is accomplished precisely by gravity
itself rather than by introducing extraneous scalar fields.

\section {Numerical analysis of specific relaxation
scenarios}\label{sec:numerical}

Let us now consider the detailed numerical analysis of the
$\F(R,\G)$-cosmology based on a $F=1/B$-term with $B$ the polynomial
given in equation~(\ref{eq:B}). We will consider a separate
discussion of the matter epoch, the radiation epoch and the late
time epoch. It is also interesting to study the future behavior, in
particular to analyze if it is a pure de Sitter phase, as in the
concordance $\CC$CDM model, or there are some significant deviations
from it. In this section, we will perform a numerical analysis of
the exact equations presented in section~\ref{action} and we will
compute the precise evolution of the following basic quantities:
deceleration parameter, the EOS parameter and the density parameters
for the various energy densities, i.e.\
\begin{equation}\label{eq:vector}
q=q(z),\ \ \ \weff=\weff(z), \ \ \
\Omega_n(z)=\frac{\rho_n(z)}{\rho_c(z)}\,, \ \ \
(\rho_n=\rR,\rM,\rLe)\,,
\end{equation}
where $\rho_c(z)=3\,H^2(z)/8\pi G_N$. In all cases we will present
the evolution as a function of the cosmological redshift $z$ or,
equivalently, with the scale factor: $a=1/(1+z)$. For the exact
numerical study of the quantities (\ref{eq:vector}), let us recall
from the discussion presented in section \ref{EOS} that the
generalized Friedmann's equation (\ref{eq:Einstein-H2}) provides the
clue for integrating the field equations, as it determines a second
order differential equation for $H(a)$ which can be solved
numerically, and thereby all the quantities (\ref{eq:vector}) can be
accounted for too\,\footnote{The concrete examples used in
Figs.~\ref{fig:Model-1overB}-\ref{fig:Unified-model-R3-over-B2}
should suffice to illustrate the working ability of the relaxation
mechanism, even though the values of $\rLi$ in realistic GUT
theories are higher than those used in our numerical analysis. The
reason for using smaller values is simply to avoid unnecessary
numerical difficulties.}. However, in order to better understand
qualitatively the meaning of the numerical results, we will precede
our numerical analysis with an approximate analytical treatment of
the behavior obtained in the various epochs. \newfinal{As we will
see}, the model under consideration faithfully reproduces the
standard matter and dominated epochs, in contrast to the traditional
modified gravity models\,\cite{Polarski}, and it leads to an
asymptotic evolution that may effectively appear either in
quintessence-like, de Sitter or phantom-like mask.

\subsection{The matter era \newtext{and} the cosmic coincidence problem}\label{sec:EOS-matter-era}

Let us start in the matter era where $q\approx\frac{1}{2}$ and
$H^{2}\sim\rho_{m}\propto a^{-3}$. We are assuming that the matter
era under consideration is not too a recent one, i.e.\ we suppose
that the matter density dominates over the vacuum energy
($\rM>\rLe$). An exception will be discussed in section
\ref{sub:Unified-dark-sector}. To compute the EOS of the DE (i.e.\
of the effective CC) in this epoch we can obtain an analytical
approximation as follows. We apply directly the relaxation condition
$B\rightarrow0$ in equation (\ref{eq:B}), which leads to
$H^4(q-\frac{1}{2})\propto H^{2n}$. Furthermore, from
(\ref{eq:Einstein-q-1-2})
\begin{equation}\label{eq:EE}
H^{2}(q-\frac{1}{2})=4\pi G_{N}(\frac13\rR+\pLe)\propto
H^{2n-2}\propto a^{-3(n-1)}\,.
\end{equation}
Notice that this relation becomes the $\Lambda$CDM scaling law
$\rR=\rR^0\,a^{-4}$ in the radiation epoch, only if $n=\frac{7}{3}$.
From (\ref{eq:EE}) we find
$\pLe=-\frac{1}{3}\rho_{r}+c_{1}a^{-3(n-1)}$ with a
constant~$c_{1}$, which implies
\begin{equation}
\rLe=c_{2}a^{-3}-\rho_{r}+\frac{c_{1}}{n-2}a^{-3(n-1)},\,\,\,\,
c_{2}=\text{const.}\label{eq:rhoD-MatEra}\end{equation}
as a result of solving the local covariant conservation law
(\ref{conslawDE}). Therefore, the dark energy EOS for the effective
CC in the matter epoch can be approximated by
\begin{equation}\label{omegamatt}
\weff(a)=\frac{\pLe}{\rLe}=\frac{-\frac{1}{3}\rho_{r}+c_{1}a^{-3(n-1)}}{c_{2}a^{-3}-\rho_{r}+\frac{c_{1}}{n-2}a^{-3(n-1)}}\,,
\end{equation}
which is a non-trivial one. \newtext{When}  $n={7}/{3}$, it actually
interpolates between dust matter ($\weff\rightarrow0$) at late times
and radiation ($\weff\rightarrow {1}/{3}$) in the early matter era.
Depending on the integration constant $c_{2}$ a pole might occur in
$\weff$, which can be seen in some \newtext{of} our numerical
examples, see e.g.\ Fig.~\ref{fig:Model-1overB}. The term
proportional to $c_{1}$ could lead to an intermediate scaling if
$n\gtrsim2$. For larger values of $n$ it is not important. Notice
that these poles in the EOS have no physical significance since all
physical quantities (energy density and pressure) are well defined
at all redshifts. The pole appears only when $\rLe=0$, but this is
no real singularity as the description in terms of the EOS parameter
is not fundamental, it is only convenient, see e.g.\ an analogous
situation in \cite{LXCDM}.

%
\FIGURE{
\includegraphics[width=1\columnwidth]{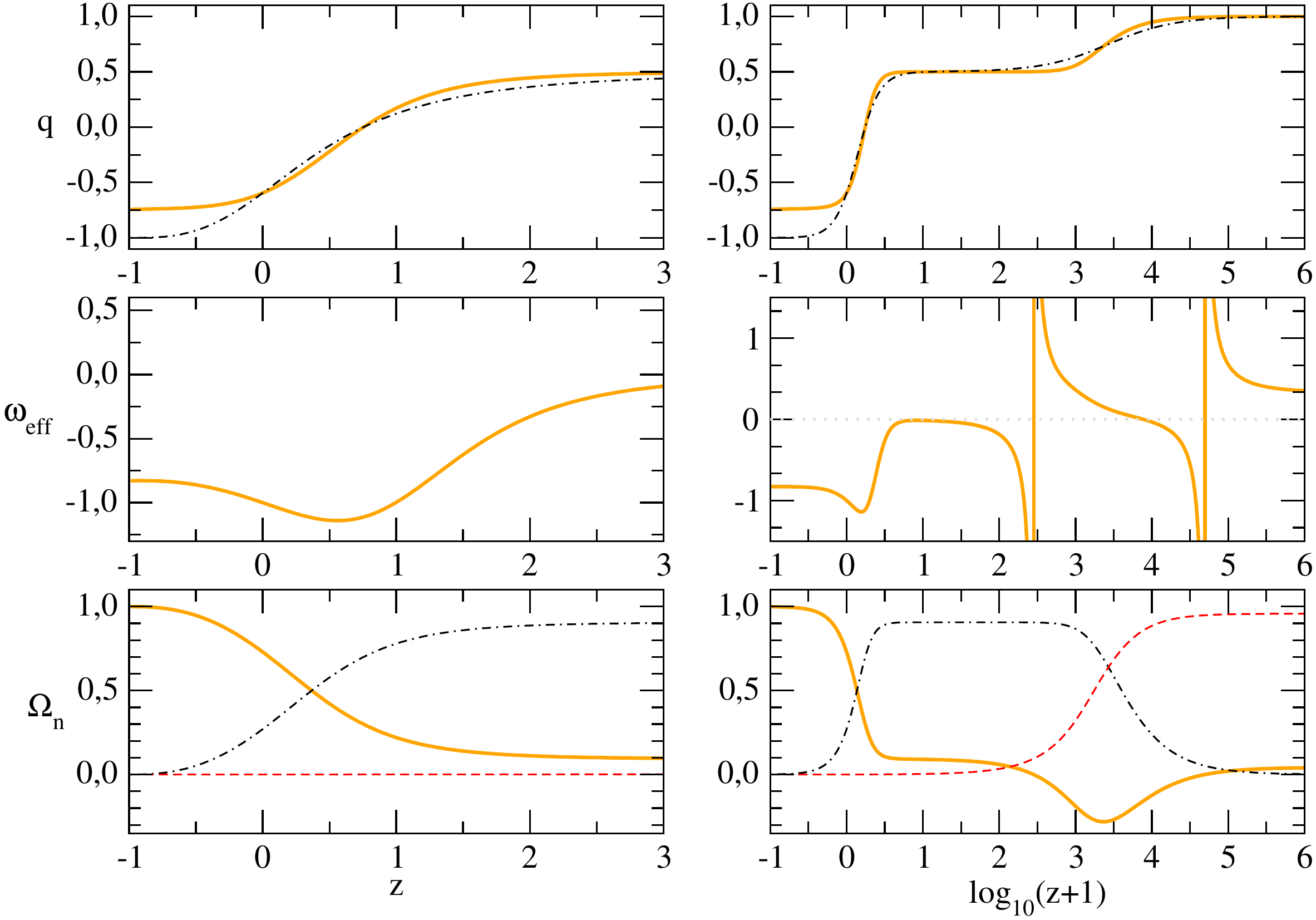}
\caption{Deceleration~$q$, dark energy EOS~$\weff=\pLe/\rLe$, and
relative energy densities~$\Omega_{n}=\rho_{n}/\rho_{c}$ of dark
energy~$\rLe$ (orange thick curve), dark matter~$\rho_{m}$ (black
dashed-dotted) and radiation~$\rho_{r}$ (red dashed) as functions of
redshift~$z$ in the model $F=1/B$ with $n=3$, $y=0.7\times
10^{-3}\,H_0^{-2/3}$, $\rLi=-10^{60}\,\text{GeV}^4$,
$\Omega_{m}^{0}=0.27$, $\Omega_{r}^{0}=10^{-4}$, $q_{0}\approx-0.6$,
$\dot{q}_{0}=-0.5\,H_0$. In the deceleration plot the thick orange
curve corresponds to the modified gravity model, and the black
dashed-dotted curve to $\Lambda$CDM.\label{fig:Model-1overB}} }

It is quite interesting to remark that the EOS analysis suggests
that dark energy behaves like dark matter in this epoch, and we will
speculate on possible applications in
Sec.~\ref{sub:Unified-dark-sector}. In addition, the tracking
relation $\rLe\propto\rM$ reflected in (\ref{eq:rhoD-MatEra}) -- in
which the last two terms on its \textit{r.h.s.}\ decay faster than
the first and should thus be comparatively negligible -- is quite
noticeable too, and can be considered as a cornerstone for solving
the coincidence problem in the $\F(R,\G)$-cosmology.
Recall that $\rho_m=\rho_m^0\,a^{-3}$ and also that $c_2/\rho_m^0$
need not be a very small number. But \newtext{even if}
$c_2/\rho_m^0\simeq 0.1$ the tracking property is remarkable because
it shows that the DE and DM densities are following parallel
evolutions at different levels, and one is not necessarily
infinitesimal as compared to the other. If so, this is an
enlightening clue to explain the coincidence problem.

The reason is that when the universe abandons progressively the
matter epoch, i.e.\ when $q$ starts to significantly deviate from
$1/2$, then the condition $B\to 0$ is no longer implemented through
$q\to 1/2$, but by a significant depletion in the value of the
expansion rate itself, which tends to a very small value $H\to
H_{*}$ near $H_0$. This breaks the approximate ``parallel''
evolutions of $\rM$ and $\rLe$ (equivalently of $\Omega_m$ and
$\Omega_{\CC{\rm eff}}$) since the universe becomes DE dominated.
Thus the two curves must cross (the DE one emerging from below
because it was subdominant earlier) and then they meet at some
point near our present. This can be seen in the numerical examples
presented in Fig.~\ref{fig:Model-1overB} and
Fig.~\ref{fig:Model-R-over-B}-\ref{fig:Unified-model-R3-over-B2},
which sustain our claim that this could be a nice possible
explanation for the cosmic coincidence problem. But, even more
remarkably, is the fact that it emerges from the same mechanism
solving the old CC problem!

\subsection{The radiation era}\label{sec:EOS-radiation-era}

Conventionally, the radiation era is considered as the cosmological
time interval which began after the phase of reheating (subsequent
to the period of primordial inflation) and ended with the beginning
of the matter era, i.e.\ when $H$ takes values in the range $H_{\rm
eq}\leqq H <H_{\rm rad}^*$. Accordingly, the deceleration is
$q\approx1$ and $H^{2}\sim\rho_{r}\propto a^{-4}$. Using again the
relaxation condition $B\rightarrow0$ in (\ref{eq:B}), which in this
case amounts to $[H^{2}(1-q)]^n\propto H^4$, we find
\begin{equation}\label{Rn}
R=6H^{2}(1-q)\propto H^{\frac{4}{n}}\propto a^{-\frac{8}{n}}\,.
\end{equation}
Note from (\ref{eq:Einstein-q-1-3}) that the above equation for
$H^{2}(1-q)$ will exactly match the scaling behavior $\sim a^{-3}$
of the $\Lambda$CDM model in the matter epoch (for negligible $\rL$)
only if $n={8}/{3}$. Therefore, we should choose $n$ close to this
value to obtain a smooth radiation--matter transition close to the
standard model. Equation (\ref{eq:Einstein-q-1-3}) leads to the
relation
\begin{equation}\label{eq:8n}
3H^{2}(1-q)=4\pi G_{N}(\rho_{m}+\rLe-3\pLe)\propto
a^{-\frac{8}{n}},\,\,\,\ \ \ (\rho_{m}=\rho_{m}^{0}a^{-3})\,,
\end{equation}
and thus to
$\pLe=\frac{1}{3}\rho_{m}+\frac{1}{3}\rLe+c_{3}a^{-\frac{8}{n}}$
with $c_{3}=\text{const}$. The dark energy density follows from
inserting this expression in the local covariant conservation law
(\ref{conslawDE}) and solving it, with the result
\begin{equation}\label{eq:lcl}
\rLe=c_{4}a^{-4}-\rho_{m}+3c_{3}\left(-4+\frac{8}{n}\right)^{-1}a^{-\frac{8}{n}}\,,
\end{equation}
where $c_{4}=\text{const.}$ The corresponding EOS of the effective
CC in this epoch interpolates between radiation ($\weff\simeq 1/3)$
at early times and dust matter ($\weff\simeq 0$) at later times:
\begin{equation}\label{EOSrad}
\weff(a)=\frac{\pLe}{\rLe}=\left(\frac{1}{3}\right)\,\frac{1+a^{(4-\frac{8}{n})}\frac{3c_{3}}
{c_{4}}(1+\left(-4-\frac{8}{n}\right)^{-1})}{1-\frac{\rho_{m}^{0}}{c_{4}}a+a^{(4-\frac{8}{n})}\frac{3c_{3}}{c_{4}}\left(-4+\frac{8}{n}\right)^{-1}}\,.
\end{equation}
Also here, a pole in $\weff$ might exist depending on $c_{4}$, which
can be seen in Fig.~\ref{fig:Model-1overB}. Again if $n\gtrsim2$,
the term proportional to $c_{3}$ could lead to an intermediate
scaling, which is absent for larger~$n$. Remarkably, we find the
tracking property~$\rLe\propto\rho_{r}\propto a^{-4}$ also in the
radiation era.

An important point to be stressed in the radiation epoch is that the
aforesaid tracking property does not endanger the primordial
nucleosynthesis process. In fact, when comparing the scaling law for
radiation $\rR=\rR^0\,a^{-4}$ with equation (\ref{eq:lcl}),
nucleosynthesis requires that the ratio $c_4/\rR^0$ is sufficiently
small, typically one order of magnitude at least. We explicitly
confirm from the numerical examples in Fig.~\ref{fig:Model-1overB}
and Fig.~\ref{fig:Model-R-over-B}-\ref{fig:Unified-model-R3-over-B2}
that this is indeed the case in the large $z$ region for $z>10^5$,
which comprises in particular the nucleosynthesis segment around
$z\sim 10^9$.

The tracking feature of the DE in the matter and radiation epochs,
when inspected in Figs.~\ref{fig:Model-1overB} and
\ref{fig:Model-R-over-B}-\ref{fig:Unified-model-R3-over-B2}, is
actually better captured if we focus on the evolution of the EOS
parameter as a function of the cosmological redshift,
$\weff=\weff(z)$, rather than looking at e.g.\ the corresponding
behavior of the density parameters $\Omega_n$. The point is that,
from the figures one might get the wrong perception that
$\Omega_{\CC{\rm eff}}$ traces $\Omega_r$ rather than $\Omega_m$ in
the matter epoch, and vice versa in the radiation epoch. This is
only due to the numerical smallness of $\Omega_{\CC{\rm eff}}$ in
both the matter and radiation epochs. Instead, when we look at the
EOS plots, we see that in the central part of the matter epoch (cf.\
at the same time the $q$-plot in the interval where $q=1/2$) the EOS
of the dark energy tends to stay around $\weff(z)\simeq 0$, whereas
deep in the radiation epoch ($q=1$) it tends to values around
$\weff(z)\simeq 1/3$.

For the rest of this work, we want to avoid fractional powers of $n$
in (\ref{eq:B}) and thus we take the value~$n=3$ implying $R\propto
H^{\frac{4}{3}}\propto a^{-\frac{8}{3}}$, which is still close to
$\Lambda$CDM.

\FIGURE[t]{
\includegraphics[width=1\columnwidth]{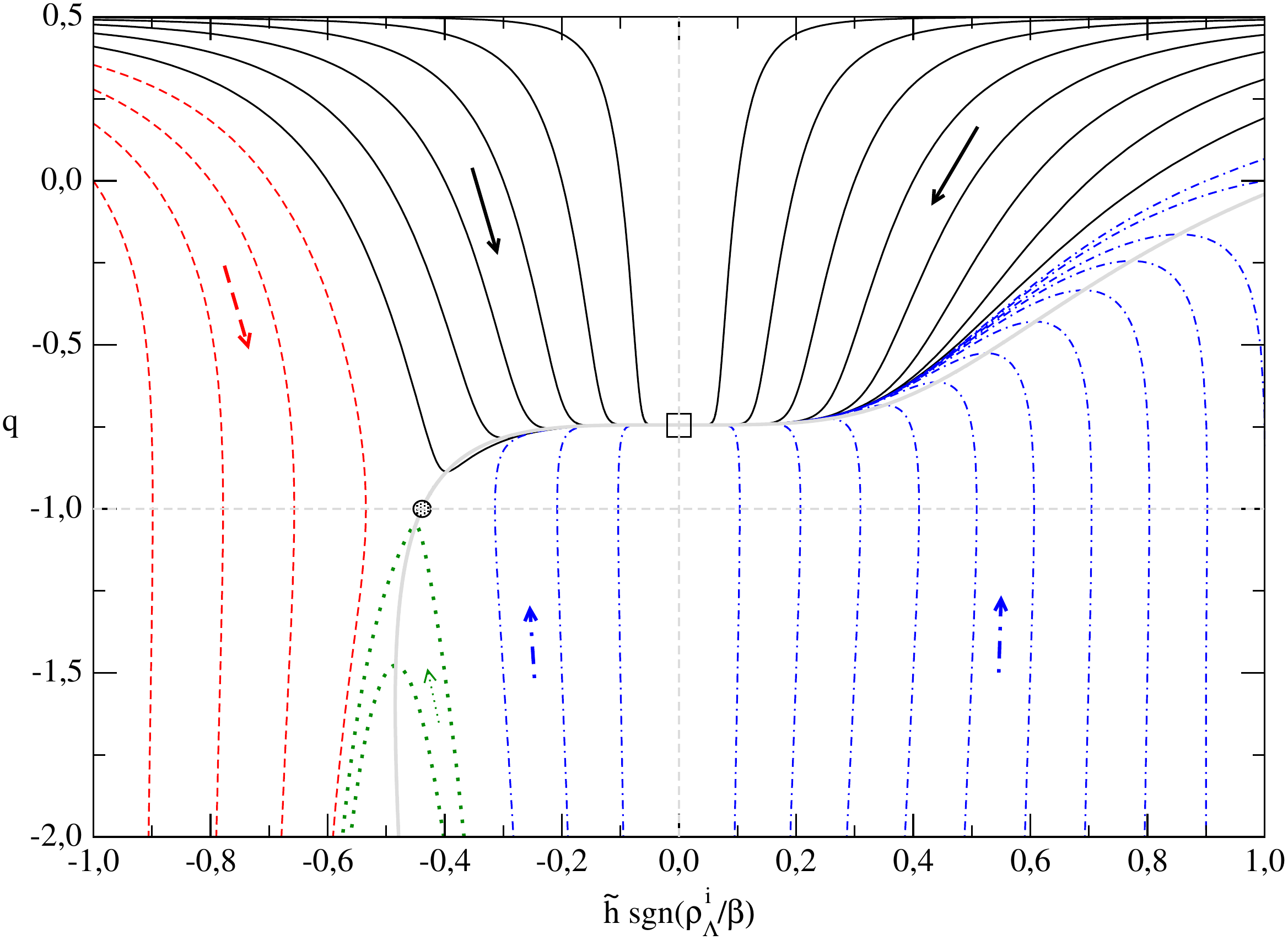}
\caption{All types of late-time solutions in the $F=1/B$ model as a
phase diagram of the deceleration~$q$ and the dimensionless Hubble
rate $\tilde{h}=\tilde{c}H>0$, $\tilde{c}=|2\rLi/\beta|^{1/4}$. The
thick grey curve corresponding to~$\dot{q}=0$ goes through the
unstable de~Sitter fixed point at $q=-1$ (circle) and the stable
fixed point at $q=a_{1}\simeq-0.74$ (square). The arrows signify the
direction of cosmic time.\label{fig:1overB-phases-latetime}} }

\subsection{The asymptotic late-time evolution}
\label{sec:asympt-late-time}

In the previous sections we have analyzed the cosmic evolution in
the matter and radiation epochs by making use of~$B\rightarrow 0$,
i.e.\ the approximate cancelation of both terms in (\ref{eq:B}).
Accordingly, the behavior of~$\rLe$ was found without solving the
complicated Einstein equations (\ref{eq:Mod-Einstein-Eqs}). However,
in the period from the recent (matter dominated) past to the
asymptotic future we cannot profit from the relation~$B\rightarrow
0$, because the Hubble rate~$H$ is too small making the $y$-term in
(\ref{eq:B}) completely negligible here, \newtext{so that in good
approximation}
\begin{equation}\label{eq:Bapprox}
B\simeq\frac{2}{3}R^{2}+\frac{1}{2}\G =
24H^{4}(q-\frac{1}{2})(q-2)\,.
\end{equation}
Obviously, the CC relaxation is supported \newtext{now} only by the
smallness of~$H$, since there are no terms in $B$ which could cancel
each other. Before we turn to numerical solutions of the Einstein
equations in Sec.~\ref{sub:Parameters}, let us discuss the
\newnewtext{very} late-time evolution analytically, \newnewtext{i.e.} the asymptotic future
regime ($t\to\infty$). For this purpose we consider first the model
$F=1/B$ with~$y=0$ -- i.e.\ with $B$ given by (\ref{eq:Bapprox}) --
and neglect all energy density contributions except for~$\rF$
and~$\rLi$. This is a good approximation at late times since the
Friedmann \newnewtext{equation} then reads
\begin{equation}\label{Friedmann2}
\rho_{c}=\frac{3H^{2}}{8\pi
G_{N}}=\rho_{m}+\rLe=\rho_{m}+\rLi+\rho_{F}\to \rLi+\rho_{F}
\end{equation}
owing to $\rho_{m}\sim 1/a^3\to 0$ for $t\to\infty$. Moreover, the
Hubble rate $H$ must be very small in order to sustain the
asymptotic late-time relaxation of the effective CC, and hence the
effective CC tends to the asymptotically vanishing value of the
critical density: $\rLe=\rho_{F}+\rLi\to \rho_c\to 0$. In practice,
this means that the equation $\rho_{F}+\rLi=0$ should be sufficient
for analyzing the background evolution
\newnewtext{in the asymptotic regime}. The induced term
corresponding to the $F=1/B$ model can be computed explicitly with
the help of (\ref{eq:E00}). After a straightforward calculation, we
find
\begin{eqnarray}
\rho_{F}=2\,E_{\,\,0}^{0}&=&\frac{864\,\beta\,H^{7}}{B^{3}}\left[H(10q^{4}-31q^{3}+15q^{2}+19q-10)-\dot{q}(4q^{2}-10q+7))\right]\nonumber\\
&=&\beta\,\left[\frac{5q^{2}-3q-5}{2H^{4}(2q^{2}-5q+2)^{2}}-\frac{\dot{q}\,(4q^{2}-10q+7)}{2H^{5}(2q^{2}-5q+2)^{3}}\right]\,.
\label{eq:rhoF-LateTime}
\end{eqnarray}
For the sake of simplifying our discussion below, it is convenient
to rewrite the previous equation
\newtext{as follows}:
\begin{equation}
\rho_{F}=\beta\left[\frac{k_{1}}{2H^{4}b^{2}}-\frac{\dot{q}}{Hb}\cdot\frac{k_{2}}{2H^{4}b^{2}}\right]\,,\label{eq:Analyt-rF}
\end{equation}
with
\begin{eqnarray}\label{k1k2b}
k_{1} & := & 5(q-a_{1})(q-a_{2})\nonumber\\
 &  & a_{1,2}=\frac{1}{10}\left(3\mp\sqrt{109}\right)\nonumber\\
k_{2} & := & 4(q-b_{1})(q-b_{2})\\
 &  & b_{1,2}=\frac{1}{4}\left(5\pm\text{i}\sqrt{3}\right)\nonumber\\
b & := & 2(q-\frac{1}{2})(q-2)\nonumber\,.\end{eqnarray}
Next, we express the asymptotic Friedmann equation $\rLi+\rF=0$ in
terms of
\newtext{a new} dimensionless Hubble rate \newtext{defined as} $\tilde{h}:=\tilde{c}H$ and
a \newtext{scaled} cosmic time $\tau:=t/\tilde{c}$, where
$\tilde{c}:=|x|^{1/4}$ and $x:=\frac{2\rLi}{\beta}$, respectively.
\newtext{Denoting} $q^{\prime}(\tau)=dq/d\tau$, we have
$q^{\prime}(\tau)=\tilde{c}\dot{q}(t)$ and hence Friedmann's
equation can be cast as
\begin{equation}\label{h4}
\mp
\tilde{h}^{4}=\frac{k_{1}}{b^{2}}-q^{\prime}(\tau)\frac{k_{2}}{b^{3}\tilde{h}},
\end{equation}
where {}``$-$'' corresponds to $x>0$ and {}``$+$'' to $x<0$,
respectively. Moreover, the time-evolution of~$\tilde{h}$ is the
same as that of the true Hubble rate,
\begin{equation}
\tilde{h}^{\prime}(\tau)=-\tilde{h}^{2}(q+1).
\end{equation}
\newtext{From} (\ref{k1k2b}) and (\ref{h4}) it follows that $q^{\prime}$
and $\tilde{h}^{\prime}$ can be expressed as functions of only $q$
and $\tilde{h}$, which fixes uniquely their evolutions independent
of~$c$. Instead of plotting a vector map, we show the integrated
solutions in Fig.~\ref{fig:1overB-phases-latetime}, where the arrows
indicate the direction of time and the thick grey curve corresponds
to~$q^{\prime}=0$. In the following we will restrict our analysis to
\newtext{late time solutions} $q<\frac{1}{2}$ and $H,\tilde{h}>0$, which implies
$b,k_{2}>0$ and $k_{1}\gtrless0$ for $q\lessgtr a_{1}\approx-0.74$,
\newtext{respectively}. Consequently, the deceleration~$q$ decreases
(\newtext{becomes more negative}) \newtext{with time} on the
left-hand side \newtext{and above} of the $q^{\prime}=0$ curve and
increases on the right-hand side \newtext{and below it}.
\newtext{Notice} from (\ref{h4}) that the de Sitter regime $q=-1$ satisfies
$\tilde{h}^4=k_1/b^2=5(1+a_1)(1+a_2)/81$, and hence
$\tilde{h}=27^{-1/4}$. As the Hubble rate decreases for~$q>-1$ and
increases for~$q<-1$, it follows that the de~Sitter fixed point
$(q=-1,\tilde{h}=27^{-1/4})$ -- see the circle in
Fig.~\ref{fig:1overB-phases-latetime} -- is unstable, and a
de~Sitter final phase would require fine-tuning.
On the other hand, the fixed point $(q=a_{1}\simeq
-0.74,\tilde{h}=0)$ -- see the square in the central part of
Fig.~\ref{fig:1overB-phases-latetime} -- corresponds to a stable
final state similar to quintessence solutions. The red dashed curves
describe universes with a decelerating past and a phantom
(\newtext{superaccelerated runaway}) future, whereas the black
curves approach the stable fixed point
\newtext{with quintessence-like behavior}. Also the blue
dashed-dotted curves end in the stable fixed point, however they are
accelerating in the far past, possibly with a transient decelerating
phase in between. Finally, the green dotted curves describe phantom
universes with $q<-1$ all the time.
%
%

Obviously \newtext{the most interesting} late time solutions
obtained from the $F(R,\G)=1/B$ models under study are those showing
stable asymptotic quintessence-like behavior. This kind of solutions
also exist for more general models, as we shall see below, although
the asymptotic future value of the deceleration~$q$ will be
different in general. \newtext{Let us look for} accelerated
solutions with constant $q>-1$ in the asymptotic future (i.e.\ with
a terminal acceleration below the pure de Sitter regime).
\newnewtext{They correspond} to power-law solutions of our field equations
in the limit $t\rightarrow \infty$, namely
\begin{equation}\label{powerlawsol}
a(t)\propto t^{r}(1+\zeta t^s)\,,
\end{equation}
where a first-order correction $|\zeta t^s|\ll1$
(with~$\zeta=\text{const.}$) has been
included\,\footnote{\newfinal{A more general approach} for finding
asymptotic solutions would be to start with the ansatz $a(t)\sim t^r
(1+f(t))$ and then linearize in the smooth function $f(t)$. In this
work, we will limit ourselves to explore the solutions of the form
(\ref{powerlawsol}) as they are already able to capture the main
features of the very late time behavior.}. We can search for these
solutions within the generalized class of models of the form
$F(R,\G)=1/B^{m}$ with $m>0$. Obviously, the model we have been
considering so far corresponds to the particular case $m=1$. Direct
calculations to first order in $\zeta t^s$ via~(\ref{eq:rF-relax})
and (\ref{eq:E00}) lead to the
\newnewtext{following results}:
\begin{eqnarray}
\rho_{F}& = & \frac{\beta}{B^{m}} \left(\rho_{F0}+\rho_{F1}\cdot
\zeta t^{s}\right)\,, \ \ \ \ \
p_{F} = \frac{\beta}{B^{m}}\left(p_{F0}+p_{F1}\cdot \zeta t^{s}\right)\,,\label{rFwF}\\
B & = & 12r^{2}(9r^{2}-9r+2r)t^{-4}
+12rs(4-27r+36r^{2}-4s+9rs)t^{-4}\cdot \zeta
t^{s}\nonumber\label{Basympt},
\end{eqnarray}
\newnewtext{where}~ $(\rho_{F0}$, $p_{F0})$ and $(\rho_{F1}$, $p_{F1})$ are (\newnewtext{dimensionless})
time-independent terms corresponding to the zeroth and first order
corrections, respectively. In particular,
\begin{eqnarray}
\rho_{F0} & = & \frac{K(r,m)}{9r^{2}-9r+2} =\left(-1-\frac{4m}{3r}\right)^{-1}p_{F0},\\
K(r,m) & = &
9r^{2}-9r+2-4m^{2}(9r-4)+3m(3r^{2}-11r+4).\label{Krm}
\end{eqnarray}
\newnewtext{However}, these terms are in fact zero because in order to fulfill
the late-time Einstein equation $\rho_{F}=-\rho_{\Lambda}^{i}$ we
need $K(r,m)=0$, otherwise $\rho_{F}\sim\rho_{F0}B^{-m}\propto
t^{4m}$ would diverge for $t\rightarrow\infty$. \newnewtext{In
addition}, we have to fix the correction term in the ansatz
(\ref{powerlawsol}). This can be achieved by the choice~$s=-4m$, as
this insures that the term $\rho_{F}\propto\rho_{F1}\cdot \zeta
t^{s} B^{-m}\sim t^{s+4m}$ remains essentially constant in time --
and this also entails $p_{F}=\text{const}$. \newnewtext{Likewise},
with this choice of $s$ we have $\zeta t^s=\zeta t^{-4m}\to 0$ with
increasing $t$ because $m\geq 1$, which means that the correction
term in the solution (\ref{powerlawsol}) becomes smaller and smaller
in the asymptotic regime, irrespective of the particular value of
the coefficient $\zeta$. Accordingly, we {find}
\begin{eqnarray}
\rho_{F1} & = & \frac{-8m^{2}(4m-1)}{r(9r^{2}-9r+2)^{2}}\,\left[2m^{2}(4-9r)^{2}-3r(7-30r+36r^{2})\right.\\
 &  & \left.-2m(-8+54r-117r^{2}+81r^{3})\right]\nonumber,\\
p_{F1} & = & \frac{8m^{2}(4m-1)}{3r^{2}(9r^{2}-9r+2)^{2}}\,
\left[8+8m^{3}(4-9r)^{2}-72r+171r^{2}-54r^{3}-162r^{4}\right.\\
 &  & \left.-2m^{2}(-80+408r-585r^{2}+162r^{3})
-4m(-16+102r-189r^{2}+54r^{3}+81r^{4})\right]\nonumber,
\end{eqnarray}
implying
\begin{equation}
\rho_{F1}+p_{F1}=\frac{16m^{2}(4m-1)(2+m(4-9r)-9r+9r^{2})}{3r^{2}(9r^{2}-9r+2)}\cdot\rho_{F0}.
\end{equation}
Since the zero-order terms~$\rho_{F0}$ and~$p_{F0}$ vanish,
\newnewtext{we obtain} the correct \newnewtext{asymptotic} EOS for the $F$-terms,
\begin{equation}
\wF=\frac{p_{F}}{\rho_{F}}=\frac{p_{F1}}{\rho_{F1}}=-1\
+\mathcal{O}(\zeta t^{-4m})\,.
\end{equation}
\newnewtext{It follows} that the effective entity that we have
called the ``cosmon'', and which is responsible for the induced DE
-- of (modified) gravitational origin -- behaves asymptotically as a
cosmological term (up to very small corrections). Recall that this
is actually so, although in a lesser extent, during most of the
cosmological history prior our time. After all the cosmon ``duty''
is to continually neutralize the initial cosmological term $\rLi$
leaving a small dynamical remainder --  the measurable CC term. What
we have just shown here is that, ultimately, it behaves (very
approximately) as a true cosmological constant, with a value equal
(but opposite in sign) to the initial CC term:
$\rF(t\to\infty)=-\rLi$.
%
%
Finally, the condition $\rho_{F0}\propto K(r,m)=0$ determines the
leading power-law exponent~$r$ in the scale factor:
\begin{equation}\label{eq:r}
r=(6m+6)^{^{-1}}(12m^{2}+11m+3\pm\sqrt{144m^{4}+200m^{3}+81m^{2}+10m+1})\,.
\end{equation}
Being $r>0$, the evolution law (\ref{powerlawsol}) for the scale
factor is an increasing one.
For $m=1$, we find the solutions $r\approx0.43$ and $r\approx3.91$,
which are already close to the large $m$ limit solutions
$r\rightarrow\frac{4}{9}$ and $r\rightarrow4m$.
\newnewtext{Neglecting} the correction term, we have $\ddot{a}\sim r(r-1)t^{r-2}$, and hence
only the $r>1$ solutions produce acceleration ($\ddot{a}>0$), and
only these can be interpreted as asymptotic quintessence-like
solutions.

\newnewtext{Let us indeed consider} in more detail the EOS of the effective vacuum energy $\rLe=\rF+\rLi$ in the asymptotic regime.
As we know, $\rLe$ is essentially zero in this regime, which means
that the corresponding EOS function $\weff$ cannot be easily derived
from the original definition (\ref{eq:omegaeff}) because the latter
involves the ratio between the two terms $1+\wF$ and $\rF+\rLi$ both
of which are very close to zero.
\newtext{Let us thus turn to} equation~(\ref{eq:Einstein-q-1-3}) and apply it to the
very late time epoch, where $\rho_m$ can be neglected. It is then
easy to show that that equation can be recast in the form
\begin{equation}\label{OmegaLeff}
\Omega_{\CC{\rm eff}}:=
\frac{\rLe}{\rho_c}=\frac{2(1-q)}{1-3\weff^{(\infty)}}\,,
\end{equation}
where $\weff^{(\infty)}$ is the asymptotic value of $\weff$. At the
same time, we know that in this regime $\rLe=\rho_{F}+\rLi$ tends to
become arbitrarily close to $\rho_c$ (both going to zero with the
same pace). Thus $\Omega_{\CC{\rm eff}}\to 1$, and consequently
\begin{equation}\label{EOSquint}
\weff^{(\infty)}=-1+\frac{2}{3}(q+1)\,.
\end{equation}
This is the quintessence-like behavior of the EOS for the effective
vacuum energy $\rLe$ in the asymptotic limit. Notice that since the
accelerated expansion implies $\weff<-1/3$, we consistently obtain
$q<0$. Let us now compute explicitly the deceleration parameter $q$
in the asymptotic regime. Using\ $q=-1-\dot{H}/H^2$ and working out
$H=\dot{a}/a$ directly from (\ref{powerlawsol}), we find
\begin{eqnarray}
H(t)&=&\frac{r}{t}\left(1-\frac{4m}{r}\zeta t^{-4m}+{\cal O}(\zeta
t^{-4m})^2\right)\,,\label{Hasympt}\\
 q(t)&=&-1+\frac{1}{r}+\frac{4m(1-4m)}{r^2}\zeta\,t^{-4m}+{\cal O}(\zeta
t^{-4m})^2\label{qasymptotic}\,.
\end{eqnarray}
We reconfirm from the last expression that $q<0$ because we look
only for the solutions satisfying $r>1$ -- the time-dependent term
in (\ref{qasymptotic}) also respects this feature because $m>1$ and
so it contributes negatively. Inserting the previous expression for
$q(t)$ in (\ref{EOSquint}) we obtain the asymptotic EOS:
\begin{equation}\label{weffasymptotic}
\weff^{(\infty)}(t)=-1+\frac{2}{3r}+\frac{8m(1-4m)}{3r^2}\zeta\,t^{-4m}+{\cal
O}(\zeta t^{-4m})^2\,.
\end{equation}
From this formula, it becomes apparent that the solutions $r>1$ of
equation (\ref{eq:r}) lead to quintessence-like behavior since
$\weff$ then lies in the interval $-1<\weff<-1/3$ (with again a very
small time dependence which does not alter this conclusion).
Furthermore, if we take e.g. the case $m=1$, equation (\ref{eq:r})
gives $r\simeq 3.91$ for the $r>1$ solution, and correspondingly
equation (\ref{qasymptotic}) yields $q\simeq -0.74$ and
(\ref{weffasymptotic}) provides $\weff^{(\infty)}\simeq-0.83$ (up to
very small time-dependent effects) -- see figures
\ref{fig:Model-1overB} and \ref{fig:1overB-phases-latetime}. This
value of $\weff$ can be read directly from
Fig.\,\ref{fig:Model-1overB} -- it corresponds to the value of
$\weff(z)$ at the terminal point $z=-1$ (i.e. $t\to\infty$).

\newfinal{We may} finally derive the very late-time time behavior of
$\rLe$. As we know, it behaves as $\rLe(t)\to\rho_c(t)\to 0$ for
$t\to\infty$, but we can compute precisely how it decays with time
in the last stages of its evolution. Substituting the asymptotic
Hubble rate (\ref{Hasympt}) into the asymptotic Friedmann's equation
(\ref{Friedmann2}), we arrive at the desired result:
\begin{equation}\label{rLeasympt}
\rLe=\rLi+\rF=\frac{3}{8\pi
G_N}\frac{r^2}{t^2}\left(1-\frac{8m}{r}\zeta t^{-4m}+{\cal O}(\zeta
t^{-4m})^2\right)\to\frac{3 r^2}{8\pi G_N}\,t^{-2}\ \ \
(t\to\infty)\,,
\end{equation}
\newfinal{where} $m\geq 1$ for all the relaxation models
under consideration. Let us also note (a posteriori) that the
obtained result is consistent with our assumption that the
asymptotic matter density decays always faster than the vacuum
energy density, as we supposed from the very beginning in
(\ref{Friedmann2}). Indeed, being $\rho_m\sim a^{-3}\sim t^{-3r}$ --
where we can neglect here the small correction term in
(\ref{powerlawsol}) -- the matter density is always much smaller
than the effective vacuum energy $\rLe\sim t^{-2}$ because the
minimum value of $r$ (for accelerating solutions) is approximately
$r\simeq 3.91$, which is attained for $m=1$ -- cf. equation
(\ref{eq:r}). For $m>1$, $r$ is larger than the previous value and
the desired condition is always secured.

Looking at the numerical examples presented in
Figs.~\ref{fig:Model-1overB} and
\ref{fig:Model-R-over-B}-\ref{fig:Unified-model-R3-over-B2}, we can
check that the asymptotic $t\to\infty$ (i.e.\ $z\to -1$) EOS
behavior $\weff=\weff(z)$ of the effective CC is always
quintessence-like ($-1<\weff<-1/3$). However, in some cases (e.g.\
in Figs.\,\ref{fig:Model-1overB} and \ref{fig:Model-R-over-B}), it
presents an effective phantom phase $\weff<-1$ for non-asymptotic
regions, actually for regions not far away in our past. Therefore,
the accessible region to our observations could present an effective
phantom-like DE behavior.  \newnewtext{In contradistinction} to
scalar models of the DE, this phantom-like behavior has nothing to
do with negative kinetic terms, as it is of purely effective nature.
For example, since $\Omega_{\rm eff}(z)>0$ in the affected regions
of Figs.\,\ref{fig:Model-1overB} and \ref{fig:Model-R-over-B}, the
phantom behavior is just triggered by the fact that the set of terms
in the field equations that we have collected under the name of the
cosmon produces $(1+\wF(z))\rF(z)<0$ in those redshift segments --
cf. equation (\ref{eq:omegaeff}).


\subsection{Parameters and initial conditions}\label{sub:Parameters}

Our modified gravity action~(\ref{eq:CC-Relax-action}) with $F=1/B$
or more generally $F=R^{s}/B^{m}$ contains two parameters $\beta$
and $y$, where the latter is directly related to the
radiation-matter transition. To solve the Einstein equations, we
have to impose one more initial condition because $F(R,\G)$ theories
have more degrees of freedom than general relativity. In fact,
\newtext{while in the standard $\Lambda$CDM model} the initial value
of the deceleration $q_{0}$ is fixed simply by the current relative
densities $\Omega_{n}^{0}$, \newfinal{see equation} (\ref{qomegas})
(specifically $q_0=\Omega_m^0/2-\Omega_{\CC}^0$ for the present
epoch), in the relaxation model we have to provide also the value of
$\dot{q}_{0}$. To see this, note that the induced $F$-term is a
function of the type
\begin{equation}\label{rFfunction}
\rF=\rF(H,q,\dot{q}; \beta,n,y)\,,
\end{equation}
which depends in general on the three free parameters
(\ref{eq:freeparam}) and involves  $H$, $q$ and also $\dot{q}$.
Recall that $n=3$ was fixed because it provides a sufficiently
smooth transition from the radiation to the matter epochs, so only
$\beta$ and $y$ remain as free parameters to be chosen. An explicit
example is the late time form given in equation
(\ref{eq:rhoF-LateTime}), where the $\dot{q}$ dependence is
manifest.
\newtext{Therefore, if we are given} $\rLi$ and we fix the current value of the CC density,
this amounts to fix the value of
\begin{equation}\label{OmegaCCeffnow}
\Omega_{\CC{\rm eff}}^0 = \frac{\rLi+\rF(H_0,q_0,\dot{q}_0;\,
\beta,n,y)}{\rho_c^0}\,,
\end{equation}
which entails a non-trivial relation between the parameters. This
relation involves explicitly the value of $\dot{q}_0$, and in our
case $q_0$ is \newfinal{fixed from} equation (\ref{q0relax}).
Therefore, equation (\ref{OmegaCCeffnow}) fixes $\dot{q}_0$, which
is tantamount to say that $\dot{q}_0$ must be provided as an
additional
\newnewtext{input} in order that the value of the current
$\Omega_{\CC{\rm eff}}^0$ is correctly matched.

\FIGURE[t]{
\includegraphics[width=1\columnwidth]{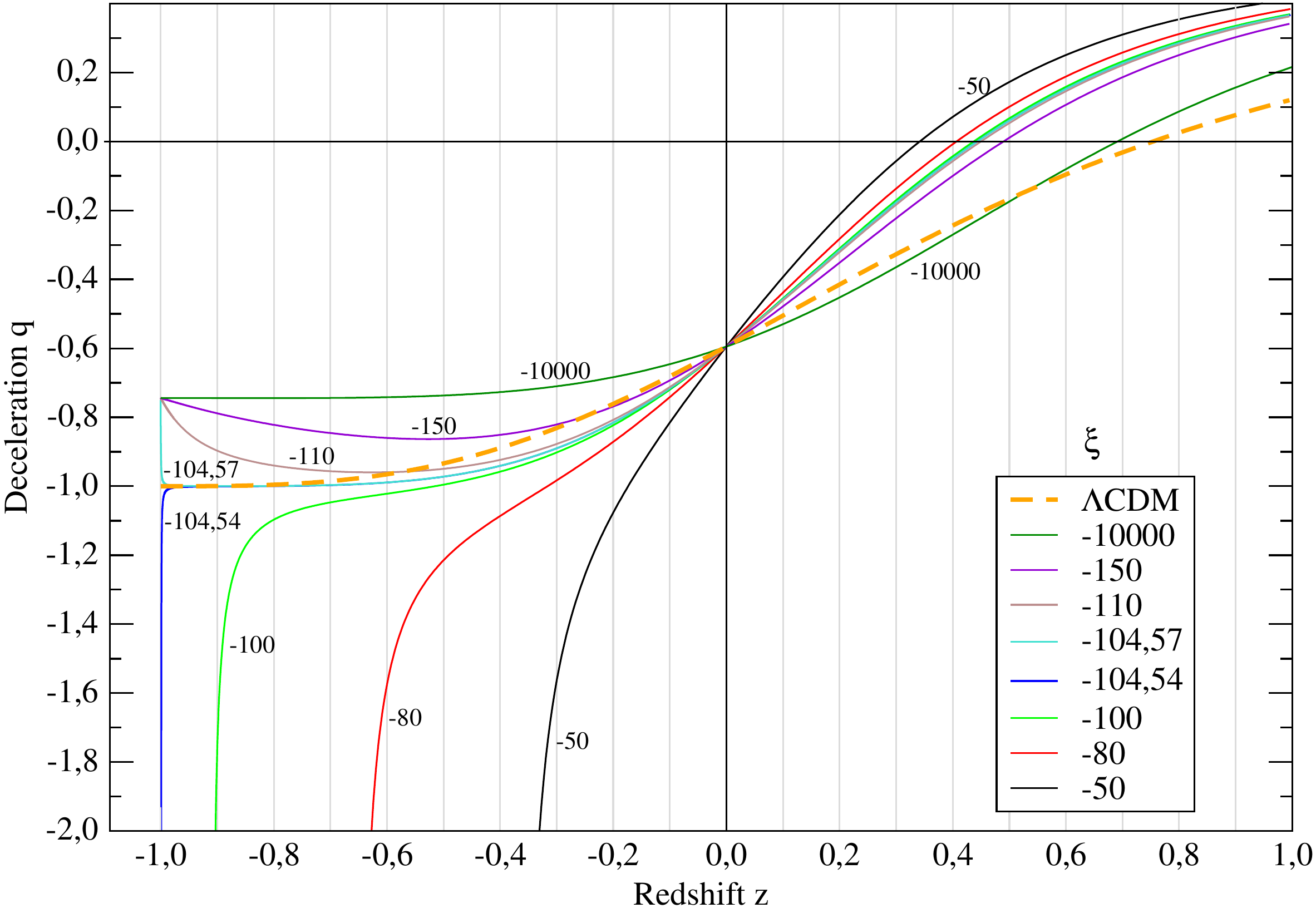}
\caption{Possible future final states in the class of $F=1/B$
models. \newtext{Shown is} the deceleration parameter versus
redshift for different values of the dimensionless parameter $\xi=
3\beta/(\rLi\,H_0^4)$, assuming $\xi<0$ in all cases. Depending on
the value of $|\xi|$, in the asymptotic limit $t\to\infty$ (i.e.\
$z\to -1$) the universe expands approximately according to a
power-law~$a\sim t^{r}$ ($r\simeq 3.91$ for this class of models) or
else it develops a future singularity characterized by a runaway
acceleration ($q\ll -1$). Notice that de Sitter space-time ($q=-1$)
lies on the boundary between the two
possibilities.\label{fig:q-Future-beta-1}} }
%
\FIGURE{
\includegraphics[width=1\columnwidth]{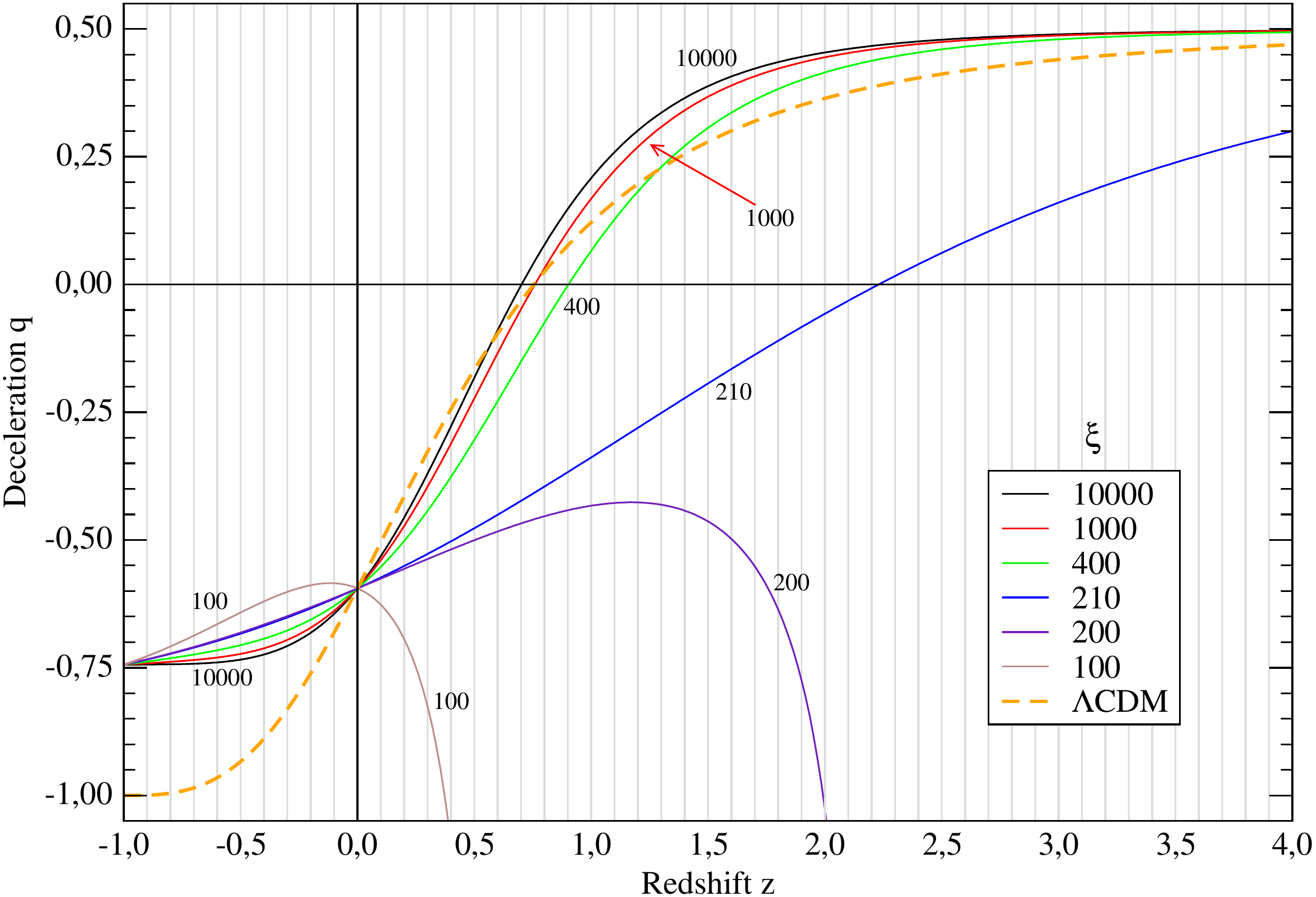}
\caption{\newtext{As in} Fig.\,\ref{fig:q-Future-beta-1}, but for
$\xi>0$. Notice that here all curves have an asymptotic power-law
expansion ~$a(t)\sim t^{r}$ for $t\rightarrow\infty$.
\newtext{However}, not all curves are admissible since $\xi$ is limited
from below ($\xi\gtrsim210$) in order to have a matter era
($q\approx\frac{1}{2}$) in the past.\label{fig:q-Future-beta-2}} }
%

\FIGURE[t]{
\includegraphics[width=1\columnwidth]{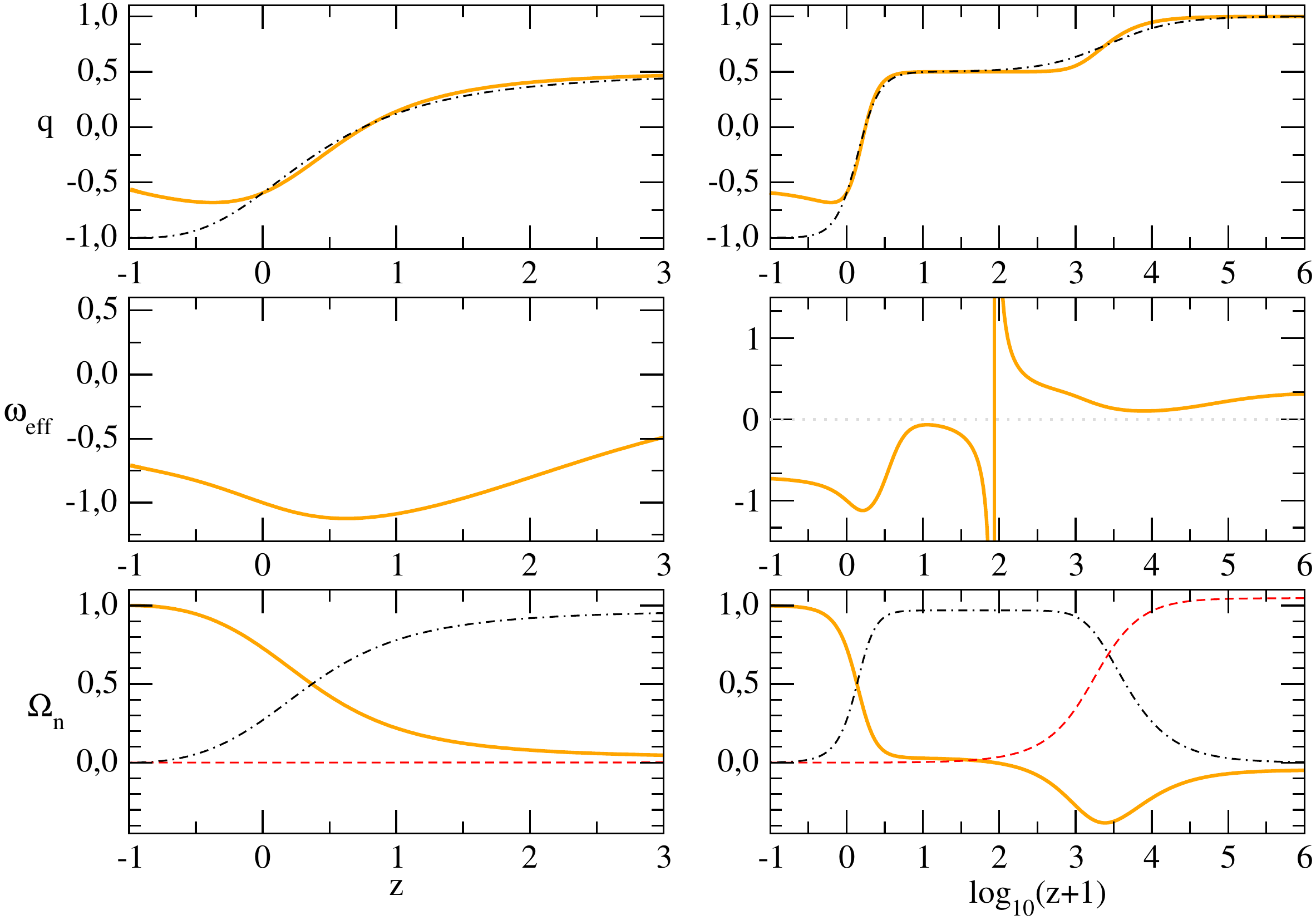}
\caption{Model $F=R/B$ with $n=3$, $y=0.7\times
10^{-3}\,H_0^{-2/3}$, $\rho_{\Lambda}^{i}=-10^{60}\,\text{GeV}^4$,
$\Omega_{m}^{0}=0.27$, $\Omega_{r}^{0}=10^{-4}$, $q_{0}\approx-0.6$,
$\dot{q}_{0}=-0.5\,H_0$. The curves have the same meaning as in
Fig.~\ref{fig:Model-1overB}. \label{fig:Model-R-over-B}} }
%
%
\FIGURE[t]{
\centering{}\includegraphics[width=1\columnwidth]{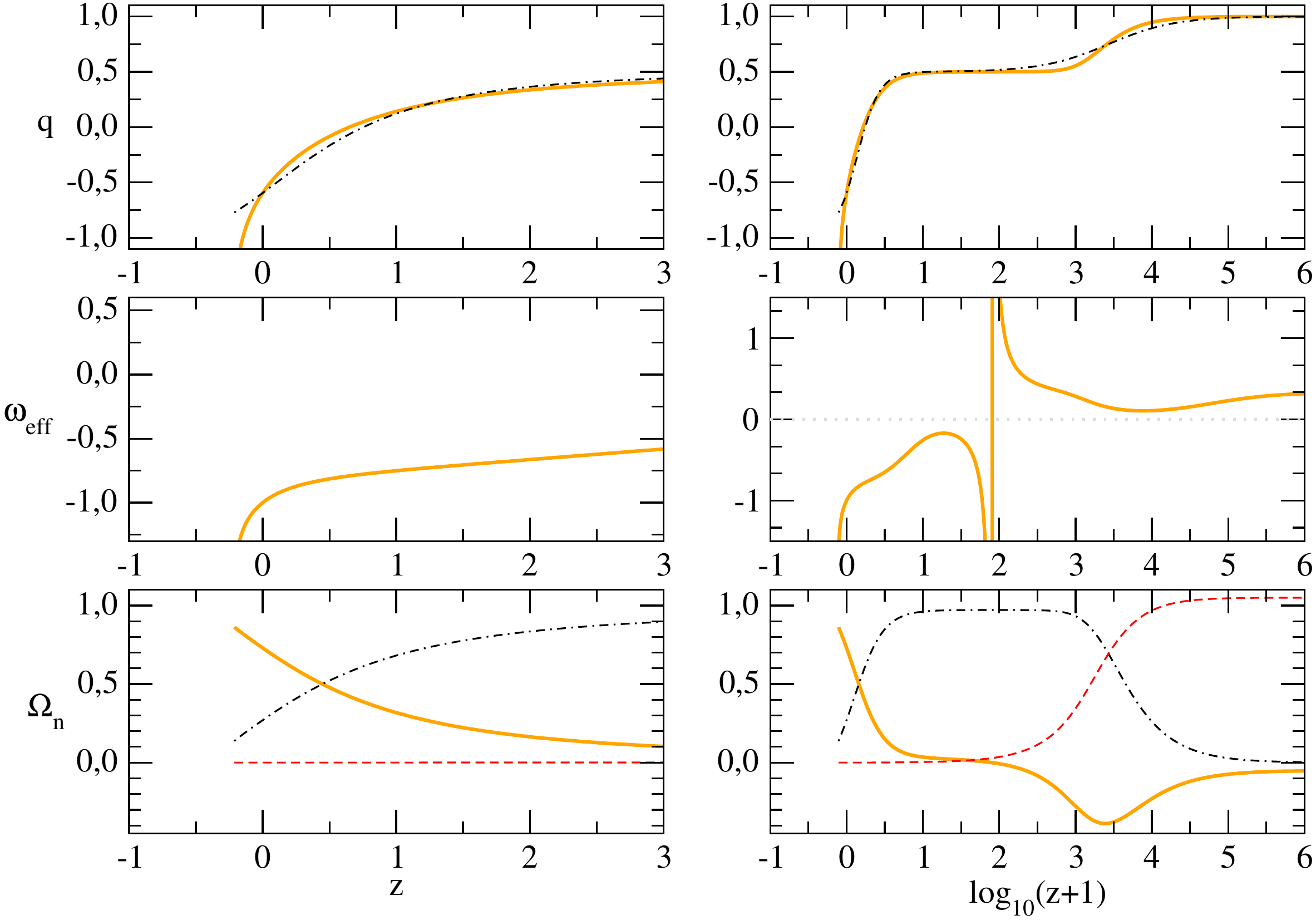}\caption{Model
$F=R^{2}/B$ with $n=3$, $y=0.7\times 10^{-3}\,H_0^{-2/3}$,
$\rho_{\Lambda}^{i}=-10^{60}\,\text{GeV}^4$, $\Omega_{m}^{0}=0.27$,
$\Omega_{r}^{0}=10^{-4}$, $q_{0}\approx-0.6$,
$\dot{q}_{0}=-1.8\,H_0$. There is a future singularity at $z>-1$.
The curves have the same meaning as in
Fig.~\ref{fig:Model-1overB}.\label{fig:Model-R2-over-B}} }

\newtext{For definiteness}, let us now focus for a while on
the canonical type of models characterized by $F=1/B$, i.e.\ our
starting class of models defined in section \ref{action}. As we
know, for these models $\beta$ has mass dimension 8 and can be
represented as $\beta\equiv{\cal M}^8$, with ${\cal M}$ some mass
scale.
\newtext{This suggests} that the numerical analysis of the $F=1/B$ models can be more
conveniently performed using the dimensionless combination
$\xi\equiv 6/(x\,H_0^4)=3\beta/(\rLi\,H_0^4)$ as an input parameter.
In Figs.~\ref{fig:q-Future-beta-1} ($\xi<0$)
and~\ref{fig:q-Future-beta-2} ($\xi>0$), we plot the deceleration
$q$ for different values and signs of $\xi$ (or equivalently of
$\dot{q}_{0}$). In these examples
\newtext{we use} $q_{0}\simeq-0.6$ for the current deceleration,
although one has some freedom here, too. The plots show a broad
range of different solutions for $q$, with some of them close to the
$\Lambda$CDM evolution. Remember that the latter makes a precise
prediction for the transition redshift from deceleration to
acceleration, given by
\begin{equation}\label{eq:transition}
z^{*}=-1+\sqrt[3]{2\,\frac{\Omega_{\CC}^0}{\Omega_m^0}}\simeq
0.75\,,
\end{equation}
where the numerical value corresponds to our choice
$\Omega_m^0=0.27$ (hence $\Omega_{\CC}^0=0.73$ for a flat universe).
We see that most of the examples plotted in
Fig.~\ref{fig:q-Future-beta-1} -- corresponding to different values
of the parameter $\xi$ and the initial conditions as in
Fig.~\ref{fig:Model-1overB} -- predict a transition redshift smaller
and hence nearer to our time, meaning that the accelerated expansion
is more recent. For large and negative $\xi$, however, one finds
$z^{*}\simeq 0.7$ and the $\CC$CDM result can almost be matched --
see e.g.\ the $\xi=-10^4$ curve in Fig.~\ref{fig:q-Future-beta-1}.
In this example, if we take as usual $|\rLi|\sim M_X^4$ with
$M_X\sim 10^{16}$~GeV, and we recall that $H_0\sim
10^{-42}\,\text{GeV}$, we find $\M=\beta^{1/8}\sim 10^{-4}$~eV,
i.e.\ a mass scale characteristic of light neutrinos. Similarly in
Fig.~\ref{fig:q-Future-beta-2}, corresponding to $\xi>0$, although
in this case we can see that there are values that give a transition
redshift much earlier in time (i.e.\ a more remote onset of the
acceleration period), although the scale $\M$ remains of the same
order of magnitude. In fact, all values of $|\xi|$ in the wide
interval $1\lesssim|\xi|\lesssim10^8$ render $\M$ in the light
neutrino range $10^{-4}\,\text{eV}\lesssim{\cal
M}\lesssim10^{-3}$~eV. However, for $\xi>0$ the values which are
below $210$ are clearly excluded
\newtext{as they correspond} to $q<0$ in the past and therefore the matter
epoch ($q=1/2$) would not have occurred.

\newtext{A few more observations} are worthwhile in regard to
Figs.\,\ref{fig:q-Future-beta-1} and \ref{fig:q-Future-beta-2}. For
negative values of $\xi$, the asymptotic evolution follows either
the power-law expansion discussed above
\newtext{(if $|\xi|\gtrsim 105$)} or a future singularity (if $|\xi|<104$).
This singularity corresponds to a situation leading to
superaccelerated cosmology ($q\ll -1$) ending in a Big Rip. The
boundary between both final states is de Sitter space-time ($q=-1$).
Larger values of $|\xi|$ lead to an expansion behavior quite close
to $\Lambda$CDM in the recent past. In the case of positive $\xi$,
only the power-law solution exists for $t\rightarrow\infty$.
However, as already mentioned, in this case $|\xi|$ is bounded from
below. Summarizing, late-time expansion histories close to
$\Lambda$CDM (as suggested by observations) require large values of
the parameter $|\xi|$ (hence of $\beta$, for a given $\rLi$), which
always imply a power-law future state. In the limit
$|\beta|\rightarrow\infty$ we find for both signs of $\rLi$ the
value $\dot{q}_{0}\approx-0.57\, H_{0}$, and the evolution of~$q$
becomes very insensitive with respect to $\beta$, see
Fig.~\ref{fig:q-Future-beta-2}. The last statement can be
\newtext{derived from (\ref{eq:rhoF-LateTime})}, \newtext{where
we see} that the terms in square brackets must vanish for
$|\beta|\rightarrow\infty$, otherwise the Einstein
equation~(\ref{eq:Einstein-H2}) is not fulfilled. Consequently,
$\dot{q}$ does not depend on $\beta$ since it can be expressed as a
function of $q$ and $H$ only.

The phantom universes with $q<-1$ at all times (green dotted curves
in Fig.~\ref{fig:1overB-phases-latetime}) are not among the
numerical solutions since we fixed the initial conditions to
$q_{0}=-0.6$. This also explains the absence of the solutions
plotted as blue dashed-dotted curves in \newtext{the case} $\xi<0$.
\newnewtext{Finally}, let us also remark that although the numerical examples
displayed in Figs.~\ref{fig:Model-1overB} and
\ref{fig:Model-R-over-B}-\ref{fig:Unified-model-R3-over-B2}
correspond to $\rLi<0$ (basically because, as we have discussed in
section \ref{CCfinetuning}, this is the situation in the important
case of the SM of particle physics), a similar set of relaxation
curves would be obtained for $\rLi>0$.

\subsection{Generalized relaxation models: $F^s_m=R^{s}/B^{m}$ \newtext{and beyond}}\label{sec:GeneralModels}

Without much effort the results in Sec.~\ref{sec:Realistic-model}
can be carried over to many models which have a factor $B^{-1}$ in
$\rLe$ and $\pLe$. \newtext{The natural generalization} is to
consider models of the form
\begin{equation}\label{Fsm}
F^s_m :=\frac{R^{s}}{B^{m}}=
\frac{R^{s}}{\left[\frac{2}{3}R^{2}+\frac{1}{2}\G+(y\,
R)^{n}\right]^m}\,,\ \ \ \ (s\geq 0, m>0;\,n>2)\,.
\end{equation}
Recall that we usually choose $n=3$ to smooth the transition between
radiation and matter eras. The canonical model with which we have
started in section \ref{action} is just the particular case $F^0_1$,
and the models considered in equation (\ref{rFwF}) are the $F^0_m$
ones. The cosmic evolution of $q(z)$ in the matter and radiation
epochs for $F^s_m$ does not show any relevant differences with
respect to $F^0_1$ because only the structure of $B$ is important
there, in the sense that the relaxation relation~$B\rightarrow0$
holds. However, the relative energy densities~$\Omega_{n}(z)$ of the
matter/energy components and the EOS~$\weff$ may evolve differently,
which is demonstrated in Figs.~\ref{fig:Model-R-over-B},
\ref{fig:Model-R2-over-B} and \ref{fig:Model-R3-over-B2}. Moreover,
at late times also the deceleration $q$ may differ significantly
among the models.
%
\FIGURE[t]{
\includegraphics[width=1\columnwidth]{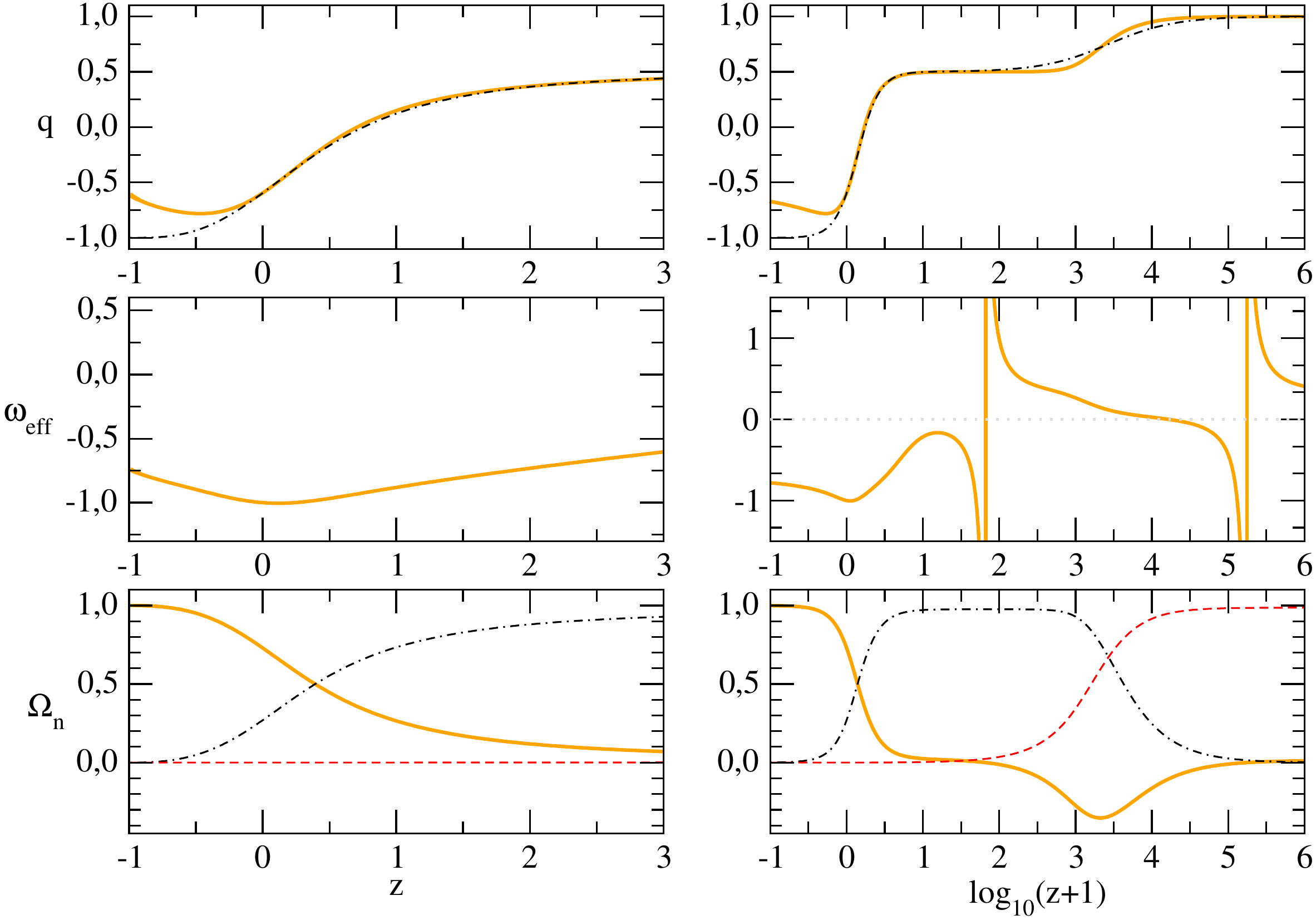}
\caption{Model $F=R^{3}/B^{2}$ with $n=3$, $y=0.75\times
10^{-3}\,H_0^{-2/3}$, $\rho_{\Lambda}^{i}=-10^{60}\,\text{GeV}^4$,
$\Omega_{m}^{0}=0.27$, $\Omega_{r}^{0}=10^{-4}$, $q_{0}\approx-0.6$,
$\dot{q}_{0}=-0.8\,H_0$. The curves have the same meaning as in
Fig.~\ref{fig:Model-1overB}.\label{fig:Model-R3-over-B2}} }
%
\FIGURE{
\includegraphics[width=1\columnwidth]{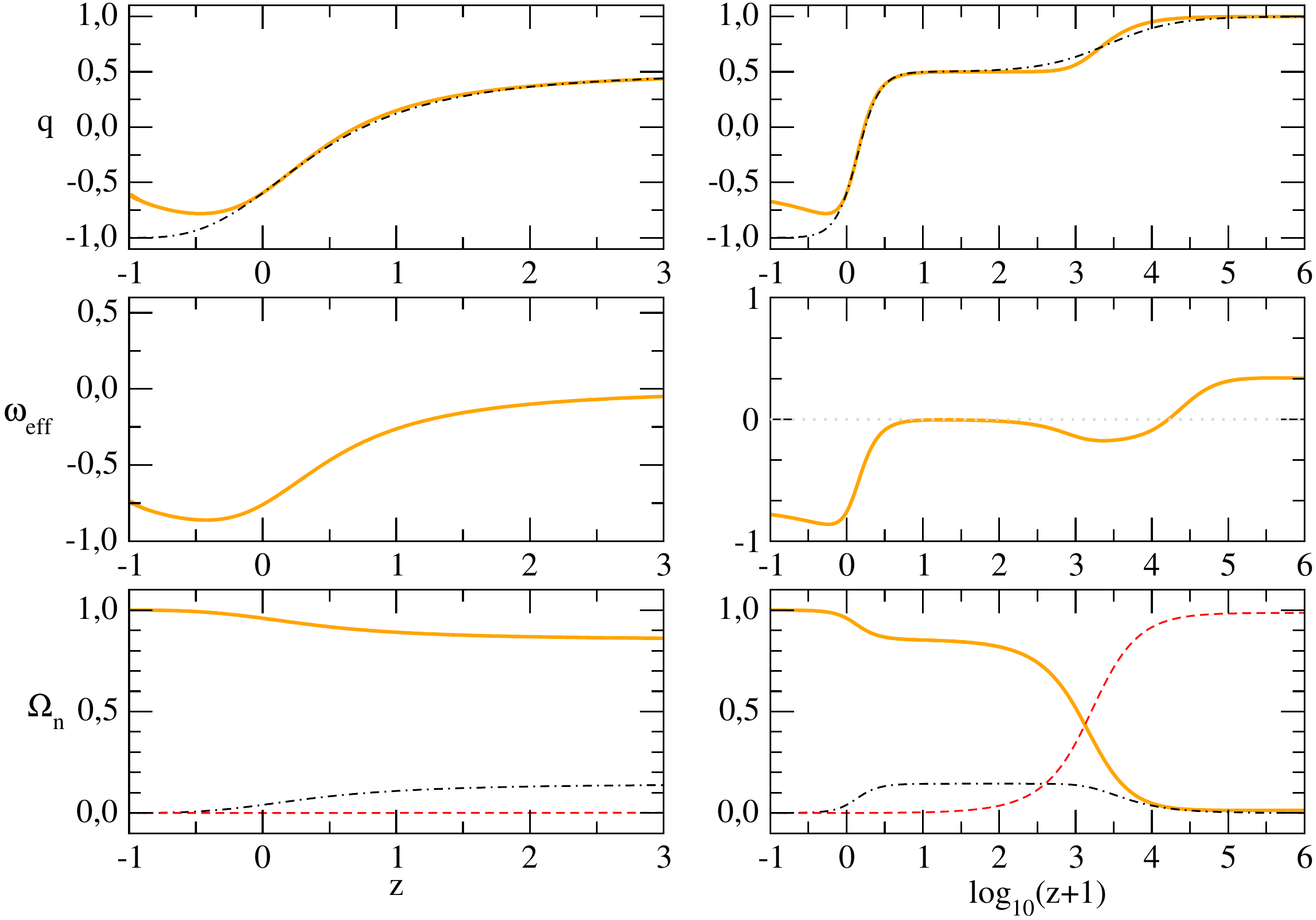}
\caption{The unified dark matter/dark energy scenario from
Sec.~\ref{sub:Unified-dark-sector} in the model $F=R^{3}/B^{2}$ with
$n=3$, $y=0.75\times 10^{-3}\,H_0^{-2/3}$,
$\rho_{\Lambda}^{i}=-10^{60}\,\text{GeV}^4$, $\Omega_{m}^{0}=0.04$,
$\Omega_{r}^{0}=10^{-4}$, $q_{0}\approx-0.6$,
$\dot{q}_{0}=-0.8\,H_0$. The curves have the same meaning as in
Fig.~\ref{fig:Model-1overB} except for $\Omega_{m}^{0}$ (black
dashed-dotted), which represents the baryons only. The unified dark
component exhibits a positive energy density $\rLe>0$ at all
times.\label{fig:Unified-model-R3-over-B2}} }

Apart from dynamical properties, the dimensionality and the value of
the parameter~$\beta$ depend on the particular generalized class
$F^s_m$ of $F(R,\G)$ functional, too. \newtext{One} can easily show
that the parameter~$\beta$ is related to a power of a mass scale $M$
as
\begin{equation}\label{Mpower}
|\beta|=M^{4-2s+4m}\,.
\end{equation}
\newtext{All these models}
yield induced DE terms of the form $\rho_{F},p_{F}\propto B^{-p}$
with $p>0$ as a result of equations (\ref{eq:E00}) and
(\ref{eq:Eij}), where the derivatives of $F$ and $\G$ introduce more
factors of $B$ in the denominators. One can readily show that the CC
relaxation and tracking properties in the matter and radiation eras
follow again from the relaxation condition.

For estimating $\beta$ let us assume that each time derivative in
the expression for  $E_{\,\,0}^{0}$ -- see equation (\ref{eq:E00})
-- yields a factor $H_{0}$ today, and no accidental cancelations are
present so that $B\sim H_{0}^{4}$. Consequently, in the model
$F^0_1=1/B$, we find $\rho_{F}^0\sim \beta/H_0^{4}$
\newtext{for the current value} of the induced term, as in the toy
model example of section \ref{sec:toy-model}. Indeed, in the late
epoch, where $q$ ceases to play a significant role, the realistic
model based on (\ref{eq:B}) produces qualitatively the same picture
as in the simple one (\ref{eq:warmup}). Therefore, within one order
of magnitude, equation (\ref{Hstar}) holds again and we can estimate
the value of the parameter $\beta$ using the current value of the
Hubble rate, $H_0\sim 10^{-33}$ eV. Since $\rF^0\sim -\rLi$ and
$|\rLi|\sim M_X^4$, we get
\begin{equation}\label{eq:estim}
|\beta|\equiv\M^{8}\sim |\rF^0|\,H_0^4\sim M_X^4\,H_0^4\,,
\end{equation}
where $\beta$ is expressed as a power of an energy scale~$\M$ as in
(\ref{eq:betaM}). Taking the standard GUT scale $M_X\sim 10^{16}$
GeV, we obtain once more a mass value for ${\cal M}$ of the order of
a light neutrino mass:
\begin{equation}\label{Mcanonical}
{\cal M}\sim \sqrt{M_X\,H_0}\sim10^{-4}\,\text{eV}\,.
\end{equation}
Incidentally, this value for $M$ is the geometric mean of the two
most extreme mass scales available in our universe below the Planck
mass.
Such value is by the way not far from the mass scale associated to
the current value of the CC:
$m_{\CC}\equiv\left(\rLo\right)^{1/4}\sim 10^{-3}$ eV, and it should
therefore be considered quite natural. Actually, the value of ${\cal
M}$ obtained in a given model could be larger than our rough
estimation, especially if cancelations in $E_{\,\,0}^{0}$ occur, but
as we shall see there is a wide spectrum of possible values for $\M$
in the class of $F^s_m$ models, and all of them within a reasonable
particle physics range.

In fact, applying the same arguments for the models $F^1_1=R/B$ and
$F^3_2=R^{3}/B^{2}$ we find in both cases
$|\beta|=\M^{6}\sim\rLi\,H_{0}^{2}\sim M_X^4\,H_{0}^{2}$, which
implies $\M\sim 0.1\,\text{GeV}$. This is again a reasonable mass
scale in particle physics, as it is of the order of the
characteristic QCD scale of the strong interactions, $\M\sim
\Lambda_{QCD}\simeq 100$ MeV. Finally, the model $F^2_1=R^{2}/B$
yields $|\beta|=\M^{4}=|\rLi|\sim M_X^4$, and in this case $\M$
would be close to the initial vacuum energy scale $M_X$ of the
GUT\,\footnote{\newtext{Notice} that although for this model the
function $R^{2}/B$ itself does \textit{not} increase with $H\to 0$
(it remains constant in this limit), the derivatives of that
function in equation (\ref{eq:E00}) do indeed increase for $H\to 0$
and hence they are entirely responsible for the relaxation process
in this case.}. This situation is also perfectly reasonable inasmuch
as $M_X$ is another natural scale of the problem, just the one
placed at the starting point of the evolution. Remarkably, the mass
parameter~$M$ of the relaxation mechanism is not only completely
free from fine-tuning problems, it also lies in a perfectly
reasonable range of particle physics masses, possibly related to
neutrinos, QCD or even GUT models. In \textit{all} these situations
\newtext{the relaxation mechanism} takes care automatically that the
observable CC scale of the vacuum energy density is
$\rLe(H)\sim\rLo$ at $H=H_0$.

In summary, the energy scale $\M$ of the CC relaxation mechanism
could be very well related to reasonable mass scales in Particle
Physics, thereby avoiding tiny energy scales $H_{0}\sim 10^{-33}$ eV
which so often appear in many dark energy models (for example, in
quintessence models\,\cite{Quintessence}). Once the correct $F$-term
has been identified, the mass scale $\M$ should be considered a sort
of ``constant of Nature'', and in this sense we expect it should
bear direct relation with a particle physics scale of the SM or of a
typical GUT scale.

\revisedtext{We remark} that the fact that the initial cosmological
term $\rLi$ was considered constant (but otherwise arbitrary), is
not a real limitation for the efficiency of our dynamical adjustment
mechanism. This mechanism works for virtually any cosmological model
in the general class of $\CC$XCDM models\,\cite{LXCDM,LXCDM2}. In
particular, we cannot exclude that the initial vacuum energy density
can be evolving with time through some cosmological quantity
$\xi=\xi(t)$, i.e.\ $\rL=\rL(\xi(t))$. For example, it has been
suggested in the literature that $\rL$ could be a quadratic function
of the expansion rate: $\rL(H(t))=n_0+n_2\,H^2(t)$, where $n_0$ and
$n_2$ are both non-vanishing coefficients, see e.g.\ the recent
papers \cite{ShapSol09,BFLWard09,Maggiore10} and the older ones
\cite{oldCCstuff1,oldCCstuff2}. The reason for this ``running'' of
the vacuum energy stems from the expanding background, and these
papers suggested that the quantum effects provide precisely this
running behavior. \newnewtext{For a summary} of alternative
proposals of evolving vacuum energy that have recently been tested
in the light of the most recent cosmological data, see e.g.
\cite{BPS09a}.

In the present framework, the effective vacuum energy $\rLe$ is
anyway a mildly evolving quantity, even in the absence of the
aforementioned running contributions. But if one accepts that the
running terms are also there, nothing essential is changed in our
relaxation mechanism. One has to keep $\dot{\rho}_{\CC}\neq 0$ in
the covariant conservation law (\ref{conslawDE2}) and the
$F$-functional will adjust automatically the contribution of the
additional terms. Technically, however, let us clarify that if these
contributions are added directly at the level of the field
equations, then in order to keep matter conservation we have to
allow for a ``running gravitational coupling $G_N$'' as well. This
possibility has also been contemplated in the literature and it
corresponds to the so-called type-II $\CC$XCDM
models\,\cite{LXCDM2}, in which the energy density conservation law
involves $\dot{\rho}_{\CC}\neq 0$ and $\dot{G}_N\neq 0$
simultaneously. We refer the reader to the aforesaid reference for
details.

\newtext{Finally} there is still another modification of our
relaxation functional that may be necessary in order to make our
cosmological model a bit more realistic: we have to insure that the
modified gravitational action (\ref{eq:CC-Relax-action}), once it is
written in a metric amenable for Solar System tests (namely, the
Schwarzschild metric in the presence of a cosmological term), is
able to pass these tests and at the same time keeps its ability to
reduce dynamically the value of the cosmological constant within
observations. While a detailed treatment of this problem will be
presented elsewhere\,\cite{BSS:Solar}, in
Appendix~\ref{sec:SolarSystem} we briefly describe the kind of
modification that we have to introduce on the functional $\F$ such
that it fulfills all these requirements without altering in any
essential way the dynamical relaxation mechanism that we have
studied so far.

\subsection{Special scenario: unified dark matter and dark energy}\label{sub:Unified-dark-sector}

In this section we discuss a special scenario of unified dark matter
and dark energy. At least for the background evolution this seems to
be possible because the dark energy density~$\rLe$ behaves like dust
matter in the matter era, see equation (\ref{eq:rhoD-MatEra}).
Therefore, it might replace dark matter completely. Whether the
effective dark energy/matter in our setup provides the correct
clustering properties to successfully seed structure formation will
be studied in a future work together with its evolution on solar
system and galactic scales\,\footnote{In fact, there exists already
some recent work considering the clustering effects of the vacuum
energy\,\cite{BPS2010a}.}. If it works, we would find a MOND-like
theory since our modified gravity model had to simulate the effect
of standard dark matter.

In the following, we concentrate on the cosmic background expansion.
The model shown in Fig.~\ref{fig:Unified-model-R3-over-B2} describes
a universe with only baryons $\Omega_{m}^{0}=0.04$, dark energy and
radiation $\Omega_{r}^{0}=10^{-4}$. Since standard dark matter is
absent, the relative energy densities of dark energy and the baryons
sum up to unity in the matter era. At late times and during the
radiation epoch there is no big difference between this special
relaxation scenario \newtext{and} the previously studied relaxation
models with dark matter.

Finally, we mention two advantages of this approach. First, we do
not need the parameters related to dark matter ($\Omega_{dm}^{0}$)
because $y$ fixes the radiation--matter transition. Second, it is
possible to have $\rLe>0$ all the time, which leads to an effective
EOS for $\rLe$ without \newtext{spurious} divergences. In fact,
these divergences should not represent a real problem, they just
indicate that $E_{\,\,0}^{0}$ in (\ref{eq:E00}) vanishes. However,
in this special scenario the full $_{\,\,0}^{0}$-component in the
Einstein equation remains always positive.

\section{Conclusions}

In this work, \newnewfinal{we have made a thorough attempt} to deal
with the old cosmological constant (CC) problem\,\cite{weinberg89}
-- the toughest cosmological conundrum of all times. It suggests
that our vacuum should contain a huge energy density~$\rLi$ coming
from quantum zero-point energies and phase transitions in the early
universe. This formidable problem cannot be solved by just finding a
source for the late-time acceleration of our cosmos. Instead, it
requires a powerful mechanism capable of disarming the large CC
during \textit{all} known stages of the cosmological evolution, not
just at the last stages. Scalar field models, for instance,
completely fail to solve the old CC problem, as they actually
introduce two severe theoretical diseases: 1) extreme fine-tuning of
the parameters (through e.g.\ an \textit{ad hoc} counterterm in the
potential that cancels the initial CC), and 2) an extremely tiny
mass scale needed to match the ground state potential with the
present value of the CC. Such mass scale is usually as small as
$H_0\sim 10^{-33}$ eV, i.e., very many orders of magnitude smaller
than the natural mass scale $m_{\CC}\equiv(\rLo)^{1/4}\sim 10^{-3}$
eV associated to the observationally measured value of the vacuum
energy density. Quintessence models, therefore, do not really solve
the problem since they input two highly unnatural ingredients that
completely spoil the credibility of the proposal. \newtext{If that
is not enough}, they simply put the vacuum energy under the rug as
if it would not exist, and then focus exclusively on the special
properties of the newly invented scalar field.

\newnewtext{Clearly}, a radically different new approach is required.
In an attempt to make a first step in this direction, we have
proposed a self-adapting
\newtext{(\textit{dynamical})} relaxation mechanism based on a pure
modification of gravity. Thanks to it, we can avoid fine-tuning at
all stages of the cosmological evolution. More specifically, we have
demonstrated that an \textit{arbitrarily large} initial CC can be
relaxed automatically by complementing the Einstein-Hilbert action
with a class of action functionals of the Ricci scalar and the
Gau\ss-Bonnet invariant, $\F(R,\G)$. The \newtext{two} most
remarkable achievements of our relaxation mechanism is the complete
absence of fine-tuning in the parameters of the model, and at the
same time the absence of tiny mass scales. The modified gravity
action induces a dynamical dark energy component, $\rF$, associated
to the $\F$-functional. This induced DE density acts effectively as
a ``dynamical counterterm'', i.e.\ one that self-adjusts to the
necessities of the universe at any given moment, and is able to
efficiently neutralize any exceedingly large vacuum energy $\rLi$
left over near the radiation epoch by the primeval inflation
mechanism, thus preventing the universe from the disastrous effects
that would ensue otherwise (as e.g.\ the disruption of the
primordial nucleosynthesis epoch). Indeed, by becoming sufficiently
large to compensate for the initial CC, the \newnewtext{effective
energy density} $\rF$ generated by the $\F$-functional self-adapts
automatically to the initial conditions of the universe \textit{and}
to the subsequent radiation and matter epochs, and enforces the
universe to follow the standard FLRW cosmic expansion history in
each one of these epochs.

As a result, the ``residual'' dark energy of our model (or effective
vacuum energy density~$\rLe$) appears, at every stage of the
cosmological evolution, as a {mildly evolving} function of the
expansion rate, $\rLe(H)=\rLi+\rF(H)$. \newnewfinal{This function}
takes values much smaller than the initial $\rLi$ because there is a
large (and automatic) compensation of it with the values taken by
$\rF(H)$, which have opposite sign. Moreover, $|\rLe(H)|\ll |\rLi|$
holds good for \textit{all} values of $H$ at (and below) the
radiation epoch, and for any given $\rLi$. In addition, the current
value of $\rLe$ can be of the order of the measured one by the
modern cosmological data, i.e.\ $\rLe^0\equiv\rLe(H=H_0)\simeq
\rLo\simeq 10^{-47}\,\text{GeV}^4$, without introducing fine-tuning
or any unnaturally small parameter. In fact, the small $\rLe^0$ is
the \textit{observable cosmological term} in the present framework,
whereas the large initial CC value $\rLi$ (and the corresponding
dynamical counterterm $\rF$) cannot be resolved individually from
the usual CC measurements.

All in all, our model -- the \textit{Relaxed Universe} -- appears
very similar to the standard Big Bang model and resembles the late
universe \newnewtext{near our time} (concordance $\CC$CDM model) in
most respects, irrespective of the initial vacuum conditions.
Therefore, the $\F(R,\G)$-cosmology appears all the time as a FLRW
model with essentially constant and tiny cosmological term. In
addition, the vacuum energy $\rLe$ exhibits remarkable tracking
properties because it evolves like radiation or dust matter in the
corresponding cosmological epochs. This feature of the model sheds
considerable light also on the cosmic coincidence problem inasmuch
as it provides a \textit{raison d'\^{e}tre} for the fact that the DE
and DM densities are still close at present. Indeed, one finds that
the DE and DM were cosmic companions with similar EOS and
approximately proportional densities, and this was so all the time
until very recently, namely until the matter epoch extinguished
(meaning that $q$ departed much from $1/2$) and the universe --
which is constantly fueled with the big initial $\rLi$ -- was forced
to use the last resort of the relaxation mechanism available under
these circumstances, which is to take a very low value of $H$ in
order to compensate for $\rLi$ and still keep showing a small
observable $\rLe$. That crucial event decoupled once and forever the
tracking behaviors of matter/radiation and DE, and since then the DE
dominates the universe expansion. Such breakdown of the tracking
property at the end of the matter epoch, on the other hand, pushed
the subdominant DE density curve upwards until crossing the decaying
DM one at some point after the matter epoch and hence near our time.
This is the origin of the ``coincidence'' event in our late time
neighborhood. But, most remarkable of all, the primary trigger of
this crossing event (representing the startup of the DE era) is the
relaxation mechanism itself, which is therefore simultaneously
responsible for explaining -- and linking together -- the two
fundamental CC conundrums of our current Cosmos: the old CC problem
and the cosmic coincidence problem.

Worth noticing is also the fact that the \newtext{asymptotic} late
time behavior of the universe in the DE epoch mimics the
quintessential or phantom behavior. Only under special initial
conditions, the future behavior can be de~Sitter space-time, but any
perturbation of these conditions would tilt its evolution into one
of the aforementioned forms.

The dynamical solution found in this paper resembles, in a sense,
the one so longed for by scalar field model builders, except that
these kind of models always failed because they ultimately involve
fine-tuning (as proven by the famous Weinberg's no-go
theorem\,\cite{weinberg89}). Even if the $\F(R,\G)$-cosmology lacks
at the moment of a fundamental understanding of the origin of the
$\F$-functional, the situation is not very different from the
innumerable and unsuccessful attempts based on scalar field models
where one has no fundamental motivation for using a particular
potential that serves our purposes. The advantage of the
$\F(R,\G)$-cosmology, in contrast, is that the job is done by
gravity itself, and is done correctly, namely without fine-tuning
inconsistencies. This was never achieved with scalar field models
nor with any other cosmological model that we know of.

In its current stage, we have been able to devise a relaxation
mechanism which starts working automatically in the early radiation
dominated epoch, hence right after the reheating process that
follows primordial inflation. It was not our purpose here to
identify the fundamental physics that may have triggered this
mechanism. It could have been caused by an ``effective inflaton''
reaching the minimum of its potential, especially if the minimum
lies below zero \newtext{and having} an arbitrary value $|\rLi|$.
\newtext{This value} fixes the initial (usually large) CC of the universe
just before starting the radiation epoch. The latter case appears
much more likely than just assuming (as usually done in the
literature) that the inflaton potential has been fine-tuned to
almost vanishing ground state energy \newtext{at the radiation
epoch}. Furthermore, since our relaxation $\F$-functional is
constructed from a modified gravity theory, it is natural to
complement it with the standard $R^2$ terms of the renormalizable
effective action of gravity. These terms do not alter the relaxation
mechanism in any significant way, as we have seen, but they can just
take care of the UV behavior of the theory and even be responsible
for inflation itself.

In the framework of modified gravity, the successful relaxation of
the CC requires an action functional $\F(R,\G)$ with a special
structure, namely one which leads to an induced dark energy density
being able to compensate dynamically the initial CC. However, our
approach is not limited to a very specific model of this kind;
actually, a whole class of working relaxation models
\newtext{$F^s_m$} has been uncovered. On the basis of several
analytical and exact numerical examples, we have shown that they
differ mostly at late times in accordance with the requirement of
the \newtext{current} cosmos being close to $\Lambda$CDM.
Interestingly enough, our action functional $\F$ contains a mass
scale $\M$ whose natural value can either be a typical particle
physics mass of the standard model of strong and electroweak
interactions (say, a neutrino mass, the QCD scale etc) or even a
typical Grand Unified mass $M_X$. But $\M$ is never an extremely
tiny mass scale of order of~$H_{0}\sim 10^{-33}$ eV, which so often
appears in the DE models of the literature.
\newnewtext{In summary}, we can say that the typical values for
the fundamental mass scale $\M$ that are needed to solve the CC fine
tuning problem in our framework range from the present CC mass
scale, $m_{\CC}\equiv\left(\rLo\right)^{1/4}\sim 10^{-3}$ eV, to the
value of the initial CC mass scale, $\left(\rLi\right)^{1/4}\sim
M_X\sim 10^{16}$ GeV. What else could be more natural?

\newtext{The} model universe we have envisaged here falls within the
general class of the $\CC$XCDM models of the cosmic
evolution\,\cite{LXCDM,LXCDM2,LXCDMmore}, in which a new entity $X$
(the ``cosmon'') appears in interplay with the cosmological term.
Such an entity is in general \textit{not} a field, but a complicated
effective quantity. The cosmon is in fact associated here to the
induced DE density $\rF$ which is generated from the $\F$-functional
\textit{at the level of the field equations}. While in Ref.
\cite{LXCDM} the cosmon served the main purpose of solving (or
highly palliating) the cosmic coincidence problem, here the cosmon
is eventually responsible for solving the fine-tuning CC problem,
and hence it provides a first fundamental step towards solving (or
highly alleviating) the old CC problem. Amazingly enough, it
provides a clue for solving the coincidence problem as well, as we
have seen. Therefore, while in the present realization the cosmon
does not appear as an elementary particle, it does however fully
accomplish its main aim, which is to cure the CC fine-tuning
problem. Let us recall that the name ``cosmon'' was coined for the
first time in the literature in Ref.\,\cite{PSW}, and in its first
implementation it was a scalar field, hence afflicted by the
aforementioned Weinberg's no-go theorem. In the modern version, it
is not. Therefore, the new cosmon well deserves its name, for it
fully adapts to the original sense of the old literature as being an
entity which is able to \textit{dynamically adjust the value of the
cosmological constant}\,\cite{PSW}.
\newnewtext{The cosmon}, indeed, generates the dynamical counterterm
$\rF$ which is able to continuously neutralize the very large
$|\rLi|\sim M_X^4$ in \textit{all epochs} of the post-inflationary
cosmic evolution.

In summary, in this paper we have extensively  discussed the working
principle of the relaxation mechanism and the corresponding
background evolution of the $\F(R,\G)$ cosmology. In doing this we
have found an exceptional model universe, the \textit{Relaxed
Universe}, in which all the main tensions with the observational
data are relaxed and can be smoothly accommodated in its inner
dynamics. The \textit{Relaxed Universe} seems to be a promising
candidate for our real universe since it is essentially free from
the toughest cosmological puzzles, the old CC problem and the cosmic
coincidence problem. \newtext{Its explicit construction} constitutes
a ``proof of existence'' of a dynamical mechanism that may account
for the value of the current cosmological constant, starting from a
value of arbitrary size. However, although our results are very
encouraging, this is just one more step in the right direction. More
efforts are necessary to uncover all the main properties of our
model and resolve upcoming problems. In the future, we plan to
investigate further aspects of this setup, e.g.\ the evolution of
perturbations, the Newtonian limit and possible instabilities,
leading to new insights and improvements. Also, it will be
interesting to know if the unification of dark matter and dark
energy can be fully realized in this framework. And of course one
should eventually look for a fundamental theory of the underlying
action functional,
\newtext{which at the moment is lacking}.

The previous issues are certainly difficult problems of fundamental
nature, which must be addressed in the future in order to assess
whether the model (or some variation of it) can eventually become a
realistic one.
\newtext{In the meanwhile}, the very \newtext{theoretical} construction
and analysis of the \textit{Relaxed Universe} does show that it is
possible to envision a model universe in which an arbitrarily large
vacuum energy early deposited in it by quantum field theory or
string theory dynamics can be rendered innocuous in a fully
dynamical way -- hence without any fine-tuning -- with the pure work
of gravity, and still resulting in an universe whose cosmological
features (in particular the tiny value of the total vacuum energy)
resembles to a great extent the standard $\CC$CDM model. In this
sense, and in spite of the many difficulties that may lie ahead us,
one could say that this model universe should be a first step
\newfinal{towards} solving the old CC problem.

\acknowledgments
The authors have been partially supported by DIUE/CUR Generalitat de
Catalunya under project 2009SGR502; FB and JS also by MEC and FEDER
under project FPA2007-66665 and by the Consolider-Ingenio 2010
program CPAN CSD2007-00042, and HS also by the Ministry of
Education, Science and Sports of the Republic of Croatia under
contract No. 098-0982930-2864.

\appendix

\section{$F(R,\G)$ modified gravity}\label{sec:FofRSTG-action}

In this section we derive the equations of motion of the $f(R,S,T)$
and $F(R,\G)$ modified gravity theories in the metric formalism. In
the modified action the $f$ and $F$ functionals contain the Ricci
scalar $R=g^{ab}R_{ab}$ as well as $S=R_{ab}R^{ab}$ and
$T=R_{abcd}R^{abcd}$, which are scalar invariants built from the
Ricci~$R_{ab}$ and Riemann~$R_{abcd}$ tensors. $\G$ is the
Gau\ss-Bonnet invariant~$\G=R^{2}-4S+T$. In general, $f(R,S,T)$
theories will introduce new degrees of freedom, which potentially
lead to instabilities not present in general relativity. However,
some problems can be avoided by specialising to $F(R,\G)$
functionals involving only the Ricci scalar and the Gau\ss-Bonnet
invariant. \newnewfinal{For instance}, all functionals of the form
$f(R,T-4S)$ are ghost free\,\cite{Comelli05}, and in particular this
is obviously the case of the $F(R,\G)$ functionals. With respect to
the more general $f(R,S,T)$ ansatz this is not a strong restriction
on an FLRW background, since in terms of the Hubble rate $H$ and the
deceleration $q$, the invariants have the form\[
R=6H^{2}(1-q),\,\,\,\, S=12H^{4}(q^{2}-q+1),\,\,\,\,
T=12H^{4}(q^{2}+1),\,\,\,\, \G=-24H^{4}q.\] Therefore, $S$ and $T$
can be replaced by $S_{*}=\frac{1}{3}R^{2}-\frac{1}{2}\G$ and
$T_{*}=\frac{1}{3}R^{2}-\G$, respectively. Note that these
replacements may not hold on the level of the equations of motion or
in general metrics.

First, we fix the notation. Our metric~$g_{mn}$ has the signature
$(+,-,-,-)$, and the Riemann and respectively the Ricci tensors are
given by\[ R_{\,\,
bcd}^{a}=\Gamma_{bd,c}^{a}-\Gamma_{bc,d}^{a}+\Gamma_{mc}^{a}\Gamma_{bd}^{m}-\Gamma_{md}^{a}\Gamma_{bc}^{m},\,\,\,\,\
\ \  R_{bc}=R_{\,\, bca}^{a},\] where~$\Gamma_{bc}^{a}$ are the
Christoffel symbols. In short, our conventions here are
characterized by the three basic signs $(-,+,-)$ relative to the
sign conventions of Ref.\,\cite{MTW} concerning metric, Riemann
tensor and the \textit{r.h.s.}\ of Einstein's equations respectively,
following the well-known classification scheme proposed in that
reference.

For a general $f(R,S,T)$ modified gravity theory the action reads \[
\mathcal{S}=\int d^{4}x\,\sqrt{|g|}\, f(R,S,T),\] and its
variation~$\delta\mathcal{S}$ with respect to the metric
yields\begin{eqnarray}
\delta\mathcal{S}[f(R,S,T)] & = & \int d^{4}x\,\left[\delta(\sqrt{|g|})f(R,S,T)+\sqrt{|g|}(F^{R}\delta R+F^{S}\delta S+F^{T}\delta T)\right]\nonumber \\
 & = & \int d^{4}x\,\sqrt{|g|}[-\frac{1}{2}g_{ab}f(R,S,T)+R_{ab}F^{R}+F_{\,\,;ab}^{R}-g_{ab}\square F^{R}\nonumber \\
 & + & 2F^{S}R_{ac}R_{b}^{\,\, c}-g_{ab}(F^{S}R^{cd})_{;cd}+2(F^{S}R^{cd})_{;ac}g_{bd}-\square(F^{S}R_{ab})\nonumber \\
 & + & 2F^{T}R_{arst}R_{b}^{\,\, rst}-4(F^{T}R_{\,\, ab}^{n\,\,\,\, m})_{;mn}]\delta g^{ab}+(\text{surface terms}),\label{eq:delSRST}\end{eqnarray}
where~$\delta(\sqrt{|g|})=-\frac{1}{2}\sqrt{|g|}g_{ab}\delta g^{ab}$
and \[ F^{Y}=\frac{\partial f(R,S,T)}{\partial Y},\,\,\,
Y\in\{R,S,T\}.\] From $\delta\mathcal{S}[f(R,S,T)]$ one can easily
derive the variation $\delta\mathcal{S}[F(R,\G)]$ after applying in
(\ref{eq:delSRST}) the replacements
\begin{eqnarray}
 & & f(R,S,T)\rightarrow F(R,\G),\,\,\,\,
 F^{R}\rightarrow\frac{\partial \G}{\partial R}F^{\G}+F^{R}=2RF^{\G}+F^{R},
\nonumber \\
 & & F^{S}\rightarrow \frac{\partial \G}{\partial S}F^{\G}=-4F^{\G},\,\,\,
 F^{T}\rightarrow\frac{\partial \G}{\partial T}F^{\G}=F^{\G},
\end{eqnarray} where on the righthand side of the
arrows $F^{R,\G}$ correspond to the partial derivatives of
$F(R,\G)$. On a FLRW background the results of this procedure are
given in Sec.~\ref{sec:Einstein-equations}.

\section{The CC fine-tuning problem in QFT: a more detailed account}\label{sec:finetuning}

In quantum field theory the CC fine-tuning problem (discussed only
at the classical level in section \ref{CCfinetuning}) becomes much
harder. We feel that despite the widespread criticisms scattered
over the literature against the fine-tuning procedure, chiefly in
connection with the CC problem, a detailed discussion of it is
lacking, or at least is not usually made available to the general
reader (apart from some vague notions deprived of a minimum
quantitative analysis), and therefore here it seems the right place
to pay due attention to it\,\footnote{Let us clarify that the
fine-tuning problem is not at all privative of the CC approach to
the dark energy. In most papers dealing with the DE from the point
of view of quintessence, the fine-tuning problem is usually ignored
\textit{ab initio} as if this problem would not go with them. This
is however an unfortunate misconception. These models are plagued,
too, with fine-tuning problems and in fact in no lesser extent than
the traditional CC approach -- including other niceties, such as the
unwanted presence of extremely tiny mass scales.}. Let us flesh out
the new ingredients of the problem at the quantum level in this
Appendix. As a first important step, we need to renormalize the
theory because otherwise it is impossible to reach finite results.
While it is unavoidable to have a curved background in the presence
of a vacuum energy, we shall confine our renormalization discussion
to flat space in order to avoid a cumbersome presentation. Even
without curvature, it will become apparent that the presence of
quantum effects makes the fine-tuning problem extremely difficult
and demands a cleverer solution to the problem. We start by
recalling some basic issues concerning the renormalization of the
effective potential\,\cite{Coleman85}. Indeed, in QFT the field
$\varphi$ becomes a quantum field operator $\hat{\varphi}$ and
therefore its ground state value $<\varphi>$ must now be interpreted
as the vacuum expectation value (VEV) of this operator{:}
$<\varphi>\equiv\langle 0|\hat{\varphi}|0\rangle$. Notwithstanding,
the theory can still be handled as if $\varphi$ were a classical
field provided its potential (\ref{Poten}) is renormalized into the
effective potential, $V\rightarrow \EP$, where
\begin{equation}\label{Veff}
 \EP=V+\hbar\,V_1+ \hbar^2\,V_2+ \hbar^2\,V_3+...
\end{equation}
The quantum effects to all orders of perturbation theory arrange
themselves in the form of a loopwise expansion where the number of
loops is tracked by the powers of $\hbar$. Thus, at one loop we have
only one power of $\hbar$, at two loops we have two powers of
$\hbar$ etc. For $\hbar=0$, however, there are no loops and the
effective potential just reduces to the classical potential, $V$,
given by equation (\ref{Poten}) in the electroweak standard model.
On the other hand, the loop terms in (\ref{Veff}) can be split into
two independent contributions, one having loops with no external
legs ({vacuum-to-vacuum} parts $V_P^{(i)}$) and the other having
loops with external legs of the quantum matter field $\varphi$
(i.e.\ the loop corrections $V_\text{scal}^{(i)}(\varphi)$ to the
classical potential):
\begin{equation}\label{splitloop}
V_1=V_P^{(1)}+V_\text{scal}^{(1)}(\varphi)\,,\ \ \ \
V_2=V_P^{(2)}+V_\text{scal}^{(2)}(\varphi)\,,\ \ \ \
V_3=V_P^{(3)}+V_\text{scal}^{(3)}(\varphi)...\,.
\end{equation}
As a result, the effective potential (\ref{Veff}) at the quantum
level splits naturally into two parts, one which is
$\varphi$-independent and another that is $\varphi$-dependent:
\begin{equation}\label{EPZPE}
\EP(\varphi)=\ZPE+\tEP(\varphi)\,,
\end{equation}
where
\begin{equation}\label{ZPE}
\ZPE=\hbar\,V_P^{(1)}+\hbar^2\,V_P^{(2)}+\hbar^3\,V_P^{(3)}+....
\end{equation}
is the \textit{zero-point energy} (ZPE) contribution, which consists
in {the sum} of all the vacuum-to-vacuum parts of the effective
potential. The ZPE part is sourced exclusively from closed loops of
matter fields (i.e.\ vacuum loops without external $\varphi$-legs).
For instance, at one-loop order, only the first term on the
\textit{r.h.s.} of (\ref{ZPE}) contributes. In dimensional
regularization, it gives
\begin{eqnarray}\label{Vacfree2}
V_P^{(1)}&=&
-\,\frac{i}{2}\,\mu^{4-n}\,\int\frac{d^n
k}{(2\pi)^n}\,\ln\left[-k^2+m^2\right]
 = \frac12\,\mu^{4-n}\, \int\frac{d^{n-1} k}{(2\pi)^{n-1}}
   \,\sqrt{\vec{k}^2+m^2}\nonumber\\
&=& \frac12\,\beta_\Lambda^{(1)}\,\left(-\frac{2}{4-n}
-\ln\frac{4\pi\mu^2}{m^2}+\gamma_E-\frac32\right) \,,
\end{eqnarray}
where we have performed a Wick rotation ($dk_0=idk_4$) into
Euclidean space. Notice that $\mu$ is the characteristic 't Hooft
mass unit of dimensional regularization, and we used $n\rightarrow
4$ in the final result. The second equality in (\ref{Vacfree2}) --
in which we performed the contour integral on $k_4$ -- was only to
make more transparent the connection of this expression with the
more traditional form of the ZPE of the vacuum fluctuations.
Finally, the expression
\begin{equation}
\label{beta4} \beta_\Lambda^{(1)}=\frac{m^4}{32\,\pi^2}
\end{equation}
is the one-loop $\beta$-function of the vacuum
term\,\cite{oldCCstuff1}. As warned from the very beginning, these
results are valid only in flat space-time, and are given just to
illustrate in a concrete way the structure of the expansion
(\ref{ZPE}). In the presence of a curved background, the expression
(\ref{Vacfree2}) must be supplemented with contributions from the
curvature invariants. The effect of the matter fields on them is to
make their coefficients to run as effective charges\,\footnote{For a
discussion of these curvature terms and the potential implications
on the vacuum energy in an expanding background,
see\,\cite{ShapSol09}.}. In all the cases the vacuum-to-vacuum part
does \textit{not} depend on $\varphi$, but only on the {set of
parameters} $P=m,\lambda,...$ of the classical potential. The ZPE
receives {in general} contributions to all orders of perturbation
theory, except at zero loop level since $\ZPE$  is a pure quantum
effect that vanishes for $\hbar=0$. Besides, there is the
$\varphi$-dependent part:
\begin{equation}\label{loopwisepot}
V_\text{scal}(\varphi) =
V(\varphi)+\hbar\,V_\text{scal}^{(1)}(\varphi)+
\hbar^2\,V_\text{scal}^{(2)}(\varphi)+
\hbar^3\,V_\text{scal}^{(3)}(\varphi)+...
\end{equation}
This one is not purely quantum (i.e.\ it does not vanish for
$\hbar=0$) as the first term is not proportional to $\hbar$ -- {it
corresponds} to the tree-level contribution $V(\varphi)$ or
classical potential. In the electroweak standard model, it is given
by (\ref{Poten}). The above $\varphi$-dependent part of $V_{\rm
eff}$ receives {in general} also contributions to all orders of
perturbation theory, and vanishes for $\varphi=0$ since in this case
all the loops have external $\varphi$ legs, {including} the
tree-level part. Thus, $\EP(\varphi=0)=\ZPE$, which is a number. In
other words, (\ref{loopwisepot}) constitutes the quantum corrected
effective potential (excluding the ZPE number). However, the full
effective potential contains both contributions.

In the above discussion, all the field theoretical ingredients ($m$,
$\lambda$, $\varphi$ and $V_{\rm eff}$) are in reality bare
quantities ($m_0$, $\lambda_0$, $\varphi_0$ and $V_{\rm eff\, 0}$)
that require renormalization inasmuch as the loopwise expansion is
UV-divergent order by order. However, renormalization just means
that we replace all the bare quantities with renormalized ones (in
some given renormalization scheme with a specific set of
renormalization conditions) plus counterterms (which are also scheme
dependent and are partially fixed by the condition of cancelling the
UV-divergences): $m_0=m+\delta m$, $\lambda_0=\lambda+\delta
\lambda$, $\varphi_0=Z_{\varphi}^{1/2}\,\varphi=(1+\delta
Z_{\varphi}/2)\,\varphi$... Of course, a similar splitting occurs
with the vacuum term, which was originally a bare term $\rLVo$.
\newnewfinal{This parameter} is of special significance in the present
discussion. We must also split it into a renormalized piece plus a
counterterm: $\rLVo=\rLV+\delta\rLV$. The full set of counterterms
is essential to enable the loop expansion to be finite order by
order in perturbation theory. For instance, if we would renormalize
the theory in the $\overline{\rm MS}$ scheme in dimensional
regularization, the suitable counterterm reads:
\begin{equation}\label{deltaMSB}
{\delta}\rLV=\frac{m^4\,\hbar}{4\,(4\pi)^2}\,\left(\frac{2}{4-n}+\ln
4\pi-\gamma_E\right)\,.
\end{equation}

The next crucial point is to apply the basic renormalization recipe
in QFT, which says that we must equate the full bare effective
action (EA) to the full renormalized EA after performing the
aforementioned renormalization transformation. Since the effective
potential is obtained from the effective action in the limit of
constant mean field or background scalar
field\,\cite{Parker09,Coleman85}, it follows that the bare EA boils
down to
\begin{eqnarray}\label{VeffEA1}
\Gamma[\varphi_0,m_0,\lambda_0,\rLVo]&=&\int d^4 x\left[-\rLVo-\EP
(\varphi_0,m_0,\lambda_0)\right]\nonumber\\
&=& -\Omega\,\left[\rLVo+\EP (\varphi_0,m_0,\lambda_0)\right]\,,
\end{eqnarray}
where $\Omega$ is the total space-time volume. After performing the
renormalization transformation of parameters and fields, we obtain
$\Gamma[\varphi(\mu),m(\mu),\lambda(\mu),\rL(\mu),\mu]$. This
renormalized EA must be the same as the bare one. In particular, in
the constant mean field limit the two terms on the \textit{r.h.s.}
of (\ref{VeffEA1}) are replaced by $\rLV(\mu)$ and
$\EP[\varphi(\mu),m(\mu),\lambda(\mu),\mu]$ respectively, where
$\delta\rLV$ has been additively incorporated in the structure of
the latter. Since the two overall expressions must be equal, we have
\begin{equation}\label{RGpostulate}
\rLVo+\EP(\varphi_0, m_0, \lambda_0)=\rLV(\mu)+\EP(\varphi(\mu),
m(\mu), \lambda(\mu); \mu)\,.
\end{equation}
Notice that the renormalized result depends on an arbitrary mass
scale $\mu$ (say, the formerly used 't Hooft mass unit in
dimensional regularization, or an arbitrary momentum $p$ in a
momentum subtraction scheme). Furthermore, in the above expressions,
the counterterms involve some regulator (e.g.\ space-time
dimensionality $n$ in dimensional regularization, or a simple
cutoff) and can be chosen to cancel all the divergences produced by
the computation of the loop corrections to the potential. The
regulator, therefore, eventually disappears, but the mass scale
$\mu$ remains as a reflex of the arbitrariness of the subtraction
point.  For instance, \newnewfinal{in the free theory} or in the
absence of SSB the one-loop $\overline{\rm MS}$-renormalized vacuum
energy density in flat space is obtained from (\ref{ZPE}) and
(\ref{RGpostulate}) for $\varphi=0$:
\begin{equation}\label{renormEfree}
\rL=\rLVo+\ZPEo=\rLV(\mu)+\ZPE(\mu)\,,
\end{equation}
where
\begin{equation}\label{renZPE}
\ZPE(\mu)=\hbar\,V_P^{(1)}+\delta\rLV
\end{equation}
is the one-loop renormalized ZPE, which, as we see, necessarily
requires the counterterm associated to the vacuum term to become
a finite quantity. From (\ref{Vacfree2}) and (\ref{deltaMSB}) the
renormalized vacuum energy density for $\varphi=0$ can be easily
derived:
\begin{equation}\label{renormZPEoneloop}
\rL=\rLV(\mu)+\frac{m^4\,\hbar}{4\,(4\,\pi)^2}\,\left(\ln\frac{m^2}{\mu^2}-\frac32\right)\,.
\end{equation}
We see that it is explicitly dependent on $\mu$, but this dependence
represents only an internal parameterization of its structure, which
is helpful to track the various kinds of quantum
effects\,\cite{ShapSol09}; the overall $\mu$-dependence, however,
must eventually cancel. In fact, the renormalized parameters are
finite quantities which are also functions of $\mu${:}
$\varphi=\varphi(\mu)$, $m=m(\mu)$, $\lambda=\lambda(\mu)$,
$\rLV=\rLV(\mu)$, and since the vacuum energy cannot depend on the
arbitrary scale $\mu$, the sum of the renormalized vacuum term and
the renormalized potential must be globally scale-independent (i.e.\
$\mu$-independent). This is obviously so because the bare vacuum
term and bare effective potential were scale-independent to start
with. Thus, from (\ref{RGpostulate}) we have
\begin{equation}\label{RGequation}
\mu\frac{d}{d\mu}\left[\rLV(m(\mu), \lambda(\mu);
\mu))+\EP(\varphi(\mu); m(\mu), \lambda(\mu); \mu)\right]=0\,.
\end{equation}
This relation implies that the full effective potential is actually
\textit{not} renormalization group (RG) invariant (contrary to some
inaccurate statements in the literature), but it becomes so only
after we add up to it the renormalized CC vacuum part $\rLV$. In
reality, the structure of the effective potential (\ref{EPZPE}) is
such that the previous relation splits into two independent RG
equations:
\begin{equation}\label{RGequation1}
\mu\frac{d}{d\mu}\left[\rLV(m(\mu), \lambda(\mu); \mu))+\ZPE(m(\mu),
\lambda(\mu); \mu)\right]=0
\end{equation}
and
\begin{equation}\label{RGequation2}
\mu\frac{d}{d\mu}\, V_\text{scal}(\varphi(\mu); m(\mu),
\lambda(\mu); \mu)=0\,.
\end{equation}
Equation (\ref{RGequation1}) shows that it is only the strict
vacuum-to-vacuum part (i.e.\ the ZPE) the one that needs the
renormalized vacuum term $\rLV$ to form a finite and RG-invariant
expression, whereas the renormalized $\varphi$-dependent part of the
potential (i.e.\ the tree-level plus the loop expansion with
external $\varphi$-tails) is finite and RG-invariant by itself. This
is of course the essential message from the renormalization group --
see Ref.\,\cite{ShapSol09,ShapSol0608} for an appropriate physical
interpretation. Explicitly, equation (\ref{RGequation2}) reads
\begin{equation}
\left\{ \mu\frac{\partial }{\partial \mu}+\beta _{P}
\,\frac{\partial }{\partial P} - \gamma_{\varphi} \,\varphi\,
\frac{\partial }{\partial \varphi}\right\} \, V_\text{scal}
\left[P(\mu),\varphi(\mu);\mu \right] =0\,,\label{RGVeff}
\end{equation}
where as usual $\beta _{P} = \mu{\partial P}/{\partial\mu}$
($P=m,\lambda,...$) and $\gamma_{\varphi}=\mu{\partial \ln
Z_{\varphi}^{1/2}}/{\partial\mu}$. Similarly, equation
(\ref{RGequation1}) can be put in the form (\ref{RGVeff}), except
that the $\varphi$ term is absent.

The following interesting result now emerges. Plugging equation
(\ref{renormZPEoneloop}) in the general RG equation
(\ref{RGequation1}), we find immediately that the renormalized
vacuum term $\rLV(\mu)$ must obey the following one-loop
RG-equation:
\begin{equation}\label{RGEvacCC}
\frac{d\rLV}{d\ln\mu}=
\frac{m^4\,\hbar}{2\,(4\pi)^2}=\beta_\Lambda^{(1)}\,\hbar\,,
\end{equation}
which is of course the reason why we called the expression
$\beta_\Lambda^{(1)}$ in (\ref{beta4}) the one-loop $\beta$-function
of the vacuum term. Notice two particular things of the one-loop
result: $\lambda$ is not involved, and $m$ does not run with $\mu$.
Both things cease to be true at higher orders.

The one-loop renormalization of the effective potential is
standard\,\cite{Coleman85}, although the usual discussions on this
subject rarely pay much attention to disentangle the ZPE part from
it. Let us do it. Once more we equate the bare and renormalized
effective potentials after having performed the renormalization
transformation of parameters and fields. Sticking to the
$\overline{MS}$ scheme in dimensional regularization to fix the
counterterms, one finds:
\begin{eqnarray}\label{VOVR2}
\EPR(\varphi)=
\frac12\,m^2(\mu)\,\varphi^2+\frac{1}{4!}\,\lambda(\mu)\,\varphi^4
+\frac{\hbar\,\left(V''(\varphi)\right)^2}{4(4\pi)^2}\left(\ln\frac{V''(\varphi)}{\mu^2}-\frac32\right)\,,
\end{eqnarray}
where from (\ref{Poten}),
\begin{equation}\label{VPP}
V''(\varphi)=m^2+\frac12\,\lambda\,\varphi^2\,.
\end{equation}
This expression already contains the renormalized ZPE, as it is just
its value at $\varphi=0$:
\begin{eqnarray}\label{VeffRen2}
\EPR(\varphi=0)=\frac{\hbar\,m^4}{4(4\pi)^2}\left(\ln\frac{m^2}{\mu^2}-\frac32\right)\,.
\end{eqnarray}
Notice that this result is perfectly consistent with
(\ref{renormZPEoneloop}). As remarked, the full effective potential
is \textit{not} RG-invariant, in particular its one-loop form
(\ref{VOVR2}). The truly RG-invariant expression is the sum
$\rLV(\mu)+\EPR(\varphi)$, as indicated by (\ref{RGequation}). The
vacuum energy density is just the VEV of this expression. Therefore,
we are now ready for addressing the CC fine-tuning problem in the
context of a well-defined, renormalized, and RG-invariant vacuum
energy density in flat space.

Once the full effective potential has been renormalized, the two
loopwise expansions (\ref{ZPE}) and (\ref{loopwisepot}) become
finite. Furthermore, the basic equation (\ref{rLphclass}) remains
formally the same in the quantum theory, i.e.\ the physical energy
density associated to the CC is the sum of the vacuum part plus the
induced part. The only difference is that the induced part now
contains all the quantum effects, i.e.\ it reads $\rLI=\VEPo$, or
equivalently $\rLI=\langle V_{\rm eff}^{\rm ren}(\varphi)\rangle$,
{where} $V_{\rm eff}^{\rm ren}(\varphi)\equiv \EP(\varphi(\mu);
m(\mu), \lambda(\mu); \mu)$ is the renormalized effective potential.
{Notice that} the latter includes the (renormalized) ZPE part, which
was absent in the classical theory. Thus, the physical CC emerging
from the renormalization program (in any given subtraction scheme)
reads
\begin{equation}\label{rLphquant}
\rLP=\rLV+\rLI=\rLV^{\rm ren}+\langle V_\text{eff}^{\rm
ren}(\varphi)\rangle=\rLV^{\rm ren}+\ZPE^{\rm ren}+\langle
V_\text{scal}^{\rm ren}(\varphi)\rangle\,,
\end{equation}
where for simplicity we have obviated the $\mu$-dependence in the
renormalized parts. The first equality in~(\ref{rLphquant}) can be
considered the bare result and is formally identical to the last
equality after removing the superscript ``ren'' everywhere. Although
we have provided explicit one-loop results for the sake of
concreteness, the above formulae are completely general. They also
illustrate the systematics of the renormalized perturbative
expansion of the vacuum energy and the kind of different
contributions that are generated to all orders of perturbation
theory, not just to one-loop order. In this sense, they are
sufficiently descriptive to formulate now the severity of the CC
problem and its relation with fine-tuning.

Indeed, since the expression (\ref{rLphquant}) is the precise QFT
prediction of the physical value of the vacuum energy to all orders
of perturbation theory, it follows that it must be equal to the
observational measure value\,\cite{cosmdata,SNIa}, i.e.\
$\rLP=\rLo\simeq 2.8\times 10^{-47}\,\text{GeV}^4$. We have already
seen in section \ref{CCfinetuning} that the lowest order
contribution from the Higgs potential is $55$ orders of magnitude
larger than $\rLo$, and that this enforces us to choose the vacuum
term $\rLV$ with a precision of $55$ decimal places such that the
sum $\rLV+\rLI$ gives a number of order $10^{-47}\,\text{GeV}^4$.
The problem is that the fine-tuning game, ugly enough as it is at
the classical level, becomes actually much more perverse at the
quantum level. Indeed, recall that we have the all order expansions
(\ref{ZPE}) and (\ref{loopwisepot}). Therefore, the quantity that
must be equated to $\rLo$ is not just (\ref{eq:Higgstree}) but the
full \textit{r.h.s.}\ of (\ref{rLphquant}), which is a finite and
RG-invariant expression. In other words, instead of just requiring
to satisfy equation (\ref{finetuningclassical}), we must fulfill the
much more severe one:
\begin{eqnarray}\label{eq:finetuning}
10^{-47}\,\text{GeV}^4=\rLV-10^8\,\text{GeV}^4&+&\hbar\,V_P^{(1)}+\hbar^2\,V_P^{(2)}+\hbar^3\,V_P^{(3)}....\nonumber\\
&+&\hbar\,V_\text{scal}^{(1)}(\varphi)+\hbar^2\,V_\text{scal}^{(2)}(\varphi)+\hbar^3V_\text{scal}^{(3)}(\varphi)...
\end{eqnarray}
Clearly, since on the \textit{r.h.s.} of this equation there are two
independent perturbatively renormalized series contributing to the
observed value of the vacuum energy (viz.\ the ZPE series and the
series associated to the quantum corrected Higgs potential), the
selected numerical value for the ``renormalized counterterm'' $\rLV$
must be changed order by order in perturbation theory; specifically,
$\rLV$ must be re-tuned with $55$ digits of precision as many times
as indicated by the order of the highest loop diagram providing a
contribution to the CC that is smaller than the experimental number
on the \textit{l.h.s.}\ of equation (\ref{eq:finetuning}). Let us
assume that each electroweak loop contributes on average a factor
$g^2/(16\pi^2)$ times the fourth power of the electroweak scale
$v\equiv\langle\varphi\rangle\sim 100$ GeV (see section
\ref{CCfinetuning}), where $g$ is either the $SU(2)$ gauge coupling
constant or the Higgs self-coupling, or a combination of both. It
follows that the order, $n$, of the highest loop diagram that may
contribute to the measured value of the vacuum energy, and that
therefore could still be subject to fine-tuning, can be
approximately derived from
\begin{equation}\label{nloops}
\left(\frac{g^2}{16\,\pi^2}\right)^n\, v^4=10^{-47}\,GeV^4\,.
\end{equation}
Take e.g.\ $g$ equal to just the $SU(2)$ gauge coupling constant of
the electroweak SM, which satisfies $g^2/(16\pi^2)=\alpha_{\rm
em}/(4\pi\sin^2\theta_W)\simeq 2.5\times 10^{-3}$. This is actually
a conservative assumption because in practice there are larger
contributions in the SM associated to the big top quark Yukawa
coupling, but it will suffice to illustrate the situation. Since
$v\equiv\langle\varphi\rangle\sim 100$ GeV, we find $n\simeq 21$.
Therefore the 21th electroweak higher order loop diagrams still
might contribute sizeably to the value of the CC, and must therefore
be readjusted by an appropriate choice of the renormalized value of
the vacuum term $\rLV$. This is of course preposterous and
completely unacceptable. Even though this is nonsense, it is
implicitly accepted by everyone that admits that such ``technical
trick'' is a viable solution to the CC problem. It goes without
saying that this situation worsens even more for higher energy
extensions of the SM, such as in grand unified theories (GUT's).

We see that the presence of the bare vacuum term $\rLV$ was not
optional. While we could have dispensed with it in the classical
theory, it is no longer possible in the quantum theory. This term is
essential in order to provide both a UV-divergent counterterm to
renormalize the ZPE part of the effective potential, as well as a
finite (and fine-tuned) counterterm to compensate the large induced
CC contribution. In other words, the vacuum term is not just a
classical object but an indispensable ingredient insuring the
technical consistency of the theory at the quantum level. However,
even if technically possible, this fine-tuning game does not seem to
be a terribly convincing recipe to solve the CC problem in any
sensible way. Specially in the QFT case, where we have made evident
that it becomes a nightmare of incommensurable proportions; in fact
one which cannot be sustained on any sound theoretically ground.

\newnewtext{We hope} to have convinced the reader that a dynamical solution of
the old CC problem is absolutely mandatory, especially if we do take
the vacuum energy of QFT any seriously. Of course, one can hide the
vacuum energy of the SM of the strong and electroweak interactions
under the rug and resort to a particular scalar field
and claim that the cosmological constant adopts its value entirely
from its ground state. However, this way of proceeding does not
explain, or give any hint of why we can ``happily'' ignore the huge
vacuum energy density predicted by the most successful quantum field
theory at our disposal, viz.\ the Standard Model of the Elementary
Particles, nor does explain at all why the value of the energy
density of such field is precisely of the order of $10^{-47}$
GeV$^4$ rather than, say, $M_X^4\sim 10^{64}$ GeV$^4$, which is what
everyone would expect for a non-SM field connected with Grand
Unified Theories.
\newnewtext{Therefore}, from our point of view, it is much more
convincing to envisage an explicit mechanism dealing with the entire
value of the vacuum energy (irrespective of its size) and reducing
it (dynamically) to its present small value, even if the fundamental
origin of this mechanism is unknown. After all, in spite of the
tremendous variety of DE models (including all the scalar field
models of course), there is not a single one which can be shown to
emerge from a fundamental theory, say GUT's or strings, and none of
them provides a solution to the fine-tuning problem either.
\newfinal{Despite scalar fields} (dilaton-like ones, for example) are
usually associated with string theory, they have a priori no serious
impact on the CC problem without providing (by hand) some suitable
form of the corresponding scalar field potential, and subsequently
exerting (also by hand) a strong fine-tuning of its parameters. In
contrast, the \textit{Relaxed Universe}  presented in this paper
provides at least an efficient dynamical scape to fine-tuning,
irrespective of the size of the initial vacuum energy in the early
post-inflationary universe. Thus, hopefully, this
\newfinal{prototype model} should be a first indispensable step
towards eventually finding a fundamental theory of the cosmological
constant.


\section{Solar System environment}
\label{sec:SolarSystem} So far we have shown that the CC relaxation
mechanism works well in a cosmological setup. In the following we
briefly discuss how the mechanism operates in the Solar System
environment, which we describe by the Schwarzschild-de~Sitter (SdS)
metric in spherical coordinates, \revisedtext{i.e} the standard
Schwarzschild metric including the presence of a cosmological term
$\CC$:
\begin{eqnarray}
ds^{2} & = & A\, dt^{2}-A^{-1}\, dr^{2}-r^{2}d\theta^2-r^{2}\sin^{2}\theta\, d\phi^{2},\label{eq:SSdS}\\
A & = & 1-\frac{r_{s}}{r}-\frac13\,\CC r^2\equiv
1-\frac{r_{s}}{r}-\frac{r^{2}}{r_{dS}^2}\,,
\end{eqnarray}
where the Schwarzschild radius of a body of mass~$m$ is denoted
by~$r_{s}=2Gm$, and $r_{dS}$ is the \revisedtext{characteristic
size} of this finite universe (for a given value of $\CC$). Indeed,
for $r_s=0$ (or in practice for $r\gg r_{s}$) the above line element
corresponds to a $3$-dimensional sphere of radius
$r_{dS}=\sqrt{3/\CC}$. Obviously, for the observed value of $\CC\sim
H_0^2$ this is a very large radius of the order of the Hubble
horizon, $r_{dS}\approx H_{0}^{-1}$, and for $r$ much smaller than
this value the presence of the term $r^2/r_{dS}^2$ in the metric can
be neglected. Therefore, we have to insure that in our relaxation
model the effective $\CC$ remains within its observable value
$\CC\sim H_0^2$ also in the above SdS metric.

On the background metric (\ref{eq:SSdS}) the Ricci scalar, the
Gau\ss-Bonnet term and respectively the denominator
function~$B=\frac{2}{3}R^{2}+\frac{1}{2}\mathcal{G}$ are given by
\begin{equation}
 R=12r_{dS}^{-2},\,\,\,\,
 \mathcal{G}=24r_{dS}^{-4}+12r_{s}^{2}r^{-6},\,\,\,\,
 B=108\,r_{dS}^{-4}+6r_{s}^{2}r^{-6}.
 \label{eq:invariantsSS}
\end{equation}
\revisedtext{One can easily see} that those terms in
(\ref{eq:invariantsSS}) that do not depend on the presence of the
mass $m$ just follow from the corresponding cosmological results
(\ref{eq:invariants}) and (\ref{eq:Bapprox}) upon setting the
correspondences $H\to r_{dS}^{-1}$ and $q\to -1$, as expected from
the behavior in the de Sitter space limit.

In order to obtain the metric~(\ref{eq:SSdS}) as a solution of the
\revisedtext{generalized} Einstein's equations
(\ref{eq:Mod-Einstein-Eqs}) of our framework, we minimally extend
our action functional by \revisedtext{adding} a new term responsible
only for the CC relaxation in the Solar System. For instance,
consider the model
\begin{equation}
\mathcal{F}=\frac{\beta}{B}+\frac{\beta_{\odot}}{R}, \label{eq:BR}
\end{equation}
where the~$\beta/B$ part has been discussed before with
$\beta\sim\rho_{\Lambda}^{i}H_{0}^{4}$ given in
Eq.~(\ref{eq:estim}). With the help of Eq.~(\ref{eq:delSRST}) it is
possible to determine the extra contribution $E_{ab}^{\odot}$ to the
tensor~$E_{ab}$ in (\ref{eq:Mod-Einstein-Eqs}) following from the
term ~$\beta_{\odot}/R$ in ~$\mathcal{F}$ computed in the
metric~(\ref{eq:SSdS}). One finds
~$E_{ab}^{\odot}=g_{ab}(-\beta_{\odot}r_{dS}^{2}/16)$, and we will
show below that~$\beta/B$ induces a negligible correction in the
Solar System environment. In the Einstein
equations~(\ref{eq:Mod-Einstein-Eqs}) we safely neglect all terms
except~$E_{ab}^{\odot}$ and~$\rho_{\Lambda}^{i}$ in accordance to
our procedure in Sec.~\ref{sec:asympt-late-time}. This implies
$2E_{ab}^{\odot}+g_{ab}\rho_{\Lambda}^{i}=0$, with all corrections
suppressed by the large CC term~$\rho_{\Lambda}^{i}$. Consequently,
we find
$\beta_{\odot}\sim\rho_{\Lambda}^{i}r_{dS}^{-2}\sim\rho_{\Lambda}^{i}H_{0}^{2}$
for the Solar System parameter, which allows the comparison of both
terms in~$\mathcal{F}$. As a result, the condition
$|\beta/B|\ll|\beta_{\odot}/R|$ holds for radii~$r\ll r_{c}$ smaller
than the critical radius~$r_{c}=(r_{dS}^{2}r_{s})^{1/3}$, which is
sufficiently large due to $r_{dS}\sim H_{0}^{-1}$. Thus, neglecting
$\beta/B$ in~$\mathcal{F}$ is justified in this environment, and the
CC is successfully relaxed by the new term only. In contrast, in the
cosmological background the newly added term
$\beta_{\odot}/R\sim\rho_{\Lambda}^{i}(H_{0}/H)^{2}$ is much smaller
than the first term of (\ref{eq:BR}) in the matter era because we
have $\beta/B\sim\rho_{\Lambda}^{i}$ as a result of the relaxation
mechanism discussed in Sec.~\ref{sec:CC-relaxation}.
\revisedtext{Furthermore}, let us also remark that the late time
behavior will be controlled  once more by the original term
~$\beta/B\sim \beta/H^4 $ in the functional, due to its higher power
of~$H$ in the denominator. It means that the cosmological evolution
of the modified model (\ref{eq:BR}) will be essentially identical to
the original one for the recent past, present and future.

Finally, let us discuss the \revisedtext{potential} occurrence of
fifth forces coming from the scalar degree of freedom in the
$\mathcal{F}=\beta_{\odot}/R$ action functional. Since these forces
are part of the gravitational sector, they have to show up in the
metric that follows from solving the Einstein equations. In our
case, we have seen above that the Schwarzschild-de~Sitter metric
in~(\ref{eq:SSdS}) is a good approximate solution. Fifth forces can
only emerge from the corrections to that metric, which we found to
be strongly suppressed in the Solar System environment. Note, that
our gravitational action with a large CC term~$\rho_{\Lambda}^{i}$
differs significantly from the model
$\mathcal{F}\propto(R-\mu^{4}/R)$ from
Refs.~\cite{Capozziello:2003tk,Carroll:2003wy}, where the late-time
acceleration follows from an interplay of both terms
in~$\mathcal{F}$ in the absence of the CC. In the latter setup, the
parameter~$\mu\sim H_{0}$ is very small and the mass of the scalar
degree of freedom is of the same order of magnitude. Hence, it will
mediate a long-range fifth force which is in contrast to
observations~\cite{Chiba:2003ir}. \revisedtext{In our scenario}, the
parameter corresponding to~$\mu$ is $\beta_{\odot}$, which is much
larger than in the aforementioned situation (in fact,
$\beta_{\odot}\propto\rho_{\Lambda}^{i}\,H_0^2$) owing to the
relaxation condition operating on the big initial $\rLi$. One can
show that this also leads to a much larger value for the physical
scalar mass. For $\rho_{\Lambda}^{i}$ in the ballpark of a typical
GUT,
the range of the interaction would be very short~\cite{BSS:Solar}.
Of course, these arguments are consistent with Eq.~(\ref{eq:SSdS})
as a good solution in the Solar System.

In summary, the simple extension of the modified gravity action by
$\beta_{\odot}/R$ does not alter our previous results at the
cosmological level, but it represents a possible way to relax the CC
also in the Solar System environment without introducing long-range
fifth forces. A more detailed exposition of the last point will be
presented elsewhere~\cite{BSS:Solar}.

\newcommand{\JHEP}[3]{ {JHEP} {#1} (#2)  {#3}}
\newcommand{\NPB}[3]{{\sl Nucl. Phys. } {\bf B#1} (#2)  {#3}}
\newcommand{\NPPS}[3]{{\sl Nucl. Phys. Proc. Supp. } {\bf #1} (#2)  {#3}}
\newcommand{\PRD}[3]{{\sl Phys. Rev. } {\bf D#1} (#2)   {#3}}
\newcommand{\PLB}[3]{{\sl Phys. Lett. } {\bf B#1} (#2)  {#3}}
\newcommand{\EPJ}[3]{{\sl Eur. Phys. J } {\bf C#1} (#2)  {#3}}
\newcommand{\PR}[3]{{\sl Phys. Rep. } {\bf #1} (#2)  {#3}}
\newcommand{\RMP}[3]{{\sl Rev. Mod. Phys. } {\bf #1} (#2)  {#3}}
\newcommand{\IJMP}[3]{{\sl Int. J. of Mod. Phys. } {\bf #1} (#2)  {#3}}
\newcommand{\PRL}[3]{{\sl Phys. Rev. Lett. } {\bf #1} (#2) {#3}}
\newcommand{\ZFP}[3]{{\sl Zeitsch. f. Physik } {\bf C#1} (#2)  {#3}}
\newcommand{\MPLA}[3]{{\sl Mod. Phys. Lett. } {\bf A#1} (#2) {#3}}
\newcommand{\CQG}[3]{{\sl Class. Quant. Grav. } {\bf #1} (#2) {#3}}
\newcommand{\JCAP}[3]{{ JCAP} {\bf#1} (#2)  {#3}}
\newcommand{\APJ}[3]{{\sl Astrophys. J. } {\bf #1} (#2)  {#3}}
\newcommand{\AMJ}[3]{{\sl Astronom. J. } {\bf #1} (#2)  {#3}}
\newcommand{\APP}[3]{{\sl Astropart. Phys. } {\bf #1} (#2)  {#3}}
\newcommand{\AAP}[3]{{\sl Astron. Astrophys. } {\bf #1} (#2)  {#3}}
\newcommand{\MNRAS}[3]{{\sl Mon. Not. Roy. Astron. Soc.} {\bf #1} (#2)  {#3}}
\newcommand{\JPA}[3]{{\sl J. Phys. A: Math. Theor.} {\bf #1} (#2)  {#3}}
\newcommand{\ProgS}[3]{{\sl Prog. Theor. Phys. Supp.} {\bf #1} (#2)  {#3}}
\newcommand{\APJS}[3]{{\sl Astrophys. J. Supl.} {\bf #1} (#2)  {#3}}

\newcommand{\Prog}[3]{{\sl Prog. Theor. Phys.} {\bf #1}  (#2) {#3}}
\newcommand{\IJMPA}[3]{{\sl Int. J. of Mod. Phys. A} {\bf #1}  {(#2)} {#3}}
\newcommand{\IJMPD}[3]{{\sl Int. J. of Mod. Phys. D} {\bf #1}  {(#2)} {#3}}
\newcommand{\GRG}[3]{{\sl Gen. Rel. Grav.} {\bf #1}  {(#2)} {#3}}




\begin{thebibliography}{50}

\bibitem{cosmdata}  D.N.~Spergel
\textit{et al.}, \APJS {170}{2007}{377}.

\bibitem{SNIa} R.~Knop \textit{ et al.}, \APJ {598}{2003}{102};
A.~Riess \textit{et al.} \APJ {607}{2004}{665};

\bibitem{weinberg89} S.~Weinberg, \RMP {61}{1989}{1}.

\bibitem{CCproblem} P.J.E.~Peebles and B.~Ratra, \RMP {75}{2003}{559};
T.~Padmanabhan, \PR {380}{2003}{235}; \, V. Sahni, A. Starobinsky,
\IJMP {A9} {2000} {373}; S.M. Carroll, \textsl{Living Rev.  Rel.}
{\bf 4} (2001) 1; E.J. Copeland, M. Sami, S. Tsujikawa, \IJMP {D15}
{2006} {1753}.

\bibitem{Zeldovich67} \newtext{Y. B. Zeldovich}, \textit{Cosmological constant and elementary
particles}, Sov. Phys. JETP Lett {\bf 6} (1967) 3167.

\bibitem{Steinhardt} P.J.~Steinhardt, \textit{Cosmological Challenges for the
21st Century},  in: \textit{Critical Problems in Physics}, edited by
V.L. Fitch, D.R. Marlow and M.A.E. Dementi (Princeton Univ. Pr.,
Princeton, 1997); P.J. Steinhardt, \textit{Phil. Trans. Roy. Soc.
Lond.} {\bf A361} (2003) 2497.

\bibitem{Perivolaropoulos08a} \newtext{L. Perivolaropoulos}, \textit{Six Puzzles for LCDM
Cosmology}, e-Print: arXiv:0811.4684 [astro-ph].


\bibitem{SUSY} \newtext{P. Fayet} and  S. Ferrara,
\PR {32}{1977}{249}; H.P Nilles, \PR{110}{1984}{1}; H.E. Haber and
G.L. Kane, \PR{117}{1985}{75}.

\bibitem{WZ74} \newtext{J. Wess} and B. Zumino, \NPB {70}{1974}{39}.

\bibitem{Dine85} M. Dine, R. Rohm, N. Seiberg and E. Witten, \PLB
{156}{1985}{55}.

\bibitem{Susskind03} L. Susskind, \textit{The anthropic landscape of string theory},
in: \textit{Universe or multiverse?}, B. Carr ed., Cambridge
University Press (2007), pg. 247-266, hep-th/0302219.

\bibitem{OldScalar} A.D.~Dolgov, \textit{An Attempt To Get Rid Of The
Cosmological Constant}, in: \textit{The very Early Universe}, Ed.
G.~Gibbons, S.W.~Hawking, S.T.~Tiklos (Cambridge U., 1982);
\newtext{L.F. Abbott}, \PLB {150}{1985}{427}; L.H.~Ford, \PRD
{35}{1987}{2339}; R.D.~Peccei, J.~Sol\`{a} and C.~Wetterich, \PLB
{195}{1987}{183}; S. M. Barr, \PRD {36}{1987}{1691}; S. M. Barr and
D. Hochberg, \PLB {211}{1988}{49}; J.~Sol\`{a}, \PLB
{228}{1989}{317}; \IJMP {A5}{1990}{4225}.

\bibitem{Quintessence} C.~Wetterich, \NPB
{302}{1988}{668}; P.J.E.~Peebles and B.~Ratra, \APJ
{325}{1988}{L17}; B.~Ratra and P.J.E.~Peebles, \PRD
{37}{1988}{3406}; P.G.~Ferreira and M.~Joyce, \PRD
{58}{1998}{023503}; R.R.~Caldwell, R.~Dave and P.J.~Steinhardt, \PRL
{80}{1998}{1582}; P.J.~Steinhardt, L.M.~Wang and I.~Zlatev,
\PRD{59}{1999}{123504}; V.~Sahni and L.M.~Wang, \PRD
{62}{2000}{103517}.

\bibitem{ShapSol09} I. L. Shapiro,  J. Sol\`a, \PLB{682}{2009}{105}, arXiv:0910.4925; see also the detailed review arXiv:0808.0315
[hep-th] on the quantum field theory of the CC term, and references
therein.

\bibitem{BFLWard09}
B.F.L. Ward, Mod. Phys. Lett. {\bf A25} (2010) 607, arXiv:0908.1764
[hep-ph],  and arXiv:0910.0490; Int. J. Mod. Phys. {\bf D17} (2008)
627, hep-ph/0610232; and Mod. Phys. Lett. {\bf A23} (2008) 3299,
arXiv:0808.3124 [gr-qc].


\bibitem{oldCCstuff1} I.L. Shapiro, J.~Sol\`a, \textit{Phys. Lett.} {\bf 475B} (2000)
236, hep-ph/9910462; JHEP {\bf 02} (2002) 006, hep-th/0012227;
\textit{Nucl. Phys. Proc. Suppl.} {\bf 127} (2004) 71,
hep-ph/0305279.


\bibitem{oldCCstuff2} A. Babic, B. Guberina, R. Horvat, H. \v{S}tefan\v{c}i\'{c},
\textit{Phys. Rev.} {\bf D65} (2002) 085002; B. Guberina, R. Horvat,
H. \v{S}tefan\v{c}i\'{c}, \textit{Phys. Rev.} {\bf D67} (2003)
083001; I.L. Shapiro, J. Sol\`{a}, H. \v{S}tefan\v{c}i\'{c}, JCAP
{\bf 0501} (2005) 012, hep-ph/0410095; F.~Bauer, Class.\ Quant.\
Grav.\ {\bf 22} (2005) 3533; F. Bauer, Ph.d. Thesis, hep-th/0610178;
gr-qc/0512007;  \newtext{J. Fabris}, I.L. Shapiro, J. Sol\`a,\,
\JCAP {02}{2007} {016}, gr-qc/0609017.

\bibitem{newCCstuff} \newtext{J. Grande}, J. Sol\`a, J. C. Fabris, I. L. Shapiro, \textit{Class.
Quant. Grav.} {\bf 27} (2010) 105004, arXiv:1001.0259.


\bibitem{ShapSol0608} I. L. Shapiro,  J. Sol\`a, J. Phys. {\bf A40} (2007) 6583, gr-qc/0611055, and references therein.

\bibitem{oldvarCC1}
O. Bertolami, \textit{Nuovo Cimento}, {\bf 93B}, 36, (1986); M. Ozer
M. and O. Taha, Nucl. Phys., {\bf B287}, 776, (1987); O. K. Freese
K., et al., \textit{Nucl. Phys}., {\bf 287}, 797, (1987); J. C.
Carvalho, J. A. S. Lima and  I. Waga, \textit{Phys. Rev}.  {\bf
D46}, 2404, (1992).

\bibitem{overduin98} J. M. Overduin and F. I. Cooperstock, Phys. Rev. D.,
{\bf 58}, 043506, (1998), and references therein.

\bibitem{BPS09a}  S. Basilakos, M. Plionis and J. Sol\`a, \textit{Phys. Rev.} {\bf
D80} (2009) 083511, arXiv:0907.4555 [astro-ph.CO].

\bibitem{CCfit}
I.L. Shapiro, J. Sol\`{a}, C. Espa\~{n}a-Bonet, P. Ruiz-Lapuente,
Phys. Lett. {\bf 574B} (2003) 149; JCAP {\bf 0402} (2004) 006; I.L.
Shapiro, J. Sol\`{a}, JHEP proc. AHEP2003/013, {astro-ph/0401015}.

\bibitem{LXCDM} J.~Grande, J.~Sol\`a and H.~\v{S}tefan\v{c}i\'{c},
\JCAP {08}{2006} {011}, gr-qc/0604057.


\bibitem{LXCDM2} J.~Grande, J.~Sol\`a and H.~\v{S}tefan\v{c}i\'{c}, \PLB
{645}{2007}{236}, gr-qc/0609083.

\bibitem{LXCDMmore} J.~Grande, J.~Sol\`a and H.~\v{S}tefan\v{c}i\'{c},
J. Phys. A {40}\,{(2007)}\,{6787}; J. Grande, R. Opher, A. Pelinson
and J. Sol\`a, \JCAP {0712}{2007}{007}, arXiv:0709.2130 [gr-qc]; J.
Grande, A. Pelinson, J. Sol\`{a}, \PRD{79}{2009}{043006},
arXiv:0809.3462 [astro-ph]; arXiv:0904.3293 [astro-ph.CO].

\bibitem{Stefancic08} H.~\v{S}tefan\v{c}i\'{c}, \PLB {670}{2009}{246}.

\bibitem{BSS09a} F.~Bauer, J.~Sol\`{a}, H.~\v{S}tefan\v{c}i\'{c}, Phys. Lett.  {\bf B 678} (2009)
427, arXiv:0902.2215.

\bibitem{Odin} S.~Nojiri, S.~D.~Odintsov, Phys.\ Rev.\ D {\bf 72} (2005) 023003.

\bibitem{Barr:2006mp}
  S.~M.~Barr, S.~P.~Ng and R.~J.~Scherrer,
  \textit{Phys.\ Rev}.\   {\bf D73} (2006) 063530.

\bibitem{Batra:2008cc} P. Batra, K. Hinterbichler, L. Hui, D.N. Kabat, \PRD{78}{2008}{043507}.

\bibitem{Demir09} D.A. Demir, Found. Phys. 39 (2009) 1407; 
J. Bernab\'eu, C. Espinoza, N. E. Mavromatos, Phys.\ Rev.\  D {\bf 81} (2010) 084002; 
J. Beltran Jim\'enez, A. L. Maroto, \JCAP {0903} {2009} {016},
arXiv:0811.0566 [astro-ph].

\bibitem{Maggiore10}  \newtext{M. Maggiore}, arXiv:1004.1782
[astro-ph.CO]; N. Bilic, arXiv:1004.4984 [hep-th].

\bibitem{Bauer2009} F.~Bauer, Class.\ Quant.\ Grav.\  {\bf 27} (2010)
055001
  [arXiv:0909.2237 [gr-qc]].

\bibitem{BSS09b}
  F.~Bauer, J.~Sol\`{a}, H.~\v{S}tefan\v{c}i\'{c}, \PLB {688}{2010}{269}, arXiv:0912.0677
  [hep-th].


\bibitem{SotiriouFaraoni08}
  S. Nojiri and S.D. Odintsov,
  eConf {\bf C0602061} (2006) 06
  [Int.\ J.\ Geom.\ Meth.\ Mod.\ Phys.\  {\bf 4} (2007) 115];
 T.P. Sotiriou, V. Faraoni, \RMP {82}{2010}{451}, arXiv:0805.1726 [gr-qc];
R. Woodard, \textit{Lect. Notes  Phys.} 720 (2007) 403.

\bibitem{Polarski} L. Amendola, D. Polarski, S. Tsujikawa, \PRL
{98}{2007}{131302}, astro-ph/0603703; L. Amendola, R. Gannouji, D.
Polarski, S. Tsujikawa, \PRD {75}{2007}{083504}, gr-qc/0612180.

\bibitem{Parker09} \newtext{L.E. Parker} and D.J. Toms, \textit{Quantum Field
Therory in Curved Spacetime: quantized fields and gravity}
(Cambridge U. Press, 2009).

\bibitem{Shapiro:2008sf} I. L. Shapiro, Class. Quant. Grav. {\bf 25}, 103001,
(2008).

\bibitem{Starobinsky80} A.A. Starobinski, \PLB {91} {1980} {99};
A. Vilenkin, \PRD {32} {1985} {2511}.

\bibitem{AII1} I. Antoniadis, E. Mottola, \PRD
{45}{1992}{2013}; I. Antoniadis, P.O. Mazur, E. Mottola, \PLB {394}
{1997} {49}; \PRD {55}{1997}{4770}; \PLB {444}{1998}{284}; I.
Antoniadis, E. Mottola, \textit{New J.\ Phys.}\, {\bf 9} (2007) 11,
\texttt{arXiv:gr-qc/0612068}, and references therein.

\bibitem{AII2} I.L. Shapiro, J. Sol\`a, \PLB {530}{2002}{10},
\texttt{hep-ph/0104182}; A.M. Pelinson, I.L. Shapiro, F.I. Takakura,
\NPB{648}{2003}{417}; N. Bilic, B. Guberina, R. Horvat, H. Nikolic,
H.~\v{S}tefan\v{c}i\'{c}, \PLB {657}{2007}{232}.

\bibitem{Fossil07} J. Sol\`a, {J. of Phys.} {\bf A41} {(2008)} {164066},
arXiv:0710.4151 [hep-th].

\bibitem{Carroll04} S.M. Carroll, V. Duvvuri, M. Trodden, M.S. Turner,  \PRD{70}{2004}{043528};
S. Nojiri, S.D. Odintsov, \PRD{68}{2003}{123512}; S.M. Carroll, A.
de Felice, V. Duvvuri, D.A. Easson, M. Trodden, M.S. Turner, \PRD
{71} {2005} 063513.

\bibitem{Hindawi:1995cu} A.~Hindawi, B.A.~Ovrut, D.~Waldram, Phys.\ Rev.\ {\bf D53} (1996) 5597;
A.~De Felice, M.~Hindmarsh, M.~Trodden, JCAP {\bf 0608} (2006) 005.

\bibitem{Nojiri:2005jg} S. Nojiri and S. D. Odintsov, \PLB {631}{2005}{1},
hep-th/0508049; G. Cognola \textit{et al.}, \PRD {73}{2006}{084007},
hep-th/0601008; \PRD {75}{2007}{086002}, hep-th/0611198.

\bibitem{Comelli05} I. Navarro
and K. van Acoleyen, \JCAP {0603}{2006}{008}, gr-qc/0511045; D.
Comelli, \PRD {72}{2005}{064018}, gr-qc/0505088.

\bibitem{BambaOdintsov} K. Bamba, S. D. Odintsov, L. Sebastiani and S. Zerbini,
\textit{Eur. Phys. J} {\bf C67} (2010) 295, arXiv:0911.4390
[hep-th].

\bibitem{SS12} J. Sol\`a, H. \v{S}tefan\v{c}i\'{c}, \MPLA {21} {2006}
{479}, astro-ph/0507110; \PLB {624}{2005}{147}, astro-ph/0505133.

\bibitem{Quartin:2008px}
  M.~Quartin, M.~O.~Calvao, S.~E.~Joras, R.~R.~R.~Reis and I.~Waga,
  JCAP {\bf 0805} (2008) 007.


\bibitem{PSW} R.D.~Peccei, J.~Sol\`{a} and C.~Wetterich, \textit{ Adjusting the Cosmological
Constant Dynamically: Cosmons and a New Force Weaker Than Gravity.} \PLB
{195}{1987}{183}.


\bibitem{BPS2010a} S. Basilakos, M. Plionis and J. Sol\`a,
arXiv:1005.5592 [astro-ph.CO].


\bibitem{MTW} C.W. Misner, K.S. Thorn, J.A. Wheeler, \textit{Gravitation}
(Freeman, San Francisco, 1973).

\bibitem{Coleman85} \newtext{S.R Coleman}, \textit{Aspects of Symmetry}
(Cambridge U. Press, 1985); P. Ramond, \textit{Field Theory. A
Modern Primer} (Benjamin/Cummings Publ. Comp., 1981); L. H. Ryder,
\textit{Quantum Field Theory} (Cambridge U. Press, 1985).


\bibitem{Capozziello:2003tk}
  S.~Capozziello, S.~Carloni and A.~Troisi,
  Recent Res.\ Dev.\ Astron.\ Astrophys.\  {\bf 1} (2003) 625
  [arXiv:astro-ph/0303041].
\bibitem{Carroll:2003wy}
  S.~M.~Carroll, V.~Duvvuri, M.~Trodden and M.~S.~Turner,
  Phys.\ Rev.\  D {\bf 70} (2004) 043528

\bibitem{BSS:Solar} F.~Bauer, J.~Sol\`{a}, H.~\v{S}tefan\v{c}i\'{c}, {\it in preparation}.


\bibitem{Chiba:2003ir}
  T.~Chiba,
  Phys.\ Lett.\  B {\bf 575} (2003) 1





\end{thebibliography}
\end{document}